\documentclass[lettersize,journal]{IEEEtran}
\usepackage{mathtools}
\usepackage{amsmath,amsfonts}
\usepackage{algorithm}
\usepackage{array}
\usepackage[caption=false,font=normalsize,labelfont=sf,textfont=sf]{subfig}
\makeatletter
\newcommand{\MakeTitleInner}[3]{%
    \begin{center}
        {\fontsize{24pt}{28pt}\selectfont #1 \par}  
        \vspace{0.5cm}
        {\fontsize{11pt}{13pt}\selectfont #2 \par}
        \vspace{0.3cm}
        {\fontsize{9pt}{11pt}\selectfont #3 \par}
    \end{center}
    \par
    \vskip 1.5em
}

\newcommand{\suppmaketitle}{%
    \MakeTitleInner{Supplementary Materials: Structure-from-Sherds++: Robust Incremental 3D Reassembly of Axially Symmetric Pots from Unordered and Mixed Fragment Collections}%
    {Seong Jong Yoo, \textit{Student Member, IEEE}, Sisung Liu, \textit{Student Member, IEEE}, Muhammad Zeeshan Arshad, Jinhyeok Kim, Young Min Kim, Yiannis Aloimonos, Cornelia Fermüller, Kyungdon Joo, Jinwook Kim, and Je Hyeong Hong, \textit{Member, IEEE}}%
    {}
}
\makeatother

\usepackage{balance}
\usepackage{textcomp}
\usepackage{stfloats}
\usepackage{url}
\usepackage{verbatim}
\usepackage{graphicx}
\usepackage{epsfig}
\usepackage{amssymb}
\usepackage{multirow}
\usepackage{makecell}
\usepackage{bbding}
\usepackage{color}
\usepackage{enumitem}
\usepackage{array}
\usepackage{algpseudocode}
\usepackage[symbol]{footmisc}
\usepackage[accsupp]{axessibility}

\usepackage{cleveref}
\usepackage{subfiles}
\usepackage[numbers]{natbib}
\usepackage[dvipsnames]{xcolor}
\usepackage{diagbox}
\usepackage{microtype}

\algrenewcommand\algorithmicindent{0.75em}

\def\eqref#1{(\ref{eq:#1})}
\def\eqlabel#1{\label{eq:#1}}
\def\figref#1{\ref{fig:#1}}
\def\figlabel#1{\label{fig:#1}}
\def\pparagraph#1{\par{\bf #1}~~}

\def\x{{\mathbf x}}

\definecolor{DarkGreen}{rgb}{0.1,0.6,0.1}
\def\check c{{\color{DarkGreen}\checkmark}}
\def\x x{{\color{red}x}}

\def\eqref#1{(\ref{eq:#1})}
\def\eqlabel#1{\label{eq:#1}}
\def\figref#1{\ref{fig:#1}}
\def\figlabel#1{\label{fig:#1}}
\def\pparagraph#1{\par{\bf #1}~~}

\def\xcomment#1{\textcolor[rgb]{.3,.3,.1}{\text{$/\!\!/$ {\em #1}}}}
\def\comment#1{\kern-1cm\xcomment{#1}}
\def\eqcomment#1{\kern-1cm\xcomment{#1}}

\def\m#1{\ensuremath{\mathtt{#1}}}

\def\v#1{\ensuremath{\mathbf{#1}}}

\DeclareMathOperator*{\argmin}{arg\,min}

\def\real{\mathbb{R}}
\def\tr{^\top}

\def\norm#1{\left\lVert#1\right\rVert}

\def\l2#1{\norm{#1}_2}

\newcommand{\AlgName}{SfS++ }

\begin{document}

\title{Structure-from-Sherds++: Robust Incremental 3D Reassembly of Axially Symmetric Pots from Unordered and Mixed Fragment Collections}

\author{
    Seong Jong Yoo,~\IEEEmembership{Student Member,~IEEE,}
    Sisung Liu,~\IEEEmembership{Student Member,~IEEE,}
    Muhammad Zeeshan Arshad,
    Jinhyeok Kim,
    Young Min Kim,
    Yiannis Aloimonos,
    Cornelia Ferm\"{u}ller, 
    Kyungdon Joo,
    Jinwook Kim,
    and Je Hyeong Hong,~\IEEEmembership{Member,~IEEE,}
    \thanks{Manuscript received xxx xx, 20xx.}    
    \thanks{
        S.J. Yoo, Y. Aloimonos, and C. Ferm\"{u}ller are with the Department of Computer Science, University of Maryland, College Park, MD, USA.
    }
    \thanks{
        S. Liu is with the Department of Artificial Intelligence, Hanyang University, South Korea.
    }
    \thanks{
        M.Z. Arshad is with the Faculty of Health Sciences, University of Ottawa, ON, Canada.
    }
    \thanks{
    J.H. Kim is with with the Artificial Intelligence Graduate School, UNIST, South Korea.
    }
    \thanks{
        Y.M. Kim is with the Department of Electrical and Computer Engineering, Seoul National University, South Korea.
    }
    \thanks{
         J.W. Kim is with the Intelligence and Interaction Research Center, Korea Institute of Science and Technology, South Korea.
    }
    \thanks{
        K. Joo is with the Artificial Intelligence Graduate School and the Department of Computer Science and Engineering, UNIST, South Korea.
    }
    \thanks{
        J.H. Hong is with the Department of Electronic Engineering, Hanyang University, South Korea.
    }    
    \thanks{
        Seong Jong Yoo and Sisung Liu contributed equally to this work.
    }
    \thanks{
        Correspondence to J.W. Kim (jinwook.kim21@gmail.com) and J.H. Hong (jhh37@hanyang.ac.kr).
    }

}

\maketitle

\begin{abstract}
Reassembling multiple axially symmetric pots from fragmentary sherds is crucial for cultural heritage preservation, yet it poses significant challenges due to thin and sharp fracture surfaces that generate numerous false positive matches and hinder large-scale puzzle solving. 
Existing global approaches, which optimize all potential fragment pairs simultaneously or data-driven models, are prone to local minima and face scalability issues when multiple pots are intermixed.
Motivated by Structure-from-Motion (SfM) for 3D reconstruction from multiple images, we propose an efficient reassembly method for axially symmetric pots based on iterative registration of one sherd at a time, called Structure-from-Sherds++ (SfS++). 
Our method extends beyond simple replication of incremental SfM and leverages multi-graph beam search to explore multiple registration paths. 
This allows us to effectively filter out indistinguishable false matches and simultaneously reconstruct multiple pots without requiring prior information such as base or the number of mixed objects.
Our approach achieves 87\% reassembly accuracy on a dataset of 142 real fragments from 10 different pots, outperforming other methods in handling complex fracture patterns with mixed datasets and achieving state-of-the-art performance.
Code and results can be found in our project page \url{https://sj-yoo.info/sfs/}.

\end{abstract}


\IEEEdisplaynontitleabstractindextext
\IEEEpeerreviewmaketitle

\section{Introduction}\label{sec:introduction}

\IEEEPARstart{R}{estoring} ancient ceramic pots is essential for understanding human history, including insights into past lifestyles~\cite{graffArchaeologicalStudiesCooking2018}, and economies~\cite{leitchReconstructingHistoryPottery2013}.
However, they are often unearthed as fragmented pieces, or sherds, dispersed across excavation sites, intermingled with other artifacts, and degraded over time.
Consequently, pottery restoration requires multiple complex processes to recover valuable historical records.

Despite significant efforts by restoration experts—who are equipped with knowledge of geometric features and historical context—reconstructing even a single pot can take days of meticulous work.
This challenge is compounded when multiple axially symmetric pots are intermixed. 
Pieces from different pots can be indistinguishable at first glance, making the trial-and-error approach time-consuming and risky, as repeated handling can cause undesired abrasions on the fracture surfaces. 
These difficulties underscore the importance of developing an efficient framework capable of virtually reassembling multiple pots at once~\cite {diangeloReviewComputerbasedMethods2022, yePuzzleFixerVisualReassembly2023}.
\IEEEpubidadjcol

\begin{figure}[t]
	\centering
	\includegraphics[width=\linewidth]{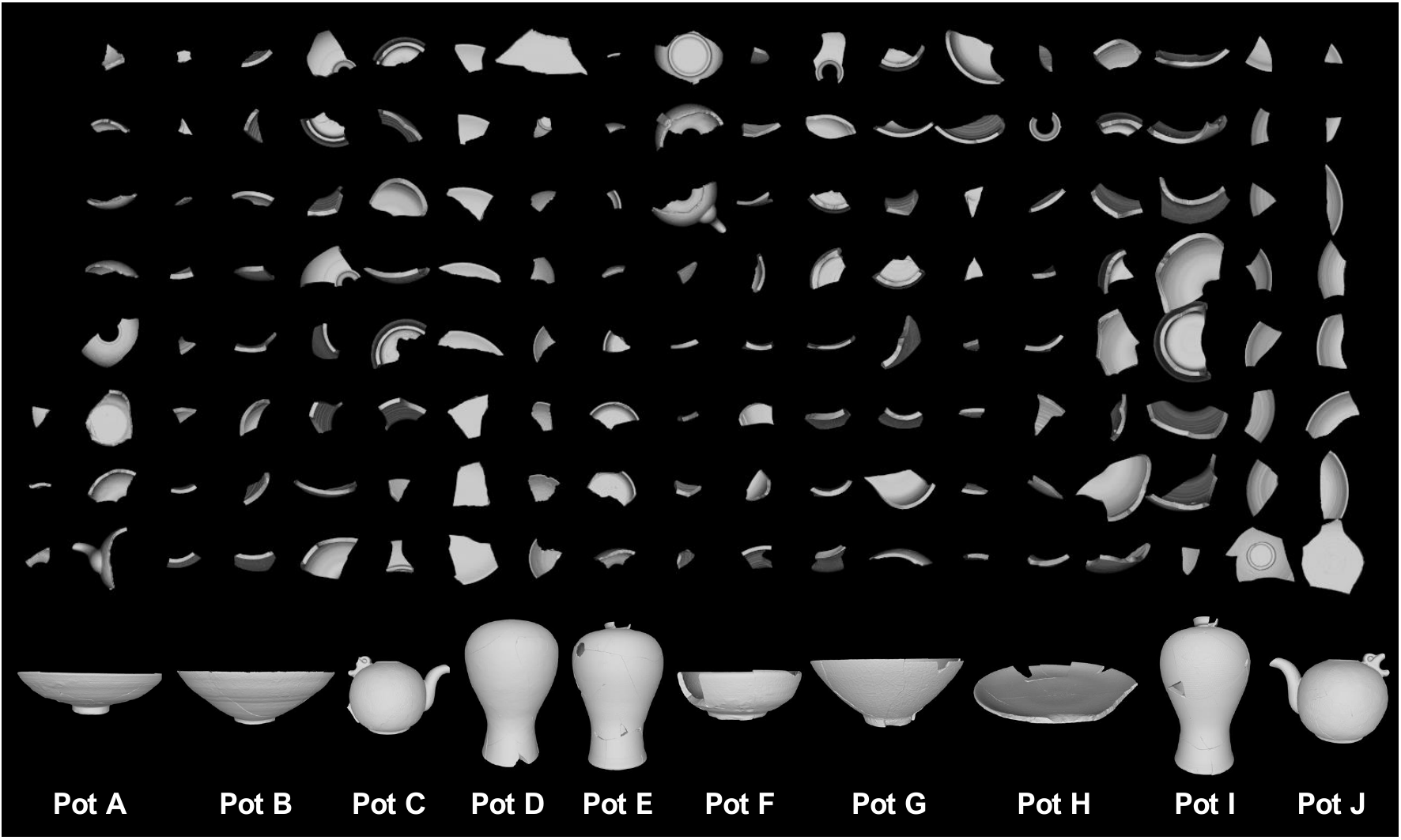}
	\caption{
	    An illustration of 10 real pots simultaneously reassembled from 142 unclustered sherds  (3D  point  clouds) using the proposed SfS++ method. The dataset includes two pairs of identical but different broken potteries \textemdash Pot (E, I) and Pot (C, J). \AlgName successfully reassembled the mixed 10 potteries with 87\% accuracy, as displayed in this figure.
	}
    \vspace{-3mm}
	\figlabel{teaser}
 
\end{figure}
\begin{figure}[t]
	\centering
	\includegraphics[width=0.95\linewidth]{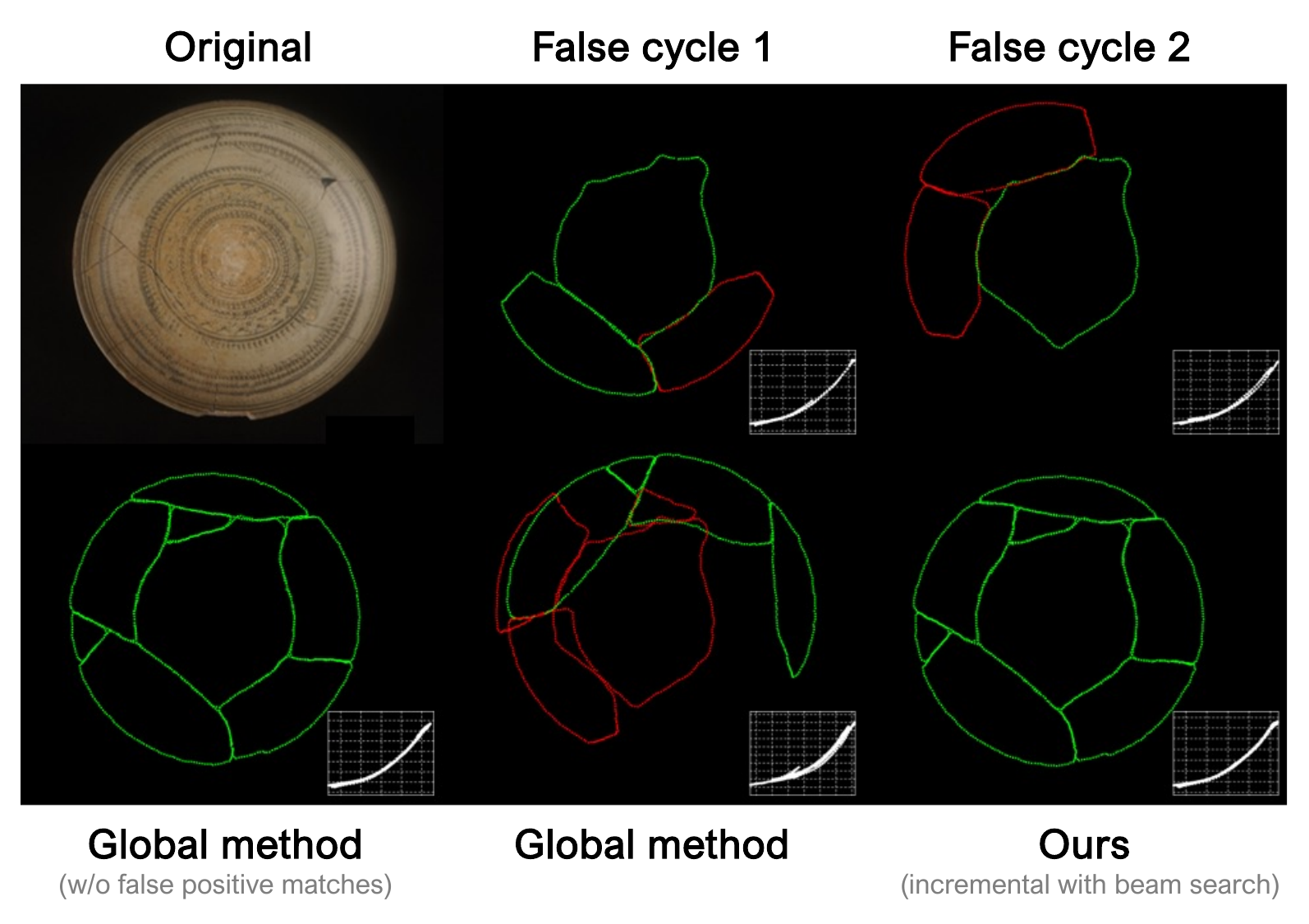}
	\caption{
        Failure cases of optimization-based global method (re-implementation of ~\citet{son13}).
	    The top row shows some exemplary false positive loops formed in pot A from Fig~\figref{teaser}. 
	    (True configurations in green and false in red).
	    Some pairwise matches are difficult to detect even at the cycle level just by looking at the edge lines and the axis profile curve (in white).
	    The global method only succeeds when false positive matches are removed.
	    On the other hand, our incremental method with beam search outputs the correct result despite false matches.
	}
	\label{fig:global_fails}
 \vspace{-3mm}
\end{figure}

Over the past decades, research in pottery restoration has generally advanced along two primary paths: optimization-based global methods and data-driven approaches (see Table~\ref{tbl:comparison}).
Optimization-based global methods~\cite{son13,papaioannou17, 9412372} typically follow a three-stage process. 
\textit{First}, extracting features from each pot sherd; \textit{second},  identifying  potential matches between each pair of sherds; and \textit{last}, solving global combinatorial optimization to determine the most likely 3D pot configuration. 
This final step simultaneously evaluates all edges in the global sherds configuration graph, allowing the identification of true positive matches.
Meanwhile, data-driven approaches~\cite{joo24, jigsaw, wang2024puzzlefusionpp} treat pottery restoration as a 3D puzzle assembly task, using deep learning models trained on large datasets to learn fragment features and matching strategies.

\begin{figure}[t]
	\centering
    \captionsetup[subfloat]{labelfont=footnotesize,textfont=footnotesize}
	\subfloat[][]{\includegraphics[width=0.25\linewidth]{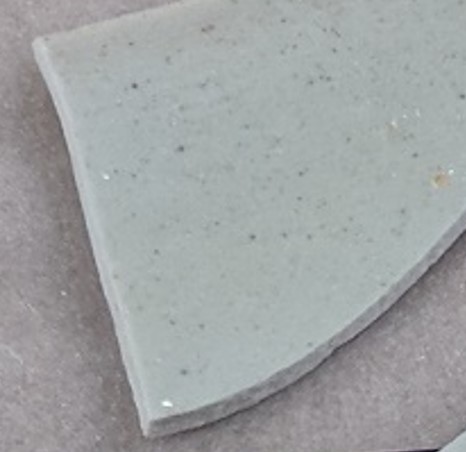}\label{sharp_edges}}
	~
	\subfloat[][]{\includegraphics[width=0.25\linewidth]{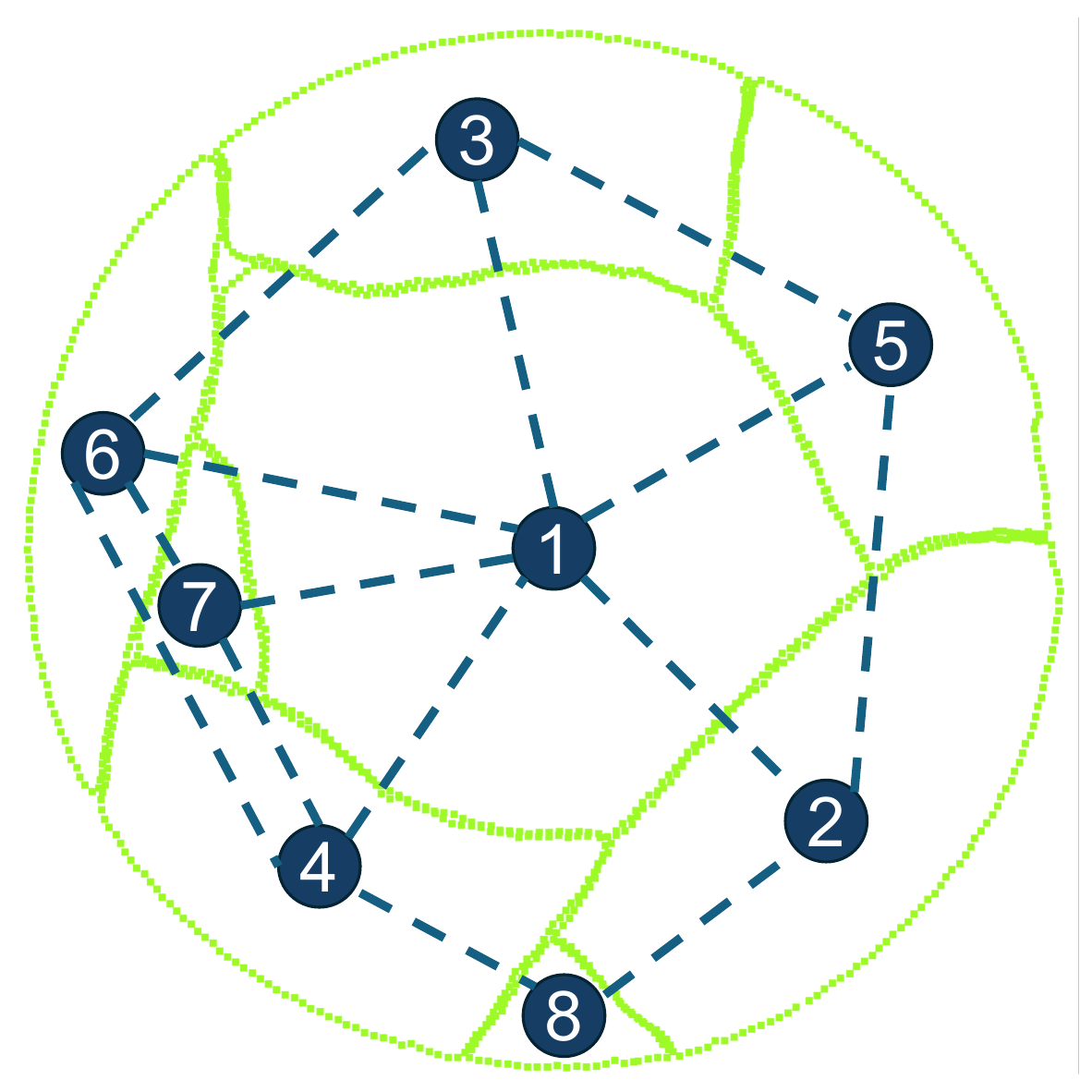}\label{intra_connectivity}}
	\subfloat[][]{\includegraphics[width=0.25\linewidth]{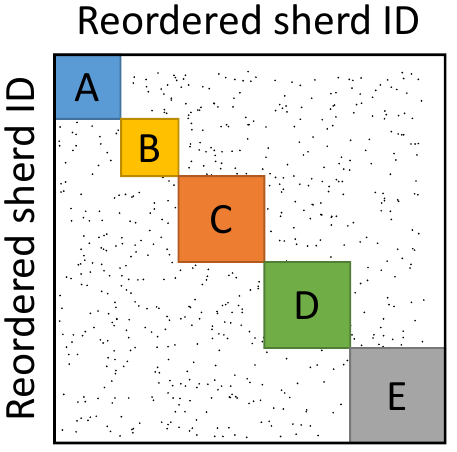}\label{inter_connectivity}}	
	\caption{
		Challenges in reassembling multiple axially symmetric pots.
		(a) Sharp, thin fracture surfaces with little distinct features often lead to false positive sherd matches. 
        (b)Within a single pot (e.g. Pot A), only physically adjacent sherd pairs are truly connected, resulting in a sparse connectivity graph with only 59\% valid connections among eight sherds.
		(c) The noisy block-diagonal sparsity structure when sherds from different pots are mixed, as discussed in Sec.~\ref{sec:challenges}. Only diagonal components are effective (colored rectangles) and otherwise are considered as false positives (black dots).
	}
	\label{fig:challenges}
 \vspace{-3mm}
\end{figure}

However, both approaches face notable limitations, particularly in handling false positive matches.
In optimization-based global approaches, these errors propagate through the reconstruction, as each sherd's placement relies on these matches.
Similarly, data-driven approaches struggle with false positives due to the challenging nature of discriminating between similar fracture patterns, even when extensive training data is available. 
These false positives are fundamentally inevitable due to the lack of distinctive features on sharply broken fracture surfaces (see Fig.~\ref{fig:challenges}(a)).

Another key challenge stems from the fundamental connectivity limitations among pottery sherds: each valid pair must be physically adjacent, resulting in a sparse connectivity graph for each pot (Fig.~\ref{fig:challenges}(b)).
When multiple pots are mixed, the overall connectivity graph typically splits into block-diagonal subgraphs—each block corresponding to a different pot—and remains noisy with numerous false positive matches (Fig.~\ref{fig:challenges}(c)).
The aforementioned issues become more problematic as the number of sherds increases, leading to an exponential rise in false positive pairs. 
Furthermore, existing approaches often assume a single-object input, which not only imposes a classification burden on other architectures but also compromises reconstruction robustness in noisy scenario, such as when sherds from different pots are included.

Recognizing these limitations, we turn our attention to Structure-from-Motion (SfM), which has achieved tremendous success in reconstructing large-scale 3D scenes from thousands of images.
The core of SfM's success lies in its incremental approach, which iteratively registers new camera views, refines pairwise matches, and robustly discovers initially undetected correspondences while reducing false positives.

Inspired by above insights, we propose an optimization-based Structure-from-Sherds++ (SfS++), specifically designed for reassembling axially symmetric pots from fragments.
Our approach not only adopts the incremental concept but also introduces features tailored to pot reassembly, enhancing algorithm's robustness. Key component, called BAISER (Base-Agnostic Incremental ShErd Registration) module, includes the use of beam search to address indistinguishable false matches and multi-graph expansions to enable the simultaneous reconstruction of multiple pots. As a result, we achieve the first successful simultaneous reassembly of ten mixed, unordered potteries, as demonstrated in Fig.~\figref{teaser}.
The key contributions of our work are:
\begin{itemize}[noitemsep,topsep=0pt]
    \item[+]
    a new SfM-inspired pipeline for simultaneous reassembly of multiple potteries from 3D scans of ceramic fragments,
    \item[+]
    design of individual modules such as the axis-based edge line descriptor, mesh processing pipeline, and multi-geometric verification using rim, profile curve, and thickness to reduce false positive matches prevalent in the task of reassembling multiple potteries,
    \item[+]
    an approach called \emph{base-agnostic incremental sherd registration} (BAISER) that goes beyond replicating incremental SfM and leverages multi-graph beam search with graph merging tailored for reconstruction of multiple pots from unordered and mixed fragments, and
    \item[+]
    extensive experimental evaluations on our publicly-released dataset of 142 real fragments from 10 different pots and the public benchmark called Breaking Bad~\cite{breakingbad}, comparing against other state-of-the-art methods.
\end{itemize}

This work extends our previous conference paper on Structure-from-Sherds (SfS)~\cite{iccvsfs}.
We addressed the issue that prior work heavily relies on the accuracy of base assembly, leading to complete reconstruction failure when the base assembly is inaccurate, by introducing a robust \textbf{multi-graph beam search with graph merging} that enables reconstruction without depending on the base.
We also \textbf{improved the mesh processing pipeline} to enhance surface extraction accuracy.
For experimental validation, we further \textbf{expanded our dataset} from 80 sherds of 5 different pots to 142 sherds from 10 different pots.
Additionally, we have evaluated \AlgName not only on our dataset but also on the Breaking Bad synthetic dataset~\cite{breakingbad}, comparing it against other learning-based SoTA methods.

\begin{table*}[t]
\centering
\caption{Comparison of representative 3D fragment reassembly approaches. 
While several methods are specialized for pottery restoration, others target more general 3D puzzle tasks.}
\resizebox{2\columnwidth}{!}{
\begin{tabular}{l|l|cccc}
\Xhline{1.pt}
            \textbf{Approach}    &    \textbf{Models}  & \textbf{Real dataset} & \textbf{Multi-object}  & \textbf{Key characteristics} & \textbf{Limitations} \\ 
            \hline
\multirow{6}{*}{\makecell[c]{Optimization\\ (Incremental)}}   

& \multirow{2}{*}{Willis \& Cooper~\cite{willis04}}& \multirow{2}{*}{\check c{}}      & \multirow{2}{*}{\x x{}} &
\makecell[tl]{
\textbullet{} Extraction of T-junctions \\
\textbullet{} Matching of edge lines, normals, and profiles} 
& \makecell[tl]{
\textbullet{} {Requires the assumption of axial symmetry} \\
\textbullet{} Absence of backtracking} \\ \cline{2-6} 

& \multirow{2}{*}{Huang et al.~\cite{huang06}}  & \multirow{2}{*}{\check c{}} & \multirow{2}{*}{\x x{}} & 
\makecell[tl]{
\textbullet{} Matching of break-surface features \\ 
\textbullet{}  Penetration-free local alignment}   & 
\makecell[tl]{\textbullet{} Reliance on feature-rich fracture surfaces}                               \\  \cline{2-6}

& \multirow{2}{*}{\citet{jiaDevelopingReassemblingAlgorithm2020}}  & \multirow{2}{*}{\check c{}} & \multirow{2}{*}{\x x{}} 
& \makecell[tl]{
\textbullet{} Keypoint-based multi-scale descriptor \\ 
\textbullet{} Robust pairwise matching with RANSAC }     

& \makecell[tl]{
\textbullet{} Reliance on pairwise matching \\
\textbullet{} Potential failure of keypoints on flat or noisy surfaces}                               \\ 
\hline 

\multirow{6}{*}{\makecell[c]{Optimization\\ (Global)}}                          & 

\multirow{2}{*}{Son et al.~\cite{son13}}  & \multirow{2}{*}{\check c{}} & \multirow{2}{*}{\x x{}} & 
\makecell[tl]{
\textbullet{} Estimation of symmetry axes \\
\textbullet{} Matching of broken curves} & 
\makecell[tl]{
\textbullet{} {Requires the assumption of axial symmetry} \\
\textbullet{} Limited handling of false positives in a global approach
}                               \\ \cline{2-6} 
                               
& \multirow{2}{*}{Zhang et al.~\cite{zhang15}} & \multirow{2}{*}{\check c{}}                     & \multirow{2}{*}{\x x{}} & 
\makecell[tl]{
\textbullet{} Template guidance for initial matching \\
\textbullet{} Matching of fracture regions}           & 
\makecell[tl]{
\textbullet{} Requirement of pre-existing templates \\
\textbullet{} Reduced effectiveness with significantly different templates}                               \\ \cline{2-6} 

& \multirow{2}{*}{\citet{wangProbabilisticMethodFractured2021}} & \multirow{2}{*}{\check c{}}                     & \multirow{2}{*}{\x x{}} & 
\makecell[tl]{
\textbullet{} Probabilistic reassembly framework \\
\textbullet{} 2D/3D descriptor with global constraints}           & \makecell[tl]{\textbullet{} Reliance on distinctive fracture surfaces \\
\textbullet{} Requirement of accurate fracture-region segmentation
}                               \\  \hline
           
\multirow{8}{*}{Data-driven} &

\multirow{2}{*}{Huang et al.~\cite{huangGenerative3DPart2020}}  & \multirow{2}{*}{\x x{}}                    & \multirow{2}{*}{\x x{}}   
& \makecell[tl]{
\textbullet{} Recurrent, dynamic graph learning for pose refinement \\
\textbullet{} Part aggregation for geometrically-equivalent shapes}           &  \\ \cline{2-5}

& \multirow{2}{*}{RGL-NET~\cite{harishRGLNETRecurrentGraph2022}}  & \multirow{2}{*}{\x x{}}                    & \multirow{2}{*}{\x x{}}   

& \makecell[tl]{\textbullet{} Recurrent graph learning \\
\textbullet{} Progressive part assembly }          
&    \makecell[tl]{
\\ \textbullet{}  {Requirement of large amount of dataset}}                       \\ \cline{2-5}

&  \multirow{2}{*}{Jigsaw~\cite{jigsaw}}  & \multirow{2}{*}{\x x{}}  & \multirow{2}{*}{\x x{}}   & 
\makecell[tl]{
\textbullet{} Hierarchical feature extraction with PointNet++ \\
\textbullet{} Joint learning of segmentation and matching }  &  
\makecell[tl]{ 
\textbullet{} {Limited performance on real-world objects} \\
\textbullet{} Sensitivity to false positive matches}             \\
                                
\cline{2-5} & \multirow{2}{*}{PuzzleFusion++~\cite{wang2024puzzlefusionpp}} & \multirow{2}{*}{\x x{}}        & \multirow{2}{*}{\x x{}}  &  
\makecell[tl]{
\textbullet{} Diffusion model for alignment \\
\textbullet{} Transformer for verification}       &       \\ \cline{1-5} 

\multirow{2}{*}{Hybrid}                     &
\multirow{2}{*}{FRASIER~\cite{joo24}}  & \multirow{2}{*}{\x x{}}    & \multirow{2}{*}{\x x{}}  & 
\makecell[tl]{
\textbullet{} PointNext + GeoTransformer \\
\textbullet{} Beam search}           &                               \\ \Xhline{1.pt}

\multirow{4}{*}{\makecell[c]{Optimization\\ (Incremental)}}     
& \multirow{2}{*}{SfS (ours)~\cite{iccvsfs}}  & \multirow{2}{*}{\check c{}}                     & \multirow{2}{*}{\check c{}} & 
\makecell[tl]{
\textbullet{} Axis-based edge line descriptor \\
\textbullet{} Multi-root beam search (base)}          
& \makecell[tl]{\textbullet{} Requires the assumption of axial symmetry \\
\textbullet{} Reliance on base pieces }                               \\ \cline{2-6} 

& \multirow{2}{*}{\AlgName (ours)} & \multirow{2}{*}{\check c{}}   & \multirow{2}{*}{\check c{}} & 
\makecell[tl]{ 
\textbullet{} Axis-based edge line descriptor \\
\textbullet{} Base-agnostic multi-graph beam search}           &    
\makecell[tl]{\textbullet{} Requires the assumption of axial symmetry}                          \\ \Xhline{1.pt}
\end{tabular}
\label{tbl:comparison}
}
\vspace{-3mm}
\end{table*}

\section{Related Work}
\label{sec:related_work_motivation}
\subsection{Geometric optimization}
\label{sec:related_work}
The virtual reassembly of archaeological pottery has garnered significant attention in computer vision and graphics over the past two decades~\cite{mcbride03,huang06,willis04,son13,zhang15}. 
Early research focused on extracting geometric properties, such as the axis of symmetry~\cite{pottmann99,cao02,mara06} or profile curve~\cite{willis03, kampelProfilebasedPotteryReconstruction2003}, from individual sherds of axially symmetric pots. 
\citet{mcbride03} introduced a method for matching fragment curves via their corners, and \citet{liPairwiseMatching3D2020} proposed features incorporating boundary curves and concave-convex patches for pairwise fragment comparison. 
However, these efforts primarily addressed pairwise matching rather than complete puzzle solving.

Unlike these initial studies that focused on individual or pairwise fragment comparison, Willis and Cooper~\cite{willis04} pioneered a reassembly pipeline for axially symmetric pots, using an incremental registration approach akin to ours. 
At each step, the best candidate sherd is selected based on factors such as its matching degree of edge line points, the corresponding surface normals, and the axis profile curve.
However, their reliance on manually extracted T-junctions and the lack of a backtracking mechanism, coupled with validation on only a single 10-sherd pot, limited its applicability.

In later years, Son et al.~\cite{son13} presented a global approach to the reassembly problem, applying it to 48 fragments from 3 pots.
Their method considers all potential pairwise matches across all pot sherds and solves the combinatorial optimization problem of finding true positive fragment pairs to retrieve the 3D pot model directly. This is achieved by minimizing algebraic costs to ensure consistency in  the axis of symmetry and the profile curve, using a spectral method~\cite{loerdeanu05}.
Although this approach demonstrated state-of-the-art performance, its handling of false positive matches remains unclear ---they are sometimes indistinguishable from ground truth~\cite{willis04}, which can be detrimental for global approaches as shown in Fig.~\figref{global_fails}.

Other studies include the work of Huang et al.~\cite{huang06}, which proposes an incremental method for reassembling geometric objects. 
However, this method relies on the existence of rich, unique features on the fracture surface, which is often not observed in pot sherds (see Fig.~\ref{fig:challenges} (a)), and has only been tested for single object reassembly.
Zhang et al.~\cite{zhang15} presented a template-based matching technique, which is also limited to single objects and known pot templates.

\subsection{Data-driven approaches}
Unlike classical approaches, data-driven approaches leveraged recent developments in deep learning and large datasets~\cite{breakingbad, FantaticBreaks, RePAIR}. For example, ~\citet{huangGenerative3DPart2020} and~\citet{harishRGLNETRecurrentGraph2022} extracted fragment features using PointNet~\cite{PointNet} and generated a connectivity graph. They then used a recurrent model or graph neural network to find the pose of each input fragment from this graph. 
Similarly, ~\citet{liuPuzzleNetBoundaryAwareFeature2023} devised an attention-based feature encoder and two decoders to estimate transformation and boundary points for pairwise matching, assembling multiple pieces through iterative greedy search.

Recently, Jigsaw~\cite{jigsaw} is a framework that leverages attention-based feature extraction and jointly learns segmentation and matching to assemble fractured 3D objects.
Meanwhile, ~\citet{joo24} introduced FRASIER, which used PointNext~\cite{PointNext} for point cloud segmentation and GeoTransformer~\cite{GeoTransformer} for iterative registration, combined with a beam search strategy~\cite{iccvsfs}. 
Furthermore, ~\citet{wang2024puzzlefusionpp} proposed PuzzleFusion++, an auto-agglomerative end-to-end neural network system that employed a diffusion model for predicting 6-DoF alignment parameters with a transformer for verifying alignments.

Despite their impressive results on the Breaking Bad dataset~\cite{breakingbad}, these methods have limitations when applied to archaeological pottery.
The broken objects in the Breaking bad often differed from the typically sharp and small fracture regions found in real-world broken ceramic sherds. 
Moreover, most approaches focused on single object reconstruction~\cite{chenNeuralShapeMating2022}, leaving their efficacy in multi-object scenarios unverified. 

\begin{table*}[t]
    \centering
    \caption{A list of analogies formed between an incremental Structure-from-Motion pipeline~\cite{schoenberger16} and pot reassembly method.}
\resizebox{2\columnwidth}{!}{
    \begin{tabular}{l | l | l}
    \Xhline{1.pt}
     & \textbf{Structure-from-motion (SfM)} & \textbf{Pot reassembly} \\ 
     \hline
    Input & RGB images & 3D point cloud of pot sherds \\ \hline
    Outputs & \makecell[l]{camera poses \& sparse 3D scene} & {sherd poses \& axially symmetric pot models } \\ \hline
    Extracted features &
    SIFT~\cite{lowe04} keypoints & 
    \makecell[l]{axis, edge line descriptor, rim, thickness, base} \\ \hline 
    Pairwise matching criteria & 
    SIFT feature distance & 
    weighted sum of above features \\ \hline
    Geometric verification & 
    \makecell[l]{inliers from fundamental (or essential) \\ matrix or homography estimation} & 
    inliers from iterative closest point (ICP)
    \\ \hline
    Sanity check & cheirality constraint & overlap, thickness \& profile curve constraints \\ \hline
    \emph{Registered} quantity & camera views & sherds \\ \hline
    \makecell[l]{\emph{Triangulated} model} & 3D scene points & \makecell[l]{3D pot model (axis of symmetry \& profile curve)} \\ \hline
    Joint estimation & bundle adjustment & global sherd and axis alignment via ICP \\ 
    \Xhline{1.pt}
    \end{tabular}
    \label{tbl:analogy}
    }
    \vspace{-3mm}
\end{table*}

\section{Motivation}
\subsection{Review of problem characteristics}
\label{sec:challenges}
As discussed in Sec. \ref{sec:introduction}, reassembling axially symmetric pots presents unique challenges compared to other 3D reconstruction tasks, primarily due to the characteristics of broken pottery sherds.
The primary challenges are outlined below:

\pparagraph{Existence of numerous false positive matches.}
Fig.~\figref{challenges} (a) shows the fracture surface (often used for matching ~\cite{huang06,zhang15, liuPuzzleNetBoundaryAwareFeature2023}) is thin and sharply broken for ceramic pots.
It is, therefore, difficult to detect features and distinguish between edge lines, leading to an increase in false positive matches, especially when identical ceramic pots (pairs of C-J and E-I pots) are mixed together. 

\pparagraph{Complex pot sherds connectivity.}
Reassembling fragmented pots involves sparse intra-pot connectivity, as each pair of sherds must be physically adjacent, leading to a limited number of valid edges in the pot’s connectivity graph. 
For instance, a connectivity graph of Pot A exhibits only 59\% complete as shown in Fig.~\figref{challenges}(b).
Moreover, when multiple pots are intermixed, the connectivity graph typically becomes block-diagonal due to fragments from different pots, yet it remains noisy with numerous false positive matches (Fig.~\figref{challenges}(c)).

\subsection{Leveraging incremental Structure-from-Motion for axially symmetric pot reassembly}
\label{sec:analogy}
Considering the challenges outlined in Sec. \ref{sec:challenges}, this study looks to the Structure-from-Motion (SfM) methodology widely used in 3D scene reconstruction for a potential solution.
SfM's success, particularly its incremental approach, offers a promising avenue for mitigating the false positive match problem and gradually resolving the complex connectivity in pot reassembly.
The section compares the pot reassembly problem with SfM and discusses adapting SfM principles to develop the SfS++ methodology.

\pparagraph{Problem analogy.}
SfM is a multi-stage process that jointly estimates 3D scene points and camera poses from a set of input images.
Similarly, pot reassembly involves jointly estimating a 3D pot model (defined by its axis of symmetry and profile curve) while determining the transformations of individual sherds from 3D point cloud data.
In both cases, the task can be framed as solving an optimization problem on a sparsely connected graph (see Fig.~\figref{challenges}(c)).

\pparagraph{Procedural similarities.}
SfM and pottery reconstruction methods share procedural similarities.
In incremental SfM, features are first extracted from each image, and pairwise matches are established. 
Then, the reconstruction is built up incrementally by adding one view at a time, refining previous matches and discovering new ones.
At each step, camera positions and 3D points are updated incrementally as new views are added, while bundle adjustment iteratively refines both camera parameters and the 3D structure.
Likewise, Son et al.~\cite{son13} proposed a pot reassembly method that extracts axis- and edge line-based features from each fragment, identifies pairwise matches, and employs a global optimization for pottery reconstruction.
However, because this approach is closer to a global SfM strategy, any undetected false matches can propagate errors throughout the entire assembly (see Fig.~\ref{fig:global_fails}).

\pparagraph{Empirical differences.}
As shown in Fig.~\ref {fig:global_fails}, the pottery reassembly problem encounters a significant challenge with false positive matches that are difficult to differentiate from true matches, even upon human visual inspection. 
It is impossible to remove all the incorrect pairs using false cycle filtering techniques similar to the work proposed by~\citet{zach10}.
Furthermore, while SfM typically constructs a single unified scene, pottery reconstruction often requires working with multiple separate models.

The procedural similarities between SfM and axially symmetric pottery reassembly in feature extraction, matching, and optimization processes (see Table~\ref{tbl:analogy}) suggest the potential applicability of established SfM methodologies to pottery reconstruction.
However, pottery reassembly presents unique challenges, such as distinguishing false positive matches and handling multiple objects, {as shown in Fig.~\ref{fig:challenges}}.
This underscores the need for a modified approach addressing these domain-specific issues rather than directly applying SfM.

\subsection{Why optimization-based method in the deep learning era?}
\label{sec:why_optim}
Recently, deep learning-based methods have been actively researched for SfM \cite{vggsfm, mast3rsfm}. 
While showing progress in solving complex geometric problems with a data-driven approach, these methods face limitations in generalization, making it difficult to completely replace SfM with deep learning.
As these limitations also apply to pottery reassembly, we argue for the necessity of optimization-based methods for pottery reconstruction, despite the increasing focus on data-driven deep learning methods.

\textit{First}, as mentioned earlier, thin and sharply fractured pottery surfaces often lack distinct features, resulting in frequent false positive matches. 
While deep learning approaches attempt to learn robust fragment features from large datasets, they still struggle with highly similar break surfaces and often violate geometric constraints.
In contrast, an incremental optimization framework can systematically validates the geometric fit as new fragments are added. By discarding uncertain matches early in the process, it effectively reduces the false positives.
\textit{Second}, multi-pot reconstruction presents a significant challenge for data-driven approaches, as it is difficult to determine whether fragments belong to different pots solely based on break surface features, as demonstrated in Table~\ref{tab:BB_mix} and Fig.~\figref{fig_result_mix}.
To effectively address this challenge, the optimization-based method is required that integrates an incremental approach with multi-graph beam search. This combination allows for the exploration of all possible fragment combinations, even when multiple pots are mixed together.
\textit{Last}, acquiring large, meticulously labeled training datasets for ancient pottery is often prohibitively expensive and time-consuming, making data-driven methods impractical in many archaeological contexts.
By leveraging explicit geometric rules and constraints, optimization-based approaches can achieve reliable reconstructions with significantly less dependence on large dataset.

\begin{figure*}[t!]
	\centering
    \includegraphics[width=\linewidth]{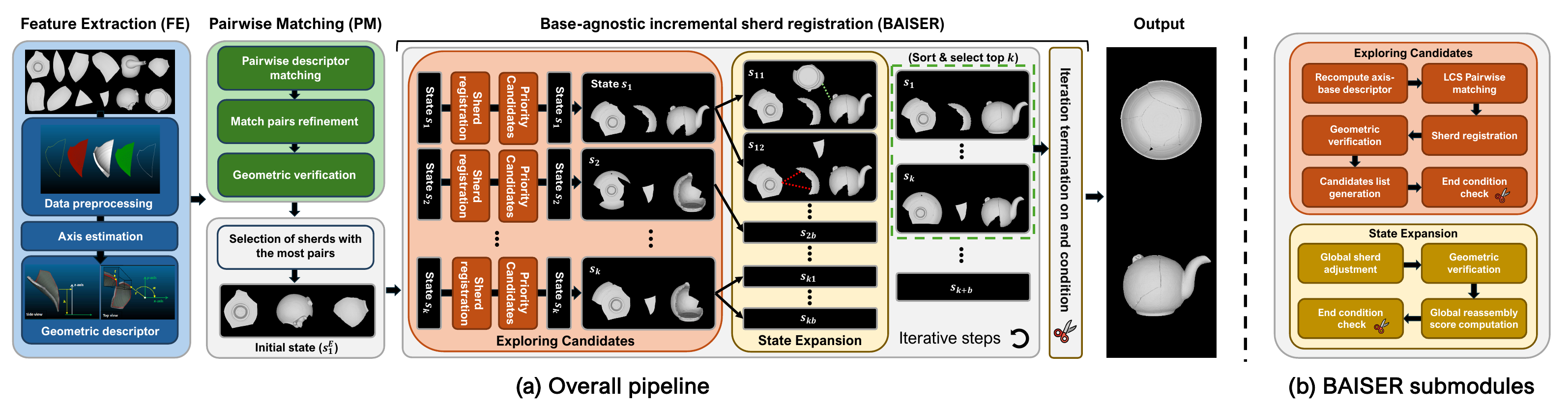}
    \vspace{-3mm}
    \caption{
	    {(a) Overall pipeline: The \AlgName pipeline consists of three modules: Feature Extraction (FE module, Sec.~\ref{sec:feature_extraction}), Pairwise Matching (PM module, Sec.~\ref{sec:pairwise_matching}), and iterative Base-Agnostic Incremental ShErd Registration (BAISER module, Sec.~\ref{sec:incremental_sherd_registration}).
        (b) BAISER submodules: The BAISER module utilizes a \textit{multi-graph} beam search algorithm, which iteratively repeats two components: Exploring candidates for expansion (Sec.~\ref{sec:prep_expansion}) and State expansion (Sec.~\ref{sec:expansion}). In the State Expansion components, the red dotted line represents graph merging.}
	}
	\label{pipeline}
    \vspace{-3mm}
\end{figure*}

\section{Proposed method}

\subsection{Algorithm overview and notations}
\label{sec:method}
\pparagraph{Algorithm overview.}
Our proposed pipeline, SfS++, for reassembling axially symmetric pots  integrates findings from Sec.~\ref{sec:analogy} and improvements from our previous conference paper~\cite{iccvsfs}. 
\AlgName consists of three main modules, as illustrated in Fig.~\ref{pipeline}.

\textit{First}, the Feature Extraction module (FE module) extracts distinctive features from individual sherds (Sec.~\ref{sec:feature_extraction}). This process includes surface segmentation, edge line extraction, symmetric axis estimation, and axis-based feature extraction. 
\textit{Second}, the Pairwise Matching module (PM module) generates possible pairwise matching candidates using the Longest Common Subsequence (LCS) algorithm (Sec.~\ref{sec:pairwise_matching}). These candidates are further refined through Iterative Closest Point (ICP) method and geometric validation.  
\textit{Finally}, the Base-Agnostic Incremental ShErd Registration (BAISER module) employs an incremental exploration approach, expanding candidate paths at each step while leveraging the multi-graph beam search method, to reconstruct pots without prior knowledge of the base (Sec.~\ref{sec:incremental_sherd_registration}).
This algorithm iteratively performs two key components: Exploring candidates for expansion (Sec.~\ref{sec:prep_expansion}) and State expansion (Sec.~\ref{sec:expansion}). 

These modules significantly enhance the algorithm's ability to robustly search possible matching configurations from noisy pairwise matches, efficiently converging toward the correct solution (see more details at Sec.~\ref{sec:justification}).

\pparagraph{Notations.}
\label{sec:notations}
We first define $P^{\mathrm{sample}}$ as the point cloud data uniformly sampled from the mesh representation. 
A point $\v {p_s} \in P^{\mathrm{sample}}$ represents a point on the surface of a sherd in 3D space and the surface normal vector at that point is defined as $\hat{\v n}_s$.
For each of the surfaces, we extract an ordered line of points around its boundary, referred to as the {edge line}. 
The \( i \)-th point on the edge line of sherd \( A \) is then defined as $\mathbf{q}^A_i = \left(\mathbf{p}^A_i, \hat{\mathbf{n}}^A_i\right) \in \mathbb{R}^6$, where \( \mathbf{p}^A_i \in \mathbb{R}^3 \) represents the position of the point, and \( \hat{\mathbf{n}}^A_i \in \mathbb{R}^3 \) represents its normal vector.

We denote $\Omega^A = \{i| \v q^A_i \in A\}$ as the set of the indices corresponding to the point cloud data on the edge line of sherd $A$. With a slight abuse of notation, we denote $\Omega^{AB}=\{(i, j)| \v q^A_i \in A, \v q^B_j \in B \}$ as the correspondence index set between sherd $A$ and sherd $B$. Lastly, we define $\Phi$ as the set of sherds participating in the optimization problem. For example, in a pairwise matching optimization problem, $n(\Phi)=2$, where $n(\cdot)$ is the cardinality of the set.

\subsection{Feature extraction from individual sherds}
The FE module extracts axis-based edge line descriptors for the initial pairwise matching candidates from the input object mesh.
Inspired by SfM, which extracts image features to support matching and optimization, this module separates surfaces into interior and exterior regions, estimates edge lines from the interior surface, and extracts axis-based edge line descriptors using parameters such as height, radius, angle, and thickness relative to the axis of symmetry.

\label{sec:feature_extraction}
\subsubsection{Data preprocessing}
\label{sec:data_preprocessing}
This section discusses the data preprocessing steps and the geometric feature descriptor. 
Firstly, we sample the point cloud data $\mathcal{S}$ from the mesh representation. 
Next, we segment the surfaces into inner and outer regions.
From the inner surface, we compute the axis of symmetry and extract the edge line, which provides a concise representation of the sherd. Finally, we compute the geometric features required for further analysis.
Compared to the previous work~\cite{iccvsfs}, we introduce cluster merging to ensure robust surface segmentation. 
\pparagraph{Interior and exterior surface extraction.}
To segment the point cloud into exterior and interior surfaces, we use the region growing algorithm~\cite{cgal20}.
This algorithm starts from the point with the minimum curvature and expands by combining the $n_b=10$ neighboring points that have a curvature value of less than $\tau_\kappa$ and a normal angle less than $\tau_\theta$ with the point’s normal.
We set the values of $\tau_\theta$=4 and $\tau_\kappa$=1 to achieve clear separation of interior, exterior, and fractured surfaces.

\pparagraph{Cluster merging.}
Following the initial segmentation, we apply a cluster merging algorithm to improve the results. 
This process starts by identifying close point pairs between clusters using the $k$-nearest neighbors method. 
We then calculate average normal vectors for local patches around these pairs to check for normal similarity and compute and compare the average curvature of boundary points for each cluster pair.
Additionally,  internal points from each cluster are selected to further verify normal and curvature similarities. 
Clusters satisfying all these criteria are merged, and the process repeats until no further merges are possible. 
This approach effectively combines parts of the same surface that were excessively separated in the initial segmentation.
Further details can be found in the supplementary material~\cite{supmat_journal}.

Subsequently, B-Spline fitting is performed for each cluster to refine surface boundaries and uniformly sample the point cloud.
Finally, we select the two largest clusters, representing the inner and outer surfaces.
Note that region-growing algorithm and clustering naturally facilitate the separation of  decorative parts, such as the nose and handle. The effect of $\tau_\theta$ and subsequent part removal results are shown in Fig.~\figref{fig:surface_segmentation} (a) and (b).

\begin{figure}[t]
	\centering
    \captionsetup[subfloat]{labelfont=footnotesize,textfont=footnotesize}
	\subfloat[][Results for different $\tau_\theta$]{\includegraphics[width=0.33\linewidth]{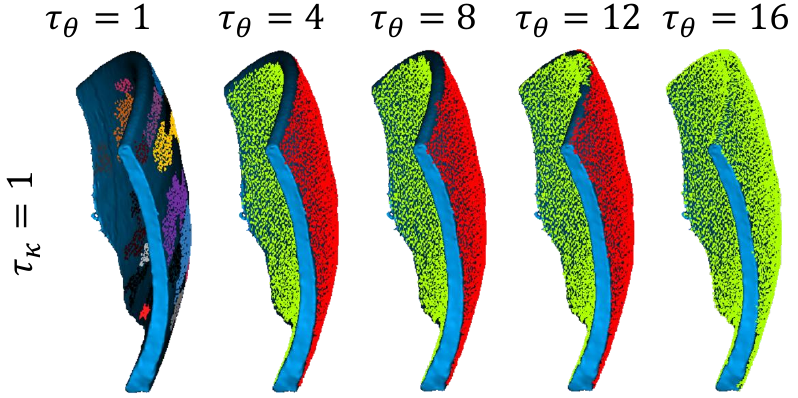}\figlabel{tau_theta}}
	~
	\subfloat[][Decorative part removal]{\includegraphics[width=0.33\linewidth]{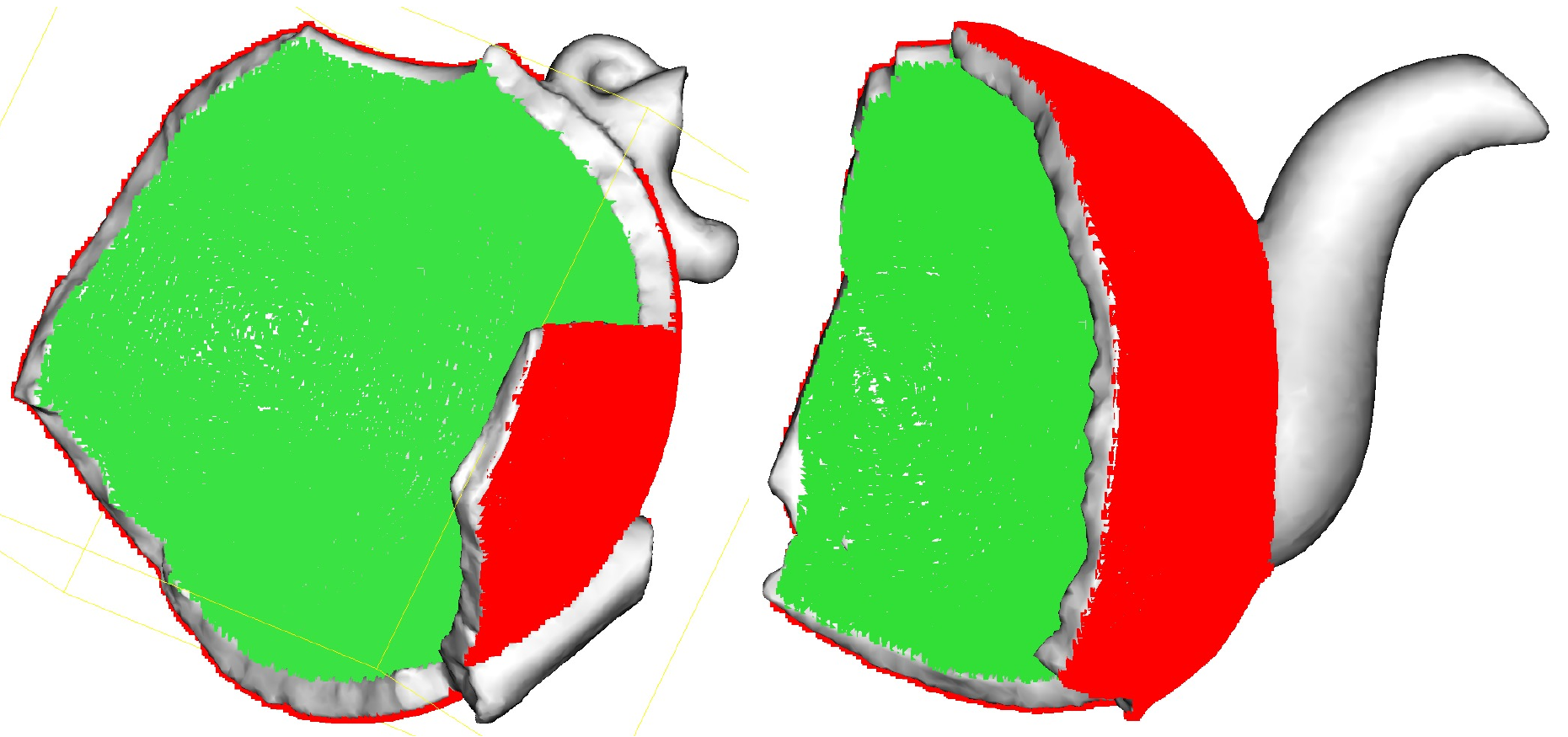}\figlabel{pot_segmentation}}	
    ~
    \subfloat[][Surface classification]{\includegraphics[width=0.3\linewidth]{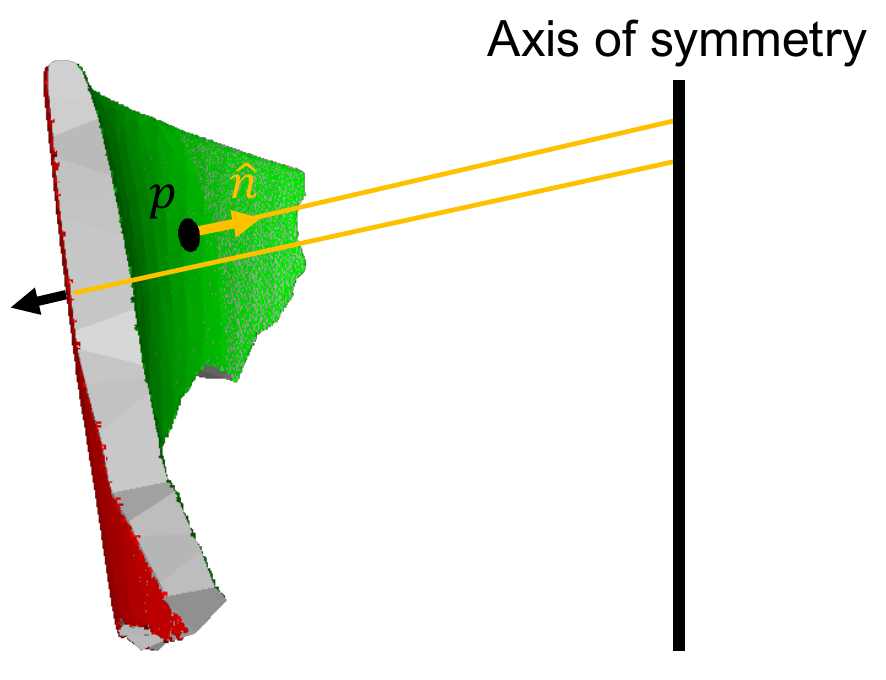}\figlabel{interior_exterior_surface_classification}}
	\caption{
        Segmentation results using the region growing algorithm in CGAL \cite{cgal20}. Each color represents a distinct segmentation clusters.
        (a) Variation of segmentation results across different $\tau_{\theta}$ values. 
        (b) Removal of non-smooth decorative parts achieved with $\tau_{\theta} = 4$ and $\tau_{\kappa} = 1$. 
        (c) Classification of interior and exterior surfaces based on the axis of symmetry.
	}
	\figlabel{fig:surface_segmentation}
    \vspace{-3mm}
\end{figure}

\pparagraph{Interior and exterior surface classification.}
To determine whether the selected surfaces are interior or exterior, we leverage geometric characteristic of axially-symmetric pots.
Specifically, for each point $\v {p_s} \in P^{\mathrm{sample}}$ on the surface, we project a ray along its surface normal direction $\hat{n}_{s}$ towards the axis of symmetry and define the point $\v {p_s}' \in \mathbb{R}^3$ where it meets the axis of symmetry.
Next, we assess whether the direction of the (near-)intersection point  $\v {p_s}'$ is in the positive or negative surface normal direction of the point $\v {p_s}$. 
If the point $\v {p_s}'$ lies on the positive surface normal, the surface to which \( p_s \) belongs is classified as the interior surface. 
On the other hand, if $\v {p_s}$ lies in the negative direction, the surface is classified as the exterior surface, as depicted in Fig.~\figref{fig:surface_segmentation} (c).

We test two configurations: one surface as interior and the other as exterior, and vice versa, and we select the configuration that had the greatest number of surface points from both surfaces that satisfied the above geometry (by checking the sign of the ray’s normal scalar coefficient).

\pparagraph{Edge line extraction and segmentation.}
The interior surface point cloud boundary makes up the edge line extracted using the point cloud boundary-estimation method from the PCL library \cite{Rusu_ICRA2011_PCL}. 
The extracted edge line is refined through B-spline surface fitting, which smooths the edge line and calculates normals for each point.
Using a K-D tree-based search, points near the edge line are identified and filtered to remove noise and outliers.
The filtered points are resampled to generate equidistant points with a point-to-point distance of {$d$ = 1.9mm}, an empirically optimized value.
The resulting edge line points are reordered counter-clockwise using their normals and a voting algorithm, as visualized in Fig.~\figref{fig:improved_breakline}. 
Finally, we segment the edge line based on corners detected by analyzing peaks in the distribution of \textit{distance scores}, which measure the deviation of each point from its local linear trend.

\begin{figure}[t]
	\centering
	\includegraphics[width=0.9\linewidth]{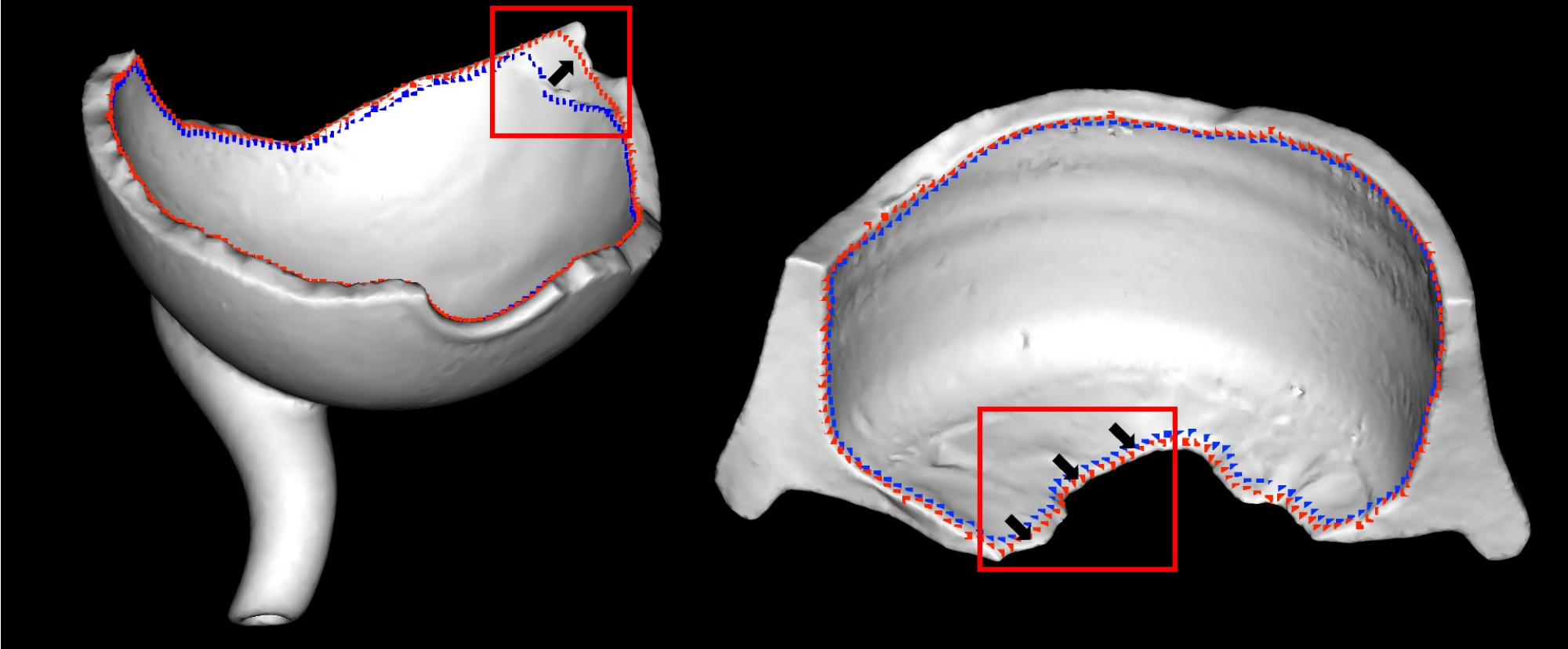}
	\caption{Improved edge line detection on 3D object meshes. B-Spline fitting method (red) extracts edge lines closer to the real edge line, demonstrating enhanced accuracy over a wider area compared to the previous approach (blue)} 
	\figlabel{fig:improved_breakline}
    \vspace{-3mm}
\end{figure}

\subsubsection{Axis-of-symmetry estimation}
\label{sec:axis_estimation}
Due to the axial symmetry of pottery, the axis estimation is a crucial step in pottery reconstruction.
Existing algorithms~\cite{cao02, mara06, son13} offer high computational efficiency but suffer from limitations such as consideration of only one surface~\cite{cao02}, or neglect of normal information, leading to a bias towards cylindrical shapes~\cite{son13}.

In contrast,  PotSAC~\cite{hong19} achieves state-of-the-art performance by robustly estimating symmetry axes from noisy 3D fragment point clouds using both exterior and interior surface information.
It initially estimates the axis by combining MLESAC-based sampling with a minimal 6-point solver that enforces Pottmann’s constraint—ensuring the offset vector, symmetry axis direction, and the cross product of the surface normal lie on the same plane. 
Subsequently,  PotSAC utilizes an extended Cao and Mumford error, applicable to both interior and exterior surfaces~\cite{hong19}, in the subsequent trust-region-based nonlinear optimization, enabling high-precision estimation even under challenging conditions.
However, since the original PotSAC outputs only a single axis even for ambiguous fragments, we propose a modified version capable of producing multiple axes.

Instead of selecting a single final axis in the MLESAC stage, the modified method selects the top 10 candidate axes. 
Then, the angles between each pair of axes are evaluated, and those with an angle of 10 degrees or less are considered duplicates, with the axis having the higher cost, defined as the robustified residual error, being removed through pruning.
To improve efficiency, we initially sample only 10\% of the surface points to generate axis candidates, and after pruning, all surface points are used for more detailed refinement of the remaining axes.

\pparagraph{Rim detection.} 
\label{sec:rim_detection}
To identify the rim of pottery sherds, we analyze the 3D coordinate information of the sherd edges, known as edge lines. 
The rim generally exhibits a flat and consistent curve in terms of height and radius and 
 transitions smoothly into the surrounding sections, unlike irregularity of a break surface.
Based on these characteristics, we identify rim candidates using the following criteria:
\textit{First}, we evaluate the consistency of height and radius, seeking sections with minimal variation, specifically where the standard deviation of both height and radius is 1.0 mm or less.
\textit{Second}, rim sections must exhibit gradual changes, defined as an average variation of 0.1 mm or less in both height and radius between adjacent points.
\textit{Last}, to minimize false positives in short edge segments, we only consider sections composed of at least 20 points that meet the previous criteria.
Edge line sections satisfying these criteria are designated as rim candidates. 
If multiple candidates are identified, the section with the most stable and consistent height and radius is selected as the final rim.

\begin{figure}[t]
	\centering
	\includegraphics[width=0.85\linewidth]{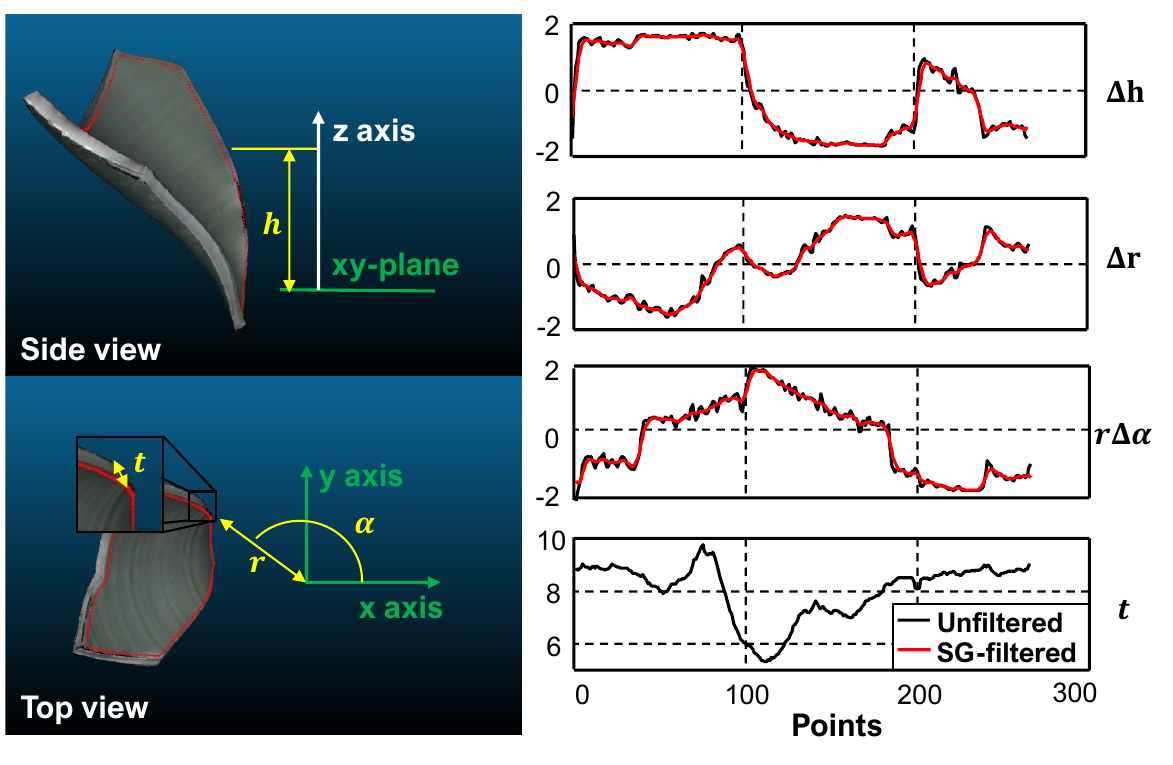}
	\caption{
	Illustration of our axis-based edge line descriptor.
	Unfiltered results are computed via the finite difference method.
	The Savitzky-Golay filter~\cite{SGmethod} is used for smoothing and differentiation, followed by  Gaussian filtering with a kernel of width 7 and $\sigma$=2.0.
	Thickness ($t$) is not filtered, as it is missing in some edge points.}
	\figlabel{edge_line_descriptor}
    \vspace{-3mm}
\end{figure}

\subsubsection{Axis-based edge line descriptor}
\label{sec:break_line_descriptor}
While~\citet{son13} have used the axis of symmetry as a constraint to identify incorrect pairwise matching, we go a step further by incorporating information about the axis to design an axis-based edge line descriptor, enhancing the pair-wise matching efficiency.

\pparagraph{Motivation.} 
Traditional SfM finds correspondences by extracting features like SIFT~\cite{lowe04} from images.
However, in our case, the fractured surfaces of sherds are extremely thin and sharply broken, making it very difficult to extract distinctive features, as mentioned in Sec.~\ref{sec:challenges}.

To address this, we observe that edge lines of neighboring sherds ideally converge into a congruent geometric shape along with the axis of symmetry. To make use of this observation, we create a feature descriptor with the height ($h$), radius ($r$), and angle ($\alpha$) relative to the axis of the symmetry coordinate system, as illustrated in Fig.~\figref{edge_line_descriptor}. Together with the thickness ($t$), this descriptor comprehensively characterizes the edge line.
This approach captures all geometric variations associated with each edge line, eliminating the need to refer back to the original point cloud (as seen in~\cite{son13}).

\pparagraph{Descriptor definition.}
Our axis-based descriptor for point $\v q_j =(\v p_j, \v n_j)\in\real^{6}$ on the edge line is defined as
\begin{align}
[\Delta h(\v q_j), \Delta r(\v q_j), r \Delta \alpha (\v q_j), t(\v q_j)]\tr
\end{align}
where $\Delta$ involves finite differentiation.
Because of noise along the edge lines, naive differentiation technique, such as finite differences, yields noisy features.
To mitigate noise along the edge lines, we apply the Savitzky-Golay digital differentiator (S-G method)~\cite{SGmethod} using 7 points and a Gaussian filter with \( \sigma = 2.0 \).
The radius \( r \) is multiplied by \( \Delta \alpha \) to obtain the geometric tangential distance.

To compute thickness \( t \), a ray is projected in the opposite direction along the surface normal from each point $\v p_{\text{in}}$ on the inner edge line to find the closest point $\v p_{\text{out}}$ on the outer surface. 
The point must satisfy:  (a) its distance from the ray is within 1 mm, and 
(b) the direction of $(\v p_{\text{in}} - \v p_{\text{out}})$ is opposite to $\hat{\v n}^{\v p_{\text{in}}}$. 
\( t \) is calculated as the distance between $\v p_{\text{in}}$ and $\v p_{\text{out}}$.

\begin{figure}[t]
	\centering
	\includegraphics[width=0.85\linewidth]{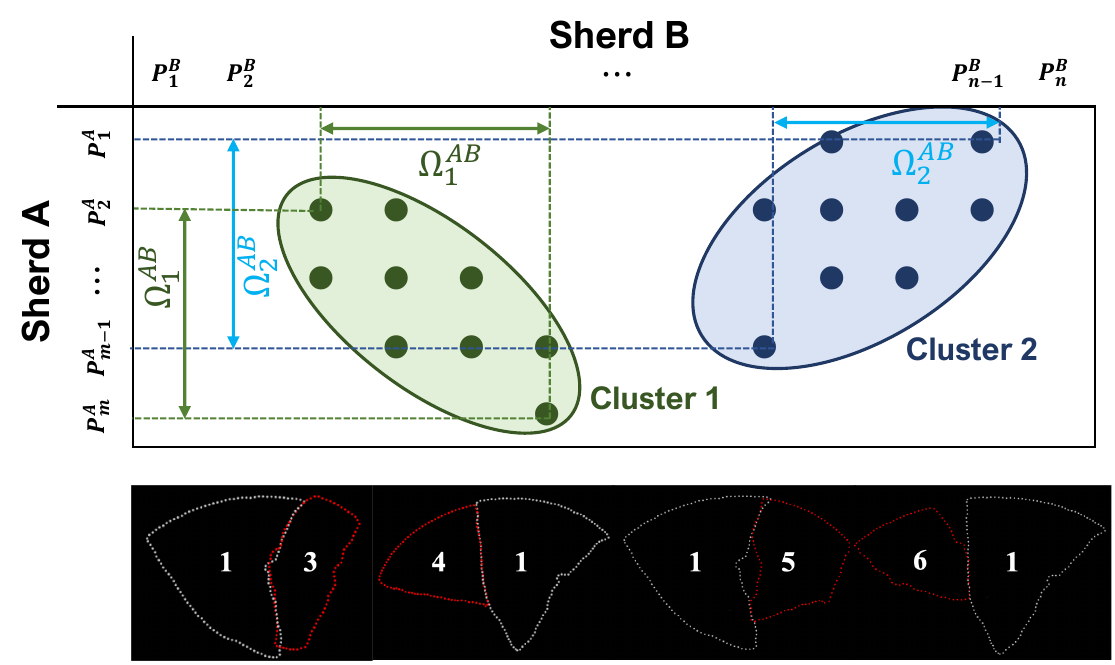}
	\caption{A conceptual diagram of pairwise feature matching. Our LCS matching scheme (above) detects multiple correspondence pairs (e.g. $\Omega^{AB}_1$ (green) and $\Omega^{AB}_2$ (blue)) of potential matches between two sherds. 
 However, only a few of these are true matches (e.g. (1-3) and (1-4) pairs in the bottom figure).} 
	\figlabel{LCS}
    \vspace{-3mm}
\end{figure}

\subsection{Pairwise matching}
\label{sec:pairwise_matching}
Incremental SfM reconstructs 3D structures through descriptor matching, RANSAC-based geometric verification, and iterative refinement. 
Similarly, our three-step Pairwise Matching (PM) module uses axis-based edge line descriptors extracted along the edge lines to perform matching via the Longest Common Subsequence (LCS) algorithm, achieving robust alignments akin to SfM's feature matching.
Matches are further refined via ICP and physical constraints, in similar spirit to the SfM's geometric verification to ensure precision.

\subsubsection{Pairwise descriptor matching}
\label{sec:descriptor_matching}
To establish the initial pairwise matching between edge lines, we employ the LCS algorithm~\cite{supmat}. 
Upon running the algorithm, we obtain multiple clusters, referred to as correspondence of sequence of points on the edge line, representing the potential regions of alignment between the matched sherds, as illustrated in Fig.~\figref{LCS}.

\pparagraph{Handling sign ambiguity and rim constraints.}
We encounter sign ambiguity stemming from the direction of the symmetric axis. To address this, we perform two rounds of matching for each pair of edge lines by inverting the descriptor of one edge line. 
Furthermore, if the matched interval contains a rim segment, we determine the match is false and discard it.

\subsubsection{Adjustment of matched sherds}
\label{sec:refining_matches}

While the above descriptor provides useful cues for finding initial correspondences, it is difficult to perfectly distinguish true matches from false matches, even with human inspection. Therefore, we adopt a two-step verification process.
\textit{First}, we run the Iterative Closest Point (ICP) on the initial pairs of edge lines.
The ICP algorithm comprises two stages: correspondence assignment followed by correspondence optimization.
\textit{Second}, we actively exploit pottery's physical properties, such as consistency of profile curve and non-overlapping, to prune out remaining incorrect matching pairs after the ICP algorithm converges.

\pparagraph{Forming correspondences.}
In the initial iteration of ICP, we use correspondence pairs from descriptor matching.
In subsequent iterations, correspondence pairs are formed by retrieving mutually closest points based on Euclidean distance, pruning pairs with normal vector differences exceeding 30$^\circ$.

\begin{figure}[t]
	\centering
	\includegraphics[width=0.95\linewidth]{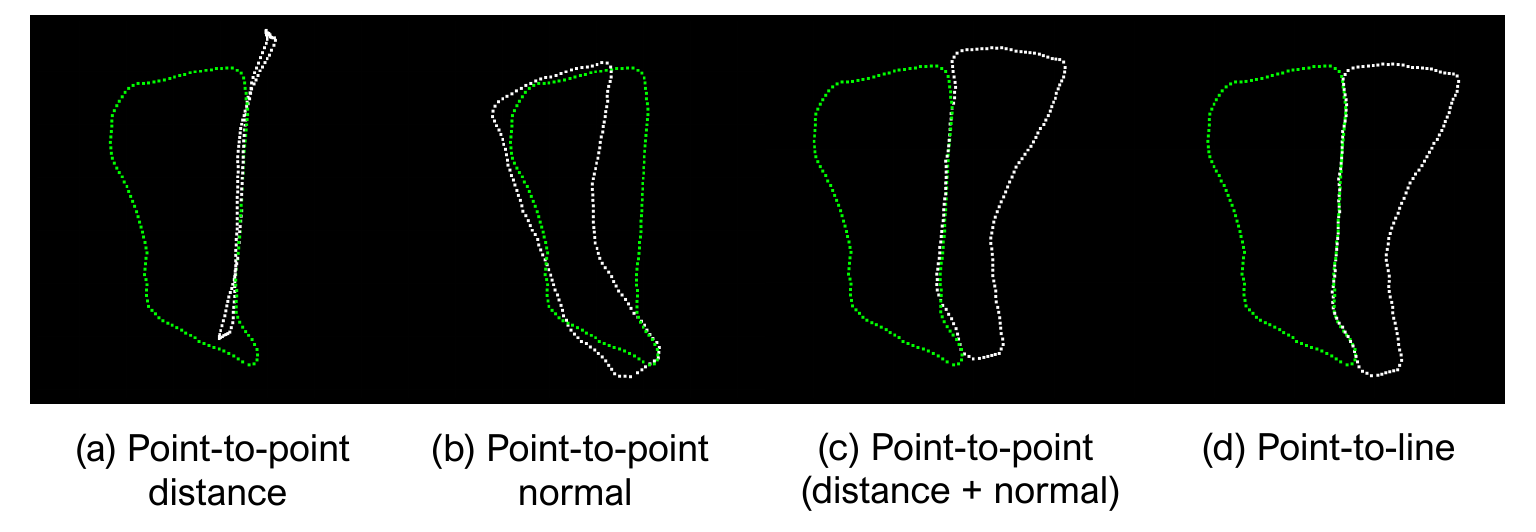}
	\caption{Pairwise ICP results based on different cost functions are as follows: Point-to-point ICP (c) provides a proper initial position, and point-to-line ICP serves as fine-tuning (d). It is important to note that if we use only distance (a) or normal (b), ICP consistently fails.}
	\figlabel{ICP_comp}
    \vspace{-3mm}
\end{figure}
\pparagraph{Minimizing correspondence distances.}

Given a set of correspondence $\Omega^{AB}$ between sherd $A$ and $B$, we refine each sherd’s transformation
$\m T^A = (\m R^A,\v t^A)$ and $\m T^B = (\m R^B,\v t^B)$, where
$\m R^A,\m R^B \in SO(3)$ and
$\v t^A,\v t^B \in \mathbb{R}^3$.
If the sherds involve a shared rim region, we also include rim radius and height with respect to the axis of symmetry, $(r^{\mathrm{rim}},h^{\mathrm{rim}})$, as optimization variables. 
Our objective consists of two main parts: a \emph{pairwise} cost $J_R(A,B)$ for aligning the fracture correspondences between two sherds, an \emph{individual} cost $J_I(A)$ for enforcing each sherd's axis or rim consistency constraints.

First, the pairwise cost is defined as
\begin{align}
    &J_R(A, B) := \sum_{\smash{\mathclap{(i,j)\in\Omega^{AB}}}} 
    \rho_d(d_{ij}(\m T^A, \m T^B)) + \lambda \rho_e(e^2_{ij}(\m T^A, \m T^B))
    \eqlabel{eq:J_R}
\end{align}
where $(i,j)$ indexes a correspondence in $\Omega^{AB}$.
$d_{ij}$ measures a geometric distance between two fracture points or between a point and a line, while $e_{ij}$ quantifies the mismatch between their surface normals.
To handle outlier correspondences effectively, we adopt robust Cauchy kernels $\rho_{d},\rho_{e}$ and scale $e_{ij}$ by a factor $\lambda$ to control its relative influence.
In practice, the initial iteration of the ICP algorithm treats $d_{ij}$ as a point-to-point (P2P) distance, providing a high-quality starting assumption.
From the next iteration onward, we switch to a point-to-line (P2L) distance, which better captures the geometric relationship between points and lines along the fracture edges, resulting in a more accurate alignment (see Fig.~\figref{ICP_comp}).

Second, to incorporate additional intra-sherd constraints, such as axis alignment or rim consistency, we define
\begin{align}
    J_{I}(A) := \mu\!\!\sum_{i\in\Omega^{A}}\!\!\rho_{f}\big(f_{i}^{2}(\m T^A)\big)
    + \nu\!\!\sum_{i\in\Psi^{A}}\!\!\rho_{g}\big(g_{i}^{2}(\m T^A,r^{\mathrm{rim}},h^\mathrm{\mathrm{rim}})\big)
    \eqlabel{eq:J_I}
\end{align}
where \(\Omega^A\) is the set of points used for axis alignment in sherd \(A\), and \(\Psi^A\) indicates the set of rim segments from the set of sherds A.  
The term \(f_i\) represents a function guiding the sherd to align with the overall axis proposed in Eq.~(9) of~\cite{hong19}, while \(g_i\) ensures that the rim conforms to the shared parameters \((r^\mathrm{\mathrm{rim}}, h^\mathrm{\mathrm{rim}})\). 
We again use Cauchy kernels $\rho_{f},\rho_{g}$ to handle potential outliers, while $\mu$ and $\nu$ control the relative importance of the axis consistency and rim consistency terms. 

Combining these two parts into a single objective, we solve:

\begin{align}
\argmin_{\m T^A, \m T^B,r^{\mathrm{rim}},h^{\mathrm{rim}}} J_{R}(A,B) + \sum_{E\,\in\, \{A, B\}} J_{I}(E)
\eqlabel{eq:PMCostFunction}
\end{align}

In the PM module, only two sherds are considered at a time. 
We employ the Levenberg-Marquardt algorithm for the robust nonlinear optimization described above, iteratively updating all variables. At each ICP iteration, the LM algorithm runs for a maximum of 100 iterations, and we set the overall maximum number of ICP iteration to 150.
By balancing \emph{pairwise} fracture matching in Eq. \eqref{eq:J_R} with \emph{individual} constraints in Eq. \eqref{eq:J_I}, we achieve stable reconstructions.
This cost functions minimize distance and normal mismatches between shards and ensure the consistent alignment of axes and rims across multiple fragments.
Detailed definitions of equations can be found in our supplementary materials~\cite{supmat_journal}.

\begin{figure}[t]
	\centering
    \captionsetup[subfloat]{labelfont=footnotesize,textfont=footnotesize}
	\subfloat[][3D edge line overlap test]{\includegraphics[width=0.45\linewidth]{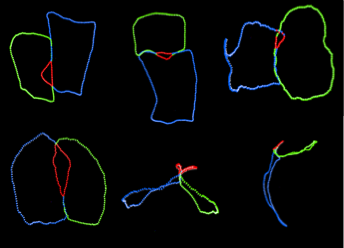}\figlabel{overlap}}
	~
	\subfloat[][Axis profile curve check]{\includegraphics[width=0.45\linewidth]{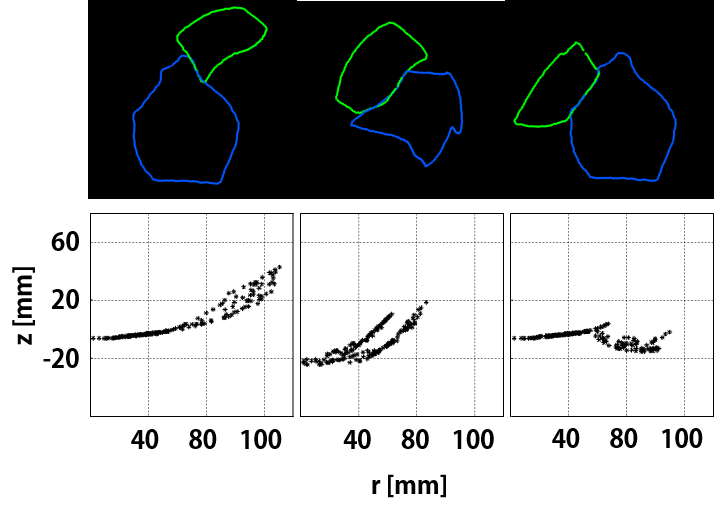}\figlabel{pc_check}}	
    ~
	\caption{
		Geometric verification filters two cases: (a) physically unrealistic overlaps between sherds, and (b) non-smooth profile curves.
	}
	\figlabel{sanity_check}
    \vspace{-3mm}
\end{figure}

\subsubsection{Geometric verification}
\label{sec:geo_verification}
LCS feature matching and ICP refinement procedures produce plausible pairwise results. 
However, these methods primarily rely on local information, which can lead to numerous false positive matches. This includes cases where physical constraints (e.g. avoiding penetration) or global pottery shapes (e.g. consistency with profile curves) are violated, as shown in Fig.~\figref{sanity_check}.
These limitations stem from the sharp geometric properties of broken ceramic sherds (see Sec.~\ref{sec:challenges}).
To address this issue, we propose a geometric verification process to further eliminate false positive matches. 

First, we employ overlapping region detection to verify the physical compatibility of sherds by identifying overlapping areas between them.
Next, we validate the axis profile curve to ensure continuity and smoothness by analyzing a 2D profile curve.
This profile curve is obtained by projecting the inner surface edge line points onto radius-height ($rz$) plane with respect to the axis of symmetry.
Conceptually, we approximate the curve from the projected edge line points and then compute the approximation error to discard configurations with high erros.
Implementation details are provided in the supplementary material~\cite{supmat_journal}.

\subsection{Base-agnostic incremental sherd registration}
\label{sec:incremental_sherd_registration}
We now describe the BAISER module, as illustrated in Fig.~\ref{pipeline}. 
As multiple sherds are involved in the optimization process, $n(\Phi) \geq 2$ at this stage, where $\Phi$ is the set of relevant sherds in the optimization process.
Our BAISER module adopts an incremental SfM-like approach, where we iteratively refine the reassembly configuration of multiple sherds and verify geometric criteria.
In doing so, we effectively prune spurious or inconsistent connections between sherds, akin to how SfM discards erroneous matches in a scene graph.

Additionally, BAISER employs \textit{multi-graph} beam search to efficiently explore potential reassembly paths, enabling simultaneous reconstruction of multiple potteries (see Algorithm~\ref{alg:beam_search}). The graphs in a state can be \textit{merged} as the search progresses, providing flexibility (see Sec.~\ref{sec:justification} for experimental results) and enabling the algorithm to prioritize the reconstruction of pottery that is easier to assemble. 
For instance, if Pot $A$ has distinct features that yield higher pairwise matching scores compared to others, the algorithm will generate states predominantly containing graphs related to Pot $A$ in its early stage. 

\subsubsection{Notations}
We define a state, $s$, as a possible configuration of sherds. Each state consists of several graphs, $g$, where each graph represents incrementally reconstructed pottery.
In the graph the nodes represent the sherds and the edges represent the sherd connectivity.
For example, the state $s=\{g_1, g_2, g_3\}$, as shown in Fig.~\figref{fig_priority}, contains three distinct graphs that do not share any sherds, each representing an individual reconstructing pottery.
Similarly, we define a set of states, $\mathcal{S}$, as a collection of such states.  In the multi-graph beam search algorithm, we use one another set of states $\mathcal{S}^O$, called set of output states. A state becomes finalized in two cases: (1) when there are no priority candidates left, although unmatched sherds could still exist, and (2) when all the sherds have been reconstructed. Finally, a 'branch` (or $b$) represents the number of states expanded from a given state, while a 'beam` (or $k$) refers to the top-ranked states carried forward to the next iteration. 

\subsubsection{Procedural overview}
During each iteration of the algorithm, $\mathcal{S}$ contains less than or equal to $k$ number of states. Each state then goes through mainly two phases: \textit{exploring candidates for expansion} (line 2-15 in Algorithm~\ref{alg:beam_search}) and \textit{state expansion} (line 16-22). 
At the end of each iteration, the set of states are pruned and sorted based on the reassembly score and only the top-$k$ states are retained for the next iteration (line 24-26). 
The iterative nature of the algorithm, which cycles back to line 2 if $\mathcal{S}$ is not empty (line 27-30), ensures comprehensive exploration of plausible configurations until no further expansions are possible.

\begin{algorithm}[t]
	\caption{Multi-graph beam search (Sec.~\ref{sec:incremental_sherd_registration})}
	\label{alg:beam_search}
	\begin{algorithmic}[1]
    \Statex {\textbf{Inputs}: Sherds and pairwise matching pairs} 
    \Statex {\textbf{Outputs}: \textit{Set of output states} $\mathcal{S}^O$}
		\State Select the top-$b$ sherds that have the greatest number of edges as the initial graphs in the state ($s_1=\{g_1 ,\cdots, g_b\})$ and append it to the \textit{set of states} ($\mathcal{S}$)
        \For{$s \in \mathcal{S}$} \algorithmiccomment{Exploring candidates for state expansion}
            \State{$s \leftarrow  \mathcal{S}$.pop()}
            \If{$n(s) < b$ or $n(L_P(s)) < \epsilon$}
                \State{Replenish the graph in the $s$ with unmatched sherds}
            \EndIf
            \State{Recompute axis-based descriptor}
            \State{Pairwise matching, including inter-graphs connections}
            \For{Possible matching pairs} 
                \State{{Sherd registration \& geometric verification}} 
            \EndFor
            \State{{Generate $L_p(s)$ (priority candidates)}} 
            \If{$s$ satisfies end condition} 
                \State{Append $s$ to \textit{set of output states} $\mathcal{S}^O$ and \textbf{continue}}
            \EndIf
            \For{Top-$b$ candidates ($\Phi\in L_P(s))$} \algorithmiccomment{{State expansion}}
                \State{Global sherds adjustment \& geometric verification} 
                \If{Expanded state is geometrically  plausible}
                    \State{Compute reassembly score}
                    \State{Append it to the $\mathcal{S}$}
                \EndIf
            \EndFor
        \EndFor
        \State{Discard redundant permutations}
        \State{Sort $\mathcal{S}$ based on the reassembly score}
        \State{Select top-$k$ states and remove others in $\mathcal{S}$}
        \If{$\mathcal{S}=\emptyset$} 
            \State{Terminate algorithm}
        \EndIf
        \State{\textbf{goto} line 2}
    \end{algorithmic}
\end{algorithm}

\subsubsection{Exploring candidates for expansion}
\label{sec:prep_expansion}
First, a state is popped out from the set ($\mathcal{S}$)\textemdash which contains either previously expanded states or the initial state from the beginning of algorithm\textemdash to explore potential candidates for expansion.
Next, using this state, axis-based descriptors and pairwise feature matching are recomputed (line 7-8).
This recomputation enhances both the geometric features and the quality of pairwise matching candidates because the previously selected top-$k$ states were globally adjusted with multiple sherds. Furthermore, potential \textit{graph merging} scenarios are explored within the state, thereby ensuring a base-agnostic property.

Similar to the PM module, the pairwise matching results are adjusted using ICP and discarded if they do not satisfy the geometric criteria (line 9-11). 
However, unlike the PM module, the \textit{sherd registration} optimizes only the newly added sherds while keeping the previously reconstructed ones fixed. 
Because many potential pairwise matching candidates are explored, the global adjustment\textemdash similar to bundle adjustment in SfM\textemdash is not carried out during the exploration stage, but only on the most probable candidates in \textit{state expansion stage} to ensure computational efficiency.

The adjusted potential candidates are then pruned and prioritized, \textit{generating priority candidates list} for effective expansion (line 12). 
If the state does not contain priority candidates, then, assuming sherd reconstruction is complete, it is appended to the set of output states ($\mathcal{S}^O$, lines 13-15).

\pparagraph{Graph merging.}
Previous work~\cite{iccvsfs} relied on explicitly defined base features to guide pottery reassembly. However, as shown in Fig.~\figref{fig_base}, reconstruction becomes challenging when base fragments are difficult to assemble.
To address this, we adopt a base-agnostic approach using the graph merging mechanism illustrated in Fig.~\figref{fig_priority}. 
Graph merging is a special case of matching candidates. By allowing graph merging, we no longer assume that \textit{different graphs within the same state necessarily represent distinct potteries}; 
consequently, the total number of graphs in the state does not directly correspond to the total number of mixed pots. For example, during the single reassembly of Pot $A$, multiple graphs may coexist within the same state, but these graphs do not share any sherds and all belong to Pot $A$. 
In Fig.~\figref{fig_priority}, while graph $g_3$ could only contain external sherds (nodes 8 and 9) in SfS, \AlgName includes both graph $g_1$ and $g_2$ as viable reconstruction candidates. 

After merging the graphs, the state is replenished with an unmatched sherd that has the highest number of pairwise matches from the previous computation, thereby opening up another potential graph to grow.

\begin{figure}[t]
	\centering
	\includegraphics[width=0.9\linewidth]{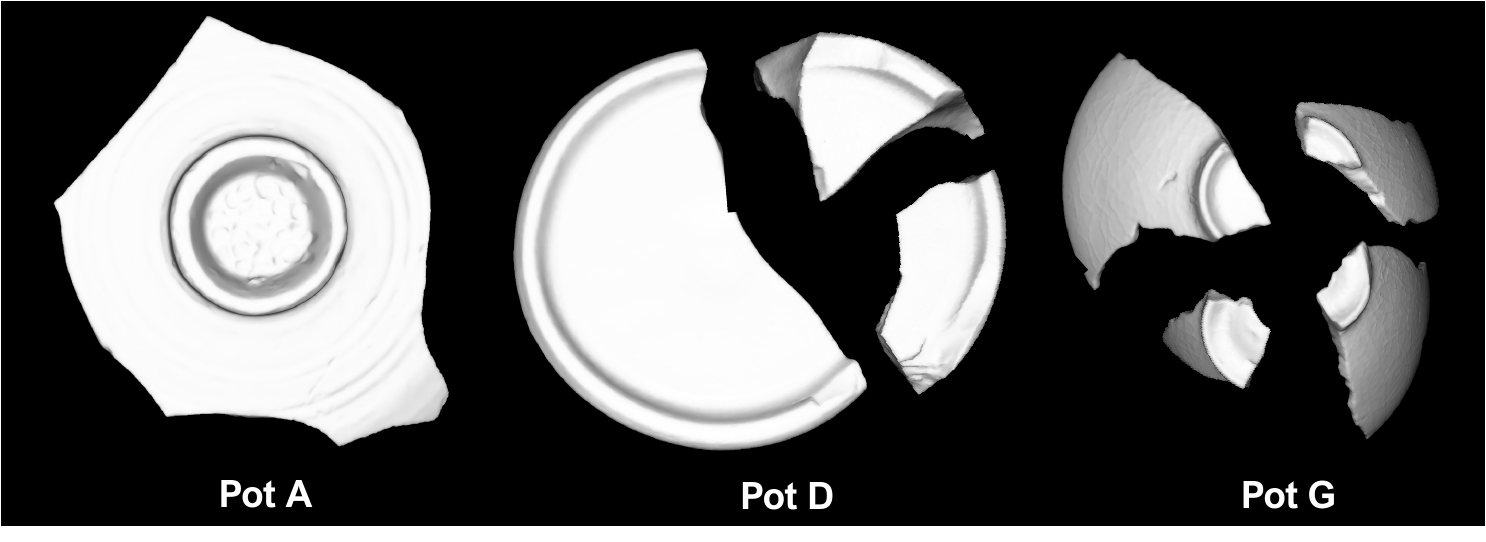}
	\caption{
	    It illustrates the distinctive base fragments of Pot A, D, and G. Pot A and D exhibit apparent base features characterized by a flat surface or a sharp transition from the base to the side surface. On the other hand, Pot G presents challenges in identifying the base fragment due to its limited flat region.
	}
	\figlabel{fig_base}
    \vspace{-3mm}
\end{figure}

\pparagraph{Sherd registration.}
Let $\Phi^{\mathrm{new}}$ represent a set of new sherds to be reassembled, and $\Phi^{\mathrm{old}}$ denote a previously reassembled set of sherds. 
The \textit{sherd registration} attempts to attach the new sherds ($\Phi^{\mathrm{new}}, n(\Phi^{\mathrm{new}}) \geq 1$) to the existing reassembled sherds ($\Phi^{\mathrm{old}}, n(\Phi^{\mathrm{old}})\geq 1$), analogous to the incremental addition of new images or views in SfM.

Without loss of generality, we assume that the axis of symmetry of $\Phi^{\mathrm{old}}$ aligns with the $z$-axis. 
The new sherds $D\in \Phi^{\mathrm{new}}$ are aligned with the $z$-axis during the registration process. Subsequently, the poses of the new sherds ($\{\m T^{\Phi^{\mathrm{new}}}\}$) are adjusted to achieve the correct configuration relative to the old sherds $\Phi^{\mathrm{old}}$. This adjustment adheres to constraints such as rim consistency and axis symmetry.
The optimization problem, described in Eq.~\eqref{eq:SherdRegistration}, is solved while keeping the sherds in $\Phi^{\mathrm{old}}$ fixed.
\begin{align}
    \argmin_{\{\m T^{\Phi^{\mathrm{new}}}\}, r^{\mathrm{rim}}, h^{\mathrm{rim}}}\sum_{D\in \Phi^{\mathrm{new}}}\left(  \sum\limits_{E \in \Phi^{\mathrm{old}}} J_R(D, E) + J_I(D) \right)
    \eqlabel{eq:SherdRegistration}
\end{align}
We use the ICP algorithm (updating correspondences at every iteration) as described in Sec.~\ref{sec:refining_matches}. 
While the default approach reassembles one sherd at a time ($n(\Phi^{\mathrm{new}})=1$), in the cases involving graph merging, 
$\Phi^{\mathrm{new}}$  may include multiple sherds that were previously reassembled. In such cases, the relative poses within $\Phi^{\mathrm{new}}$ remain unchanged\textemdash that is the multiple sherds are considered as a single reassembled large chunk. After sherd registration, infeasible configurations are filtered using the geometric criteria described in Sec.~\ref{sec:geo_verification}. 

\begin{figure}[t]
	\centering
	\includegraphics[width=0.9\linewidth]{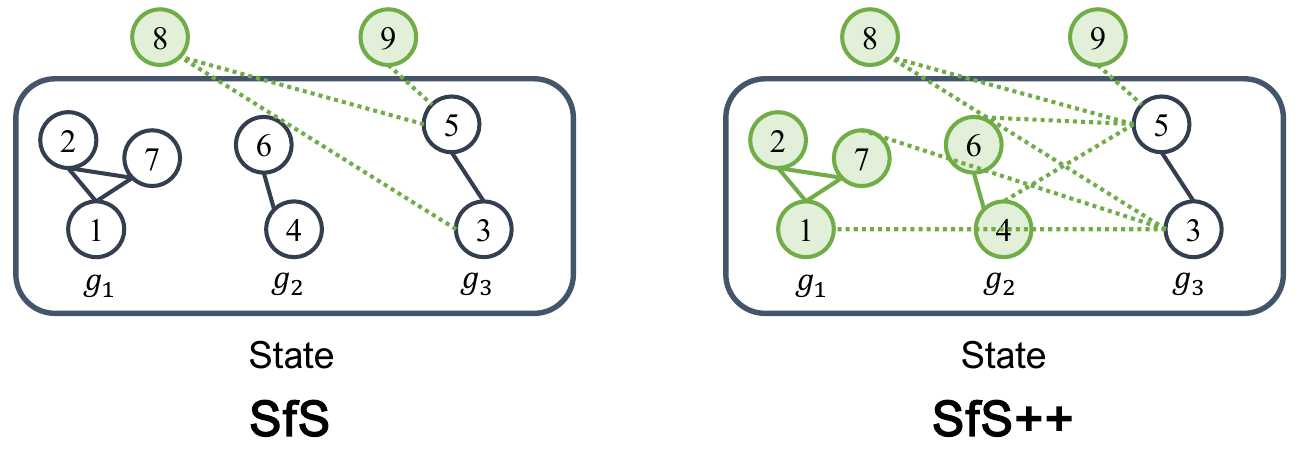}
	\caption{
	    An illustration of the priority candidates is shown in green color. Solid lines indicate nodes reconstructed in previous iteration, while dotted lines represent edges obtained from feature matching (Sec. \ref{sec:pairwise_matching}). Each state contains multiple graphs, with each individual graph representing a separate pottery reconstruction. 
	}
	\figlabel{fig_priority}
    \vspace{-3mm}
\end{figure}

\pparagraph{Generation of priority candidates list}
Even after filtering out implausible configurations through the \textit{sherd registration} process, numerous false positive, indistinguishable potential matching pairs and similar configurations remain due to the inherent properties of ceramic pots (as described in Sec.~\ref{sec:challenges}). To address this, we combine similar matching candidates and prioritize them by evaluating their likelihood. 

First, the results of the \textit{sherd registration} process are grouped based on the similarity of their transformation matrices, using thresholds of 25 $^\circ$ for rotation and 20 mm for translation.
Next, we compute the average transformation matrix of the combined pairs to serve as the initial condition for the \textit{global sherds adjustment} (Sec~\ref{sec:expansion}).
Then, a score for each combined matching pair is calculated based on the number of inliers.
Based on these scores, the candidates are sorted, resulting in a prioritized list of candidates, $L_{P}(s_i)=[\Phi_1, \Phi_2, \cdots]$, where $\Phi_j=\Phi_{j}^{\mathrm{old}} \cup \Phi_{j}^{\mathrm{new}}$ for the state $s_i\in\mathcal{S}$. 
These candidates represent the most probable reassembly configurations at the current state.

\subsubsection{State expansion}
\label{sec:expansion}
Based on the generated priority candidates (e.g. list of most probable expansion path from the current state), the algorithm expands the state with $b$ potential matching configurations (line 16-22).
Each expanded state then undergoes a global optimization step\textemdash referred to as \textit{global sherds adjustment}\textemdash followed by geometric verification (line 17). 
If an expanded state satisfies the geometric criteria, it is appended back to the set of states ($\mathcal{S}$) and a global reassembly score is computed (line 18-21). Notably, each expanded state represents a distinct potential outcome.

\pparagraph{Global sherds adjustment.}
Global sherds adjustment is conceptually similar to bundle adjustment in incremental SfM, its goal is to enhance global consistency among sherds while progressively eliminating erroneous configurations in noisy graphs.
For each state, we select the top-$b$ ranked priority candidates for global configuration adjustment. 
At this stage, we solve the similar optimization described at Eq.~\eqref{eq:SherdRegistration}; however, the set of relevant sherds now includes all sherds, defined as $\Phi = \Phi^{\mathrm{old}} \cup \Phi^{\mathrm{new}}$, where $ \Phi^{\mathrm{old}} \cap \Phi^{\mathrm{new}} = \emptyset$.
Unlike the sherd registration process, previously registered sherds in $\Phi^{\mathrm{old}}$ also participate in this adjustment. 
With a slight abuse of notation, let's define transformation matrices set of sherds in $\Phi$ as $\{\m T^{\Phi}\} = \{\m T^C| \forall C \in \Phi\}$. We then solve the following global optimization problem:
\begin{align}
\argmin_{\{\m T^\Phi\},r^{\mathrm{rim}},h^{\mathrm{rim}}} 
\sum_{A, B\in \Phi} \frac{1}{2} J_{R}(A,B) + \sum_{E\,\in\,\Phi} J_{I}(E)
\eqlabel{eq:GlobalCostFunction}
\end{align}
where $J_R$ and $J_I$ are defined in Eqs. \eqref{eq:J_R}-\eqref{eq:J_I}. As in Sec.~\ref{sec:refining_matches}, we employ the LM algorithm for nonlinear optimization and utilize the ICP algorithm to update correspondences. Finally, the globally adjusted configurations are verified against the geometric criteria.

\pparagraph{State pruning.}
During each iteration of the beam search, different paths may yield to the same reconstruction outcome. We actively identify and eliminate these redundant beams to maintain efficiency (line 24). 
The set of states ($\mathcal{S}$) is now filled with expanded state, which are then sorted based on their reassembly scores. Only top-$k$ states are retained for next iteration (line 26).

\begin{table*}[t]
\centering
\footnotesize
\caption{
    Preprocessing results of enhanced single pottery data. Initial pairwise matches were computed after pairwise feature matching results, and the geometric screening process pruned pairwise matches after pairwise ICP. The reduction ratio represents the difference between initial and pruned pairwise matches.
}
\begin{tabular}{l l || c | c | c | c | c | c | c | c | c |c }
\Xhline{1.pt}
\multicolumn{2}{l||}{\textbf{Pot ID}} & \textbf{A} & \textbf{B} & \textbf{C} & \textbf{D} & \textbf{E} & \textbf{F} & \textbf{G} & \textbf{H}& \textbf{I} & \textbf{J} \\  \hline
\multicolumn{2}{l||}{Pottery Type} & dish & dish & tea pot & vase & vase & dish & dish & dish & vase & tea pot \\ 
\multicolumn{2}{l||}{\# sherds} & 8 & 9 & 4 & 29 & 31 & 6 & 7 & 11 & 27 & 11 \\
\multicolumn{2}{l||}{\# base sherds (detected / actual)} & 0 / 1 & 1 / 1  & 1 /1  &  6 / 3  & 7 / 4  &  3 / 3 & 4 / 4  & 6 / 5  & 6 / 3 & 3 / 1\\
\multicolumn{2}{l||}{\# false negative base sherds} & 0 & 0  & 0  &  3  & 3  &  1 & 1  & 1  & 3 & 2\\
\multicolumn{2}{l||}{\# rim sherds (detected / actual)} & 5 / 6 &  7/ 7  &  0/ 3 & 1 / 1 &  0 / 1 & 1 / 6 & 4 / 6 & 1 / 8 & 0 / 2  & 0 / 3\\
\multicolumn{2}{l||}{\# Ground truth edges } &  15 &  15  &  5 &  69 & 62 & 7 & 10 & 17 & 65 & 21 \\
\hline
\multicolumn{2}{l||}{\# initial pairwise matches} & 124 & 114 & 125 & 2445 & 1912 & 36 & 51 & 485 & 1424 & 731 \\
\multicolumn{2}{l||}{\# pruned pairwise matches} & 70  & 77  & 40 & 891 & 823 & 27 &  31 & 232 & 667 & 200 \\
\multicolumn{2}{l||}{Reduction ratio (\%)}& 43.5 & 32.5 & 68.0 & 63.6  & 57.0 & 25.0 &  39.2 & 52.2 & 53.2 & 72.6 \\
\Xhline{1.pt}
\end{tabular}
\label{tbl:preprocessing_single}

    \vspace{-3mm}
\end{table*}

\section{Experiments}
We conducted comprehensive experiments to evaluate the performance of the proposed pipeline under various conditions. Through these experiments, we aim to answer the following questions: \textit{First}, how does the performance of the proposed method compare to other SoTA approaches, including the previous SfS method~\cite{iccvsfs}? \textit{Second}, what is the effect of different beam search parameters ($k$ and $b$)? \textit{Third}, how does the enhanced edge line data influence the results? \textit{Last}, what are the benefits of a base-agnostic pipeline?

\subsection{Datasets}
\label{sec:exp_datasets}
\pparagraph{Proposed real pottery dataset.}
We construct a dataset comprising 142 fragments from 10 distinct potteries, which includes an additional set of 5 potteries in addition to the previous dataset~\cite{iccvsfs}.
Pots A to E are the same as those used in the last study, while F to J are the newly introduced objects.
Specifically, we introduce new dishes (e.g. pots F, G, and H), two pairs of identical potteries but shattered differently (e.g. pair (E, I) and (C, J)), and containing missing fragments (e.g. pots F and H).
These scenarios reflect common situations encountered in real excavation sites, posing a more challenging restoration problem for experts. The profiles of these pots are identical, or missing sherds reduce the connectivity between fragments. 
Pot H shows a distinctively flat shape, adding further diversity to our dataset. For detailed information on our dataset, including the number of bases, rims, and pairwise matching edges, please refer to Table \ref{tbl:preprocessing_single} and \ref{tbl:preprocessing_multire}. For the detection of base fragment, we followed the previous work~\cite{iccvsfs} (check supplementary material~\cite{supmat_journal} for more detail).

We prepare 3 different types of datasets: non-enhanced dataset, obtained using previous method~\cite{iccvsfs}; enhanced dataset generated using an enhanced edge line method (see Sec.~\ref{sec:data_preprocessing}); and the ground truth data. 
In the dataset preparation process, we exclude very small fragments comprising fewer than 50 points in the respective surface edge lines (with a perimeter of approximately 9cm) and special parts such as the handle, nose, or ornament. This results in removing 1 piece from pot C, 2 pieces from pot D, 4 pieces from pot E, 2 pieces from pot F, 3 pieces from pot I, and 8 pieces from pot J. 

\begin{table}[t]
\centering
\footnotesize
\caption{
    Preprocessing results of enhanced multiple potteries data.  
    * indicates that the pots are identical but have different ways of breaking. Dishes contain pot IDs A, B, F, G, and H. Initial pairwise matches were computed after pairwise feature matching results, and pruned pairwise matches were pruned by the geometric screening process after pairwise ICP. The reduction ratio represents the difference between initial and pruned pairwise matches.
}
\resizebox{\columnwidth}{!}{
\begin{tabular}{l l || c | c | c | c | c  }
\Xhline{1.pt}
\multicolumn{2}{l||}{Pot ID} & CJ$^*$ & EI$^*$ &  Dishes & ABCDE & ALL \\ \hline
\multicolumn{2}{l||}{\# sherds} & 15 & 58 & 41 & 81 & 142 \\
\multicolumn{2}{l||}{\# Ground truth edges } &  26 &  127  &  64 &  166 & 286   \\
\multicolumn{2}{l||}{\# initial pairwise matches} & 1538 & 6256 & 2433 & 11602 & 35388   \\
\multicolumn{2}{l||}{\# pruned pairwise matches} & 427 & 2756 & 1150& 4183 & 12479   \\
\multicolumn{2}{l||}{Reduction ratio (\%)} & 72.2 & 55.9 & 52.7 & 63.9 & 64.7  \\
\Xhline{1.pt}
\end{tabular}
\label{tbl:preprocessing_multire}
}
    \vspace{-3mm}
\end{table}

\pparagraph{Generating ground truth data.}
We received a collection of reconstructed pot images, each labeled by a team of restoration experts. Leveraging this information, we manually positioned the sherds in their correct locations using CloudCompare\footnote{https://cloudcompare.org/}. These correctly placed sherds served as the initial states for conducting ICP on the fracture surfaces between individual sherds, resulting in precise reassembly models. Then, for each pot, we recorded the absolute transformation of each sherd with respect to its initial location, and we obtained ground truth axis information from these models. 
Also, we generated a ground truth matching matrix ($\m M$) between sherd $i$ and $j$, where $[\m M]_{ij}=1$ if they are connected, otherwise $[\m M]_{ij}=0$. 
Ground truth datasets are based on enhanced edge lines with correctly identified rim and base indexing.

\pparagraph{Breaking Bad.}
Breaking Bad~\cite{breakingbad} is a large-scale synthetic dataset containing physically realistic fracture patterns of various objects, serving as a crucial benchmark for the reassembly of fractured objects. 
In this study, we constructed a test dataset by selecting three objects from each of the Bowl, Plate, and Vase classes.
The selection criteria were limited to axially-symmetric objects with fragments containing both interior and exterior surfaces, excluding those with additional elements such as food.
Since the dataset is normalized, we scaled it to match the size of our dataset for experiments with our model, which is designed for real-world scenarios, while retaining the original scale for experiments with other deep learning models.
Additionally, these objects were taken from test datasets that the deep learning models had not trained on.

\begin{table*}[t]
\centering
\caption{
    Reassembly results of single pot  with enhanced data compared to other SoTA, SfS~\cite{iccvsfs} and \AlgName (ours). N/A indicates that the algorithm can not process the pottery due to the limited number of input fragments. We used $k=5$ and $b=3$ beam search parameters for SfS pipelines. We represent the result with respect to the pot ID and the number of fragments with parenthesis and use two different metrics: Sherd accuracy (SA) and Edge accuracy (EA).}
\begin{tabular}{l| l||  c | c | c | c | c | c | c | c | c | c }
\Xhline{1.pt}
\textbf{Metric} & \textbf{Method}  &\textbf{ A (8)} & \textbf{B (9)} & \textbf{C (4)} & \textbf{D (29)} & \textbf{E (31)} &\textbf{ F (6)} & \textbf{G (7)} &\textbf{ H (11)} & \textbf{I (27)} & \textbf{J (11)} \\  
\Xhline{1pt}
\multirow{5}{*}{SA (\%) $\uparrow$} & Jigsaw~\cite{jigsaw} & 25.0 & 22.2 & 0.0 & N/A & N/A &  0.0 & 42.9 & 0.0 & N/A & 0.0 \\ 
 & FRASIER~\cite{joo24} & 0.0 & 22.2 & \textbf{100} &  N/A &  N/A & 33.3 & 0.0 & 0.0 &  N/A & 0.0\\
 & PuzzleFusion++~\cite{wang2024puzzlefusionpp} &  62.6 & 33.3 & 50.0 & N/A & N/A & \textbf{100} & \textbf{100} & 0.0 & N/A & 0.0\\
 \cline{2-12}
 & SfS (ours)~\cite{iccvsfs} & 0.0 & \textbf{100} & \textbf{100} & \textbf{82.1} &\textbf{100} & \textbf{100} & 0.0 & 81.8 & 88.9& 18.2\\
 & \AlgName (ours)  & \textbf{100} &  \textbf{100} & \textbf{100} & \textbf{82.1} & \textbf{100} & \textbf{100} & \textbf{100} &  \textbf{90.9} & \textbf{92.6} & \textbf{72.7} \\
\hline
\multirow{5}{*}{EA (\%) $\uparrow$} & Jigsaw~\cite{jigsaw} & 6.7 & 6.7 & 0.0 & N/A & N/A &  0.0 & 20.0 & 0.0 & N/A & 0.0 \\ 
 & FRASIER~\cite{joo24} & 0.0 & 6.7 & 40.0 &  N/A  & N/A & 11.1 & 0.0 & 0.0 & N/A & 0.0\\
 & PuzzleFusion++~\cite{wang2024puzzlefusionpp} &  46.7 & 20.0 & 20.0 & N/A & N/A & \textbf{100} & \textbf{100} & 0.0 & N/A & 0.0\\
 \cline{2-12}
 & SfS (ours)~\cite{iccvsfs} & 0.0 & \textbf{100}  & \textbf{100} & 74.2 & 77.4& \textbf{100} & 0.0 & 76.5 &70.8 & 4.8\\
 & \AlgName (ours)  & \textbf{100}  & \textbf{100}  & \textbf{100}  & \textbf{83.3} & \textbf{100} & \textbf{100}  & \textbf{100} & \textbf{88.2} & \textbf{90.8} & \textbf{76.2} \\
\Xhline{1.pt}
\end{tabular}
\label{tbl:single_results}
    \vspace{-3mm}
\end{table*}

\begin{table*}[t]
\centering
\caption{Reassembly results for a single pot from the Breaking Bad dataset~\cite{breakingbad}, showing enhanced performance compared to other state-of-the-art methods. The beam search parameters used for the SfS pipelines were $k=5$ and $b=3$. Results are presented with the pot ID and number of fragments in parentheses, evaluated using two metrics: Sherd Accuracy (SA) and Edge Accuracy (EA).}
\resizebox{2\columnwidth}{!}{
\begin{tabular}{l| l|| c | c | c | c | c | c | c | c | c  }
\Xhline{1.pt}
\textbf{Metric} & \textbf{Method}  & \textbf{Plate 1 (12)} & \textbf{Plate 2 (8)} & \textbf{Plate 3 (7)} & \textbf{Bowl 1 (10)} & \textbf{Bowl 2 (11)} & \textbf{Bowl 3 (6)} & \textbf{Vase 1 (8)} & \textbf{Vase 2 (9)} & \textbf{Vase 3 (9)}\\  
\Xhline{1pt}
\multirow{4}{*}{SA. (\%) $\uparrow$} 
& Jigsaw~\cite{jigsaw}  & 16.7 & 0.0 & 28.6 & 0.0 & 0.0 & 0.0 & 50.0 & 0.0 & 0.0 \\
& FRASIER~\cite{joo24}  & 33.3 & 50.0 & \textbf{100} & 20.0 & 54.5 & 83.3 & 25.0 & 0.0 & 0.0 \\
& PuzzleFusion++~\cite{wang2024puzzlefusionpp}  & 0.0 & 50.0 & 0.0 & 20.0 & 18.2 & 50.0 & \textbf{100} & 55.6 & \textbf{100} \\
& \AlgName (ours) & \textbf{83.3} & \textbf{100} & \textbf{100} & \textbf{80.0} & \textbf{90.0} & \textbf{100} & 75.0 & \textbf{88.9} & \textbf{100} \\
\hline

\multirow{4}{*}{EA (\%) $\uparrow$}  
& Jigsaw~\cite{jigsaw}   & 4.5 & 0.0 & 12.5 & 0.0 & 0.0 & 0.0 & 25.0 & 0.0 & 0.0 \\
& FRASIER~\cite{joo24} & 9.1 & 33.3 & \textbf{100} & 9.1 & 35.7 & 27.3 & 8.3 & 0.0 & 0.0 \\
& PuzzleFusion++~\cite{wang2024puzzlefusionpp}  & 0.0 & 25.0 & 0.0 & 9.1 & 7.1 & 27.3 & \textbf{100} & 35.7 & \textbf{100} \\
& \AlgName (ours)   & \textbf{81.8} & \textbf{100}  & \textbf{100} & \textbf{38.1}  & \textbf{83.3}  & \textbf{100} & 75.0 & \textbf{71.4} & \textbf{100} \\

\Xhline{1.pt}
\end{tabular}
\label{tbl:breakingbad_results}
    \vspace{-3mm}
}
\end{table*}

\subsection{Experimental setup}
\label{sec:exp_setup}

\pparagraph{Implementation details.}
{For comparison, we used the official code for Jigsaw, Fraiser, and PuzzleFusion++, performing inference using pre-trained weights on BreakingBad dataset. 
Our algorithm (SfS++) was evaluated on a system with an AMD Ryzen 9 5900X processor and an NVIDIA RTX 3080Ti GPU, while the other deep learning models were run on a system with an Intel i9 13900K processor and an NVIDIA RTX 4090 GPU.}

\pparagraph{Evaluation metrics.}
In contrast to the previous study~\cite{iccvsfs}, this pipeline does not rely on the base fragment for reconstruction, which makes it challenging to assess success based on specific sherds (previously, we used base sherds as a reference). To address this, we introduce two metrics, edge accuracy, and sherd accuracy, similar to the previous studies (part accuracy)~\cite{huangGenerative3DPart2020, harishRGLNETRecurrentGraph2022, wang2024puzzlefusionpp}.
Although we use relative transformation difference instead of Chamfer distance, both methods ultimately assess the accuracy of relative positions and orientations between sherds, providing comparable evaluation metrics.
Specifically, we define $\m M$ as an adjacent matrix, where the element $[\m M]_{ij}$ is one if the sherd pair $(i, j)$ is connected; otherwise, it is zero. Then, the edge accuracy (EA) is defined as follows:
\begin{align}
    \text{EA}:= \frac{1}{|\m M|}\sum_{(i, j) \in \m M}  &\mathbb{I}\Big\{ \|\m R^{ij} \ominus \m R_{GT}^{ij}\|_{2}^{2} < \tau_R \\ \nonumber
    & \cap\|\v t^{ij}- \v t_{GT}^{ij}\|_{2}^{2} < \tau_t \Big\}
\end{align}
where $\m R^{ij}$ and $\v t^{ij}$ are a relative rotation and translation between sherds $i$ and $j$, respectively. $\ominus$ represents the smallest angle difference between two rotation matrices. 
The Sherd accuracy (SA) is obtained by counting the number of correctly reconstructed sherds with at least one correct edge. We set threshold values of $\tau_R = 20^\circ$ for 3D rotation (excluding axis deviation) and $\tau_t = 50$mm for translation.

\subsection{Results on the reassembly of single and mixed potteries}
\label{sec:single_multi_result}
In this experiment, we show the superior performance of the proposed method compared to other methods, including Jigsaw~\cite{jigsaw}, PuzzleFusion++~\cite{wang2024puzzlefusionpp}, FRASIER~\cite{joo24}, and SfS~\cite{iccvsfs}. 

\pparagraph{Reassembly of single pottery.}
As shown in Table \ref{tbl:single_results} and Fig. 1 in the supplementary material \cite{supmat_journal}, our method, SfS++, outperforms data-driven models such as Jigsaw, PuzzleFusion++, and FRASIER in single pottery reassembly experiments using our real dataset.
Data-driven models primarily rely on extracting features from the fracture surfaces of the sherds.
However, in our dataset, the fracture surfaces are thin, making it challenging for these models to capture sufficiently distinctive features. 
This limitation mainly affects the performance of models like PuzzleFusion++ when the surface area is small, and the features are less distinct.

\begin{figure}[t]
	\centering
	\includegraphics[width=0.95\linewidth]{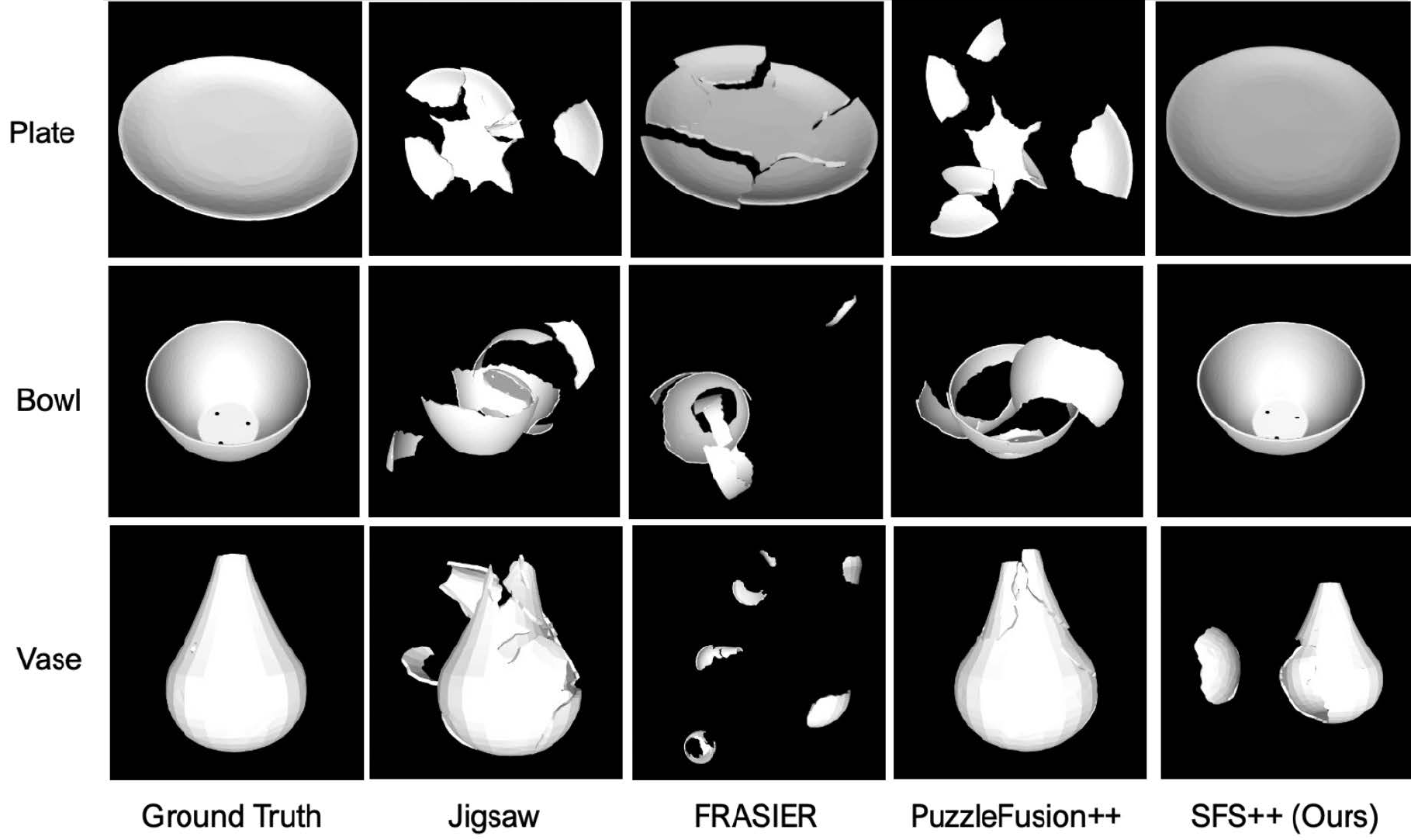}
	\caption{
	    Comparison of single-object assembly results between other deep learning methods and \AlgName with $k=5$ and $b=3$ (ours) from Breaking Bad dataset~\cite{breakingbad}.
        Only one example per category is visualized here; further detailed visualizations are provided in the supplementary material \cite{supmat_journal}.}
	\label{Result_BB_sample}
    \vspace{-3mm}
\end{figure}

In contrast, \AlgName leverages axis-symmetric information computed from sherd surfaces and edge line geometry to facilitate more robust feature matching.
As a result, it achieves superior performance in both SA (sherd accuracy) and EA (edge accuracy) compared to data-driven approaches.

Moreover, our method demonstrates strong performance on the Breaking Bad dataset (Table \ref{tbl:breakingbad_results} and Fig.~\ref{Result_BB_sample}).
While our primary focus is on real datasets, \AlgName also generalizes effectively to synthetic datasets and yields competitive results compared to data-driven models.
This highlights the method’s adaptability across diverse data types, from synthetic environments to the real world.

When comparing SfS~\cite{iccvsfs} with SfS++, the previous method is a base-dependent method that relies on the base sherd to guide the matching process. 
As shown in Fig.~\figref{fig_base}, base reassembly is relatively easy for D due to the sharp transition from the flat surface to the sides at the base. 
However, Pot G has a limited flat area and a subtle curvature change between the base and the sides, making it challenging to identify the base fragment. 
To address this issue, SfS++ eliminates the reliance on the base and adopts a more flexible matching approach, effectively handling fragments of varying sizes and shapes.
As demonstrated in Table \ref{tbl:single_results}, SfS++ outperforms SfS~\cite{iccvsfs}, especially in challenging cases such as Pots H, I, and J.

Finally, \AlgName shows high quality reassembly results, as shown in Fig.~\ref{Result_BB_sample} due to the global optimization process (see Sec.~\ref{sec:expansion} for more detail)\footnote{More results can be found on the supplementary material~\cite{supmat_journal}}.

\begin{figure*}[t]
	\centering
	\includegraphics[width=0.95\textwidth]{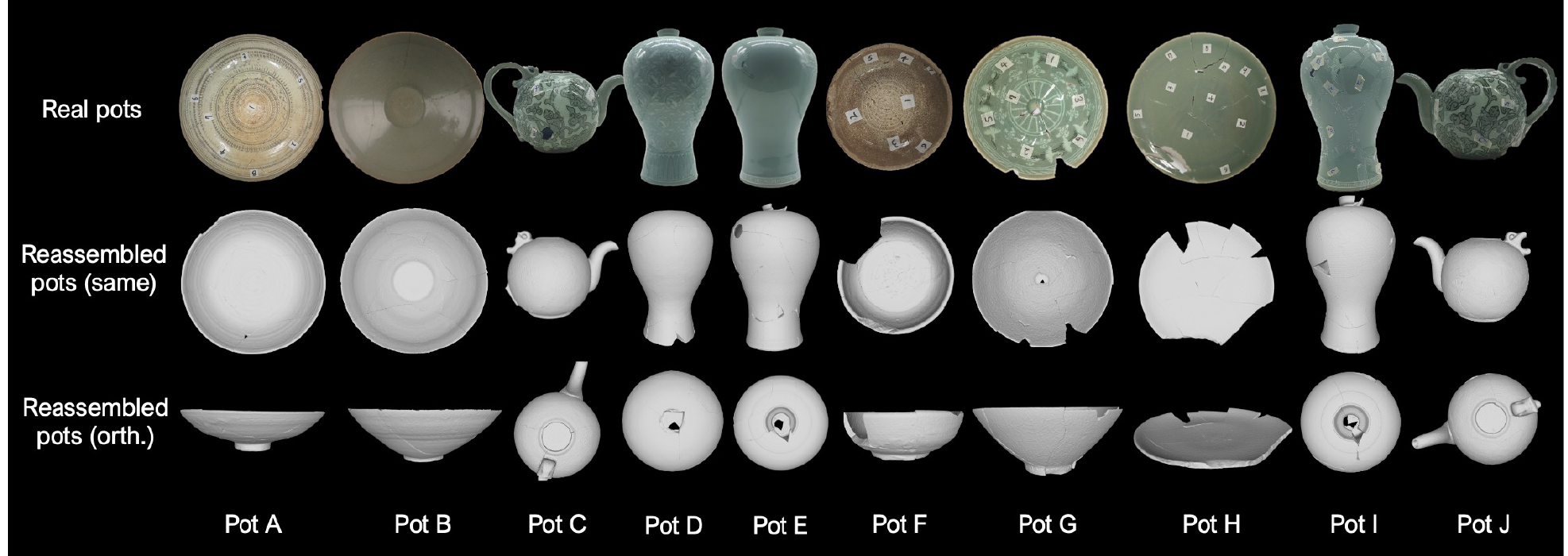}
	\caption{ 
        Visualization of the reassembly results of mixed potteries using \AlgName with $k=20$ and $b=10$ is shown in the figure. The top row displays the corresponding real pottery examples, while the middle and bottom rows illustrate the reassembled results. A total of 10 potteries, consisting of 142 fragments, were randomly mixed. The middle row represents the reassembled pots from the same perspective as their corresponding real pottery, whereas the last row provides an orthogonal view (top or side). For the 3D visualization of the results, please visit our project homepage (https://sj-yoo.info/sfs/).}

	\figlabel{fig_allresult}
    \vspace{-3mm}
\end{figure*}

\begin{table}[t]
\centering
\caption{
    Reassembly results of multiple mixed  potteries, showing enhanced performance compared to other state-of-the-art methods, SfS~\cite{iccvsfs} and \AlgName{} (ours). The beam search parameters for the SfS pipelines were set to $k=20$ and $b=10$. The number of fragments is indicated in parentheses under each pot ID, and two metrics are used for evaluation: Sherd Accuracy (SA) and Edge Accuracy (EA). Pots marked with * are identical but broken in different ways. Dishes include pot IDs A, B, F, G, and H.
}
\resizebox{\columnwidth}{!}{
\begin{tabular}{c| c||c | c | c | c | c  }
\Xhline{1.pt}
\multirow{2}{*}{\textbf{Metric}} & \multirow{2}{*}{\textbf{Method}} & \textbf{CJ}$^*$& \textbf{EI}$^*$&  \textbf{Dishes}& \textbf{ABCDE} & \textbf{ALL} \\
& & \textbf{(15)} & \textbf{(58)} &\textbf{(41)} & \textbf{(81)} & \textbf{(142)} \\     
\Xhline{1pt}
\multirow{2}{*}{SA (\%) $\uparrow$} & SfS (ours)~\cite{iccvsfs}  & 66.7 & 62.1 & 43.9  & 83.8  & 29.6 \\
 & \AlgName (ours)   & \textbf{80.0} & \textbf{96.6} & \textbf{97.6} & \textbf{92.5}  & \textbf{87.3} \\
 \hline

\multirow{2}{*}{EA (\%) $\uparrow$} & SfS (ours)~\cite{iccvsfs}   & 30.8 & 38.6 & 34.1 & 73.3 & 16.8 \\
 & \AlgName (ours)   & \textbf{80.8} & \textbf{95.3} & \textbf{97.0} & \textbf{90.8} & \textbf{85.1} \\
 \Xhline{1.pt}
\end{tabular}
\label{tbl:multi_results}
    \vspace{-3mm}
}
\end{table}

\pparagraph{Reassembly of mixed potteries.}
{In the mixed potteries setting, our \AlgName achieves a state-of-the-art result in the ABCDE mixed pottery scenario, attaining a SA of 92.5 \% (an 8.7 \% improvement over the previous state-of-the-art \cite{iccvsfs}), as shown in Table~\ref{tbl:multi_results}.}

We also achieve an SA of 87.3 \% in a total of 10 mixed pottery experiments comprising 142 fragments.
Notably, \AlgName outperforms in the CJ and EI experiments, which involved the same shapes of pots but were shattered differently. SfS demonstrates a significant performance drop compared to single pot experiments. This deterioration is due to the failure in base reconstruction caused by the strong shape ambiguity, which is particularly challenging to resolve in the early registration stages when only a few fragments are available.

Furthermore, we test PuzzleFusion++~\cite{wang2024puzzlefusionpp} and \AlgName on mixtures of multiple pots from the Breaking Bad dataset~\cite{breakingbad}, as shown in Table~\ref{tab:BB_mix}. When two objects are mixed, the number of false positive matching pairs dramatically increases, and the algorithm is required to handle them properly. However, PuzzleFusion++ is unable to reassemble the objects under noisy conditions. For example, PuzzleFusion++ successfully reassembles Vase A in a single object test but fails when it is mixed with another, as shown in Fig.~\figref{fig_result_mix}. 
On the other hand, \AlgName can handle realistic scenarios, mostly mixed fragments from multiple different objects, involving not only two objects but also more than ten objects.

\begin{table}[t]
\centering
\caption{
    Reassembly results for multiple mixed pots from the Breaking Bad dataset~\cite{breakingbad}. The parameters for the \AlgName experiment were set to $k=5$ and $b=3$, with evaluation based on two metrics: Sherd Accuracy (SA) and Edge Accuracy (EA).
}
\begin{tabular}{c| c||c | c   }
\Xhline{1.pt}
\multirow{2}{*}{\textbf{Metric}} & \multirow{2}{*}{\textbf{Method}}  & \textbf{Bowl 2 + 3} & \textbf{Vase 1 + 3} \\
& & \textbf{(16)} & \textbf{(17)} \\
\Xhline{1pt}
\multirow{2}{*}{SA (\%) $\uparrow$} & PuzzleFusion++~\cite{wang2024puzzlefusionpp} & 0.0& 23.5 \\
 & \AlgName (ours)  & \textbf{93.8} & \textbf{88.2} \\
 \hline

\multirow{2}{*}{EA (\%) $\uparrow$} & PuzzleFusion++~\cite{wang2024puzzlefusionpp} & 0.0 & 6.5 \\
 & \AlgName (ours)   & \textbf{91.3} & \textbf{90.3}\\
 \Xhline{1.pt}
\end{tabular}
\label{tab:BB_mix}
    \vspace{-3mm}
\end{table}

\subsection{Ablation study}
\label{sec:ablation}
\pparagraph{Effect of different beam search parameters.}
As shown in Table~\ref{tab:abl1_beam_param}, we conducted an ablation study on different beam search parameters within the all-pottery mixed scenario. The beam search algorithm has two main parameters: $k$, which represents the number of beams, or the most likely states at each step, and $b$, which indicates the breadth of the algorithm's search for possible expansions for the next step. Performance constantly improves as we increase the values of $k$ and $b$. We observe a significant performance jump from $k=5, b=3$ to $k=5, b=5$. This improvement is primarily due to the large number of false positive pairs in all mixed pottery scenarios, as shown in Table~\ref{tbl:preprocessing_multire}. These false positive pairs disturb early-stage reconstruction because it is challenging to distinguish between true and false configurations with only a few sherds. The beam search algorithm effectively handles this ambiguity by utilizing broader searching branches with higher values of $b$. However, there is a trade-off between higher $b$ and increased runtime. 

\begin{figure}[t]
	\centering
	\includegraphics[width=0.8\linewidth]{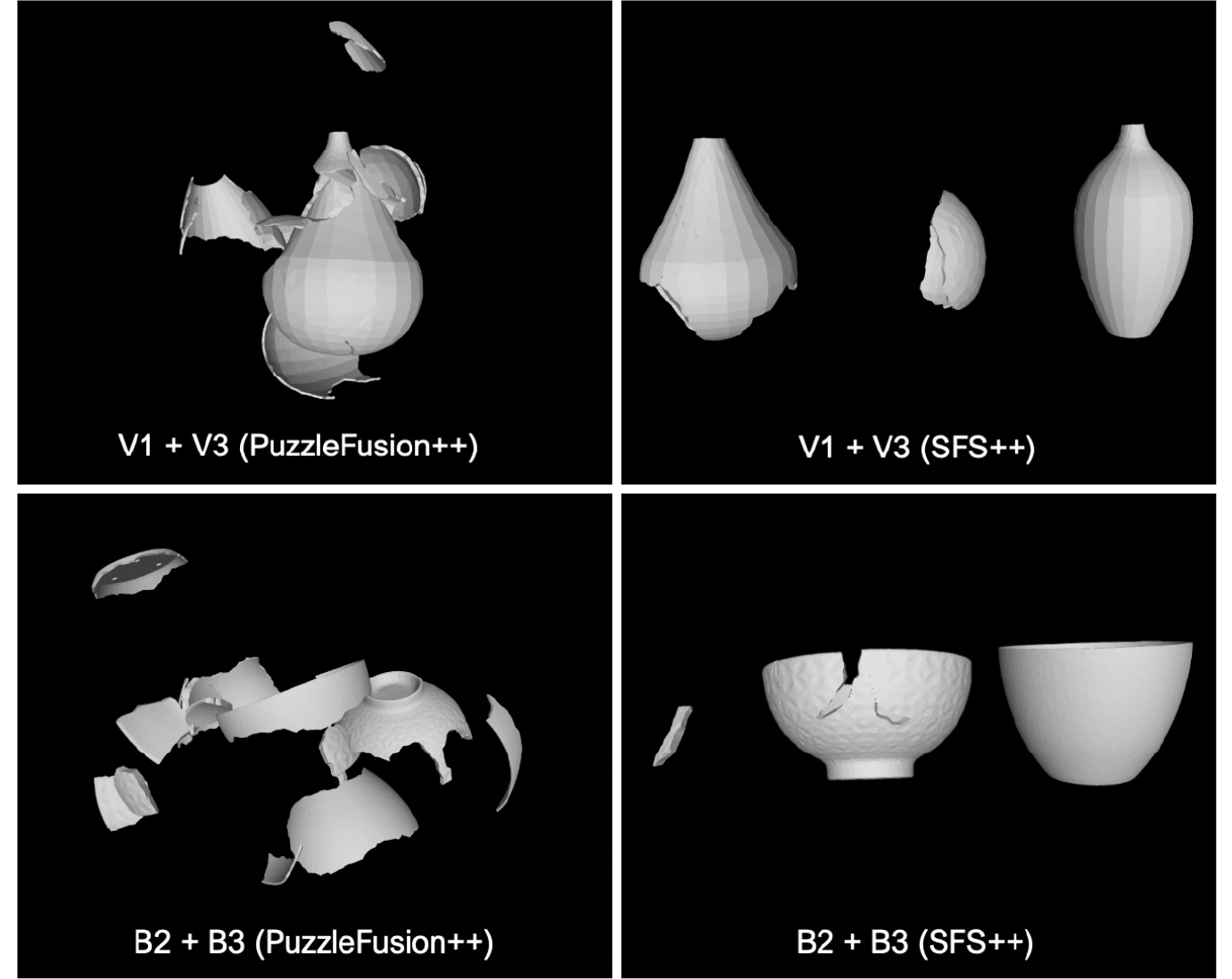}
	\caption{
	    Comparison of reassembly results of mixed potteries between PuzzleFusion++ and \AlgName with $k=5$ and $b=3$ (ours) from Breaking Bad dataset~\cite{breakingbad}.
     The left column shows the reconstruction results using PuzzleFusion++, while the right column is SfS++. 
	}
	\figlabel{fig_result_mix}
    \vspace{-3mm}
\end{figure}

\pparagraph{Effect of different data preprocessing methods.}
As shown in Fig.~\figref {fig:improved_breakline}, our enhanced dataset demonstrates better edge line quality. This naturally raises the question: how much of the performance gain can be attributed to the improved dataset? To answer this, we conducted an ablation study on the performance across different datasets, as shown in Table~\ref{tab:abl2_dataset}. We observed consistent performance improvements on better preprocessing methods from non-enhanced data to enhanced data and ground truth data. 
Note that SfS exhibits a performance drop on Enhanced data due to its reliance on a base-driven assembly approach.
As shown in Fig.~\figref{fig:improved_breakline}, previous preprocessing method extracts the edge line by positioning it slightly inward, whereas the enhanced preprocessing method yields a breakline that closely approximates the true edgeline.
This change in the edge characteristics results in altered descriptor values. However, since the existing SfS hyperparameters were optimized for the previous edge characteristics, these changes in descriptor values can trigger false positive matches during base reconstruction, thereby degrading performance as shown in Table~\ref{tbl:preprocessing_single}.
As described in Sec.~\ref{sec:why_optim}, we can reasonably expect better performance by leveraging recent data-driven methods to enhance the quality of data preprocessing since our pipeline offers a standard structure for addressing the problem.

\subsection{Benefits of base-agnostic pipeline}
\label{sec:justification}
\pparagraph{Robustness.} Compared to previous work~\cite{iccvsfs}, our proposed pipeline is independent of base constraints. 
For instance, in Fig~\figref{fig_base}, Pot G exhibits smaller base features than the others, making it challenging to correctly classify the base without having a false positive. 
As a result, SfS (base-dependent pipeline) fails to reassemble Pot A and G, as shown in Table~\ref{tbl:single_results}, because it cannot initiate the pipeline without a correctly reassembled base as the starting state. 
Furthermore, SfS suffers from false negative base classifications, as shown in Table \ref{tab:abl2_dataset}. Since Pot E was mistakenly classified as a base fragment in our enhanced data, SfS achieved even lower accuracy than when using the original non-enhanced dataset, which did not contain this misclassification result \cite{iccvsfs}. 
In contrast, the performance of our pipeline is not affected by these base classification results. 

\begin{table}[t]
    \centering
    \caption{Ablation study 1: Effect of different beam search parameters ($k$ and $b$) tested on all mixed pottery experiments.}
    \begin{tabular}{l|c|c|c|c}
    \Xhline{1.pt}
       \textbf{$(k, b)$}  & (5, 3) &(5, 5) & (10, 5) & (20, 10) \\
         \Xhline{1pt}
         SA (\%) $\uparrow$ & 40.1 & 83.1& 85.9 & 87.3  \\
         EA (\%) $\uparrow$ & 41.6 & 73.3&84.6 & 85.1 \\ 
         Runtime (hour) & 4.8& 22.5 &42.7 & 137.0 \\
         \Xhline{1.pt}
    \end{tabular}
    \label{tab:abl1_beam_param}
    \vspace{-2mm}
\end{table}

\begin{table}[t]
    \centering
    \caption{Ablation study 2: Effect of different data preprocessing methods including non-enhanced(blue in Fig.~\figref{fig:improved_breakline}), enhanced(red in Fig.~\figref{fig:improved_breakline}), and ground truth. 
    The algorithms are tested on ABFGH potteries mixed experiment with $k=20$ and $b=10$ beam search parameters and reported with edge accuracy (EA) metric in percentage.}
    \begin{tabular}{l|c|c|c}
    \Xhline{1.pt}
          \backslashbox[20mm]{\textbf{Method}}{\textbf{Data}} & Non-enhanced & Enhanced & Ground truth \\
        \Xhline{1pt}
           SfS~\cite{iccvsfs}& 73.4 & 34.1 & 95.1 \\
           SfS++ & \textbf{93.8} & \textbf{97.0} & \textbf{100}  \\
    \Xhline{1.pt}
    \end{tabular}
    \label{tab:abl2_dataset}
    \vspace{-3mm}
\end{table}

\pparagraph{Flexibility.}
Since our pipeline does not impose initial constraints on the reconstruction process, Algorithm~\ref{alg:beam_search} can choose the initial start fragment solely based on matching priority. As a result, the searching path can expand broadly from the initial stage, leading to successful reconstruction results with a small number of branches ($b$) and beams ($k$) as demonstrated in Table~\ref{tab:abl1_beam_param}. The graph merging adds further flexibility to the algorithm by avoiding the case that discards well-reconstructed graphs from previous iterations instead of utilizing them. In the previous algorithm \cite{iccvsfs}, well-developed graphs would be removed if they fell behind others. However, the current approach allows merging them with others, considering them as potential reconstruction candidates (Sec.~\ref{sec:prep_expansion}. Additionally, the proposed approach builds easy pottery first (i.e. larger pottery), which aligns with our intuition. As a result, our pipeline can successfully reassemble a large number of fragments (i.e. 10 potteries with 142 fragments), while the Base-dependent pipeline failed to scale up. 

\subsection{Limitations}
Although our model demonstrates state-of-the-art performance across various datasets, it does have some limitations. 
Firstly, applying the model requires several preprocessing steps, such as surface extraction, fracture line adjustment, and counterclockwise alignment.
These processes can add extra time depending on the complexity of the data. 
Secondly, while we have tuned the parameters for real-world scenarios, some constraints require parameter adjustments or data scaling when the dataset changes.
Moreover, the model assembles fragments based on axes, limiting its application to axis-symmetric objects. 
This makes applying the model to asymmetric objects challenging, potentially reducing its flexibility in handling a wider variety of objects.

\section{Conclusion and future work}
In this work, we have addressed the problem of virtually reassembling axially symmetric pots from 3D scanned fragments.
To tackle this challenge, we proposed Structure-from-Sherds++ (SfS++), a novel optimization-based pipeline inspired by the incremental approach of Structure-from-Motion (SfM). 
SfS++ utilizes multi-graph beam search with an incremental reassembly strategy, enabling the individual, simultaneous, and base-agnostic reconstruction of multiple intermixed potteries in complex scenarios. 
Through experiments on a dataset of 142 real fragments from 10 different potteries, we demonstrated that SfS++ achieves a reconstruction accuracy of 87\%, surpassing existing state-of-the-art methods.

{While these results highlight our approach's strengths, base information is important for improving reconstruction in real-world scenarios.
Base fragment help identify mixed potteries and reduce false positives in feature matching.
In future work, we will utilize base information to enhance pottery identification, even under imperfect base segmentation.}

\ifCLASSOPTIONcaptionsoff
  \newpage
\fi

\bibliographystyle{IEEEtranSN}
\bibliography{egbib}

\begin{thebibliography}{46}
\providecommand{\natexlab}[1]{#1}
\providecommand{\url}[1]{#1}
\csname url@samestyle\endcsname
\providecommand{\newblock}{\relax}
\providecommand{\bibinfo}[2]{#2}
\providecommand{\BIBentrySTDinterwordspacing}{\spaceskip=0pt\relax}
\providecommand{\BIBentryALTinterwordstretchfactor}{4}
\providecommand{\BIBentryALTinterwordspacing}{\spaceskip=\fontdimen2\font plus
\BIBentryALTinterwordstretchfactor\fontdimen3\font minus \fontdimen4\font\relax}
\providecommand{\BIBforeignlanguage}[2]{{%
\expandafter\ifx\csname l@#1\endcsname\relax
\typeout{** WARNING: IEEEtranSN.bst: No hyphenation pattern has been}%
\typeout{** loaded for the language `#1'. Using the pattern for}%
\typeout{** the default language instead.}%
\else
\language=\csname l@#1\endcsname
\fi
#2}}
\providecommand{\BIBdecl}{\relax}
\BIBdecl

\bibitem[Cao and Mumford(2002)]{cao02}
Y.~Cao and D.~Mumford, ``Geometric structure estimation of axially symmetric pots from small fragments,'' in \emph{Proceedings of Signal Processing, Pattern Recognition, and Applications}, 2002.

\bibitem[Chen et~al.(2022)Chen, Li, Turpin, Jacobson, and Garg]{chenNeuralShapeMating2022}
Y.-C. Chen, H.~Li, D.~Turpin, A.~Jacobson, and A.~Garg, ``Neural {{Shape Mating}}: {{Self-Supervised Object Assembly With Adversarial Shape Priors}},'' in \emph{Proceedings of the {{IEEE}}/{{CVF Conference}} on {{Computer Vision}} and {{Pattern Recognition}}}, 2022, pp. 12\,724--12\,733.

\bibitem[Di~Angelo et~al.(2022)Di~Angelo, Di~Stefano, and Guardiani]{diangeloReviewComputerbasedMethods2022}
L.~Di~Angelo, P.~Di~Stefano, and E.~Guardiani, ``A review of computer-based methods for classification and reconstruction of {{3D}} high-density scanned archaeological pottery,'' \emph{Journal of Cultural Heritage}, vol.~56, pp. 10--24, Jul. 2022.

\bibitem[Duisterhof et~al.(2024)Duisterhof, Zust, Weinzaepfel, Leroy, Cabon, and Revaud]{mast3rsfm}
B.~Duisterhof, L.~Zust, P.~Weinzaepfel, V.~Leroy, Y.~Cabon, and J.~Revaud, ``Mast3r-sfm: a fully-integrated solution for unconstrained structure-from-motion,'' \emph{arXiv preprint arXiv:2409.19152}, 2024.

\bibitem[Graff(2018)]{graffArchaeologicalStudiesCooking2018}
S.~R. Graff, ``Archaeological {{Studies}} of {{Cooking}} and {{Food Preparation}},'' \emph{Journal of Archaeological Research}, vol.~26, no.~3, pp. 305--351, Sep. 2018.

\bibitem[Harish et~al.(2022)Harish, Nagar, and Raman]{harishRGLNETRecurrentGraph2022}
A.~N. Harish, R.~Nagar, and S.~Raman, ``{{RGL-NET}}: {{A Recurrent Graph Learning}} framework for {{Progressive Part Assembly}},'' in \emph{{{IEEE}}/{{CVF Winter Conference}} on {{Applications}} of {{Computer Vision}}}, Jan. 2022, pp. 647--656.

\bibitem[Hong et~al.(2019)Hong, Kim, Wi, and Kim]{hong19}
J.~H. Hong, Y.~M. Kim, K.-C. Wi, and J.~Kim, ``{PotSAC: A Robust Axis Estimator for Axially Symmetric Pot Fragments},'' in \emph{IEEE International Conference on Computer Vision Workshops}, Oct 2019.

\bibitem[Hong* et~al.(2021)Hong*, Yoo*, Arshard, Kim, and Kim]{supmat}
J.~H. Hong*, S.~J. Yoo*, M.~Z. Arshard, Y.~M. Kim, and J.~Kim, ``Supplementary document for structure-from-sherds: Incremental 3d reassembly of axially symmetric pots from unordered and mixed fragment collections,'' \url{https://github.com/SeongJong-Yoo/structure-from-sherds}, 2021.

\bibitem[Hong et~al.(2021)Hong, Yoo, Zeeshan, Kim, and Kim]{iccvsfs}
J.~H. Hong, S.~J. Yoo, M.~A. Zeeshan, Y.~M. Kim, and J.~Kim, ``Structure-from-sherds: Incremental 3d reassembly of axially symmetric pots from unordered and mixed fragment collections,'' in \emph{IEEE International Conference on Computer Vision}, 2021, pp. 5443--5451.

\bibitem[Huang et~al.(2020)Huang, Zhan, Fan, Mo, Shao, Chen, Guibas, and Dong]{huangGenerative3DPart2020}
j.~Huang, G.~Zhan, Q.~Fan, K.~Mo, L.~Shao, B.~Chen, L.~J. Guibas, and H.~Dong, ``Generative {{3D Part Assembly}} via {{Dynamic Graph Learning}},'' in \emph{Advances in {{Neural Information Processing Systems}}}, vol.~33.\hskip 1em plus 0.5em minus 0.4em\relax {Curran Associates, Inc.}, 2020, pp. 6315--6326.

\bibitem[Huang et~al.(2006)Huang, Fl\"{o}ry, Gelfand, Hofer, and Pottmann]{huang06}
Q.-X. Huang, S.~Fl\"{o}ry, N.~Gelfand, M.~Hofer, and H.~Pottmann, ``Reassembling fractured objects by geometric matching,'' \emph{ACM Trans. Graph.}, vol.~25, no.~3, p. 569–578, Jul. 2006.

\bibitem[Jia et~al.(2020)Jia, He, Yang, Han, Chang, and Han]{jiaDevelopingReassemblingAlgorithm2020}
C.~Jia, L.~He, X.~Yang, X.~Han, B.~Chang, and X.~Han, ``Developing a {{Reassembling Algorithm}} for {{Broken Objects}},'' \emph{IEEE Access}, vol.~8, pp. 220\,320--220\,334, 2020.

\bibitem[Kampel and Sablatnig(2003)]{kampelProfilebasedPotteryReconstruction2003}
\BIBentryALTinterwordspacing
M.~Kampel and R.~Sablatnig, ``Profile-based {{Pottery Reconstruction}},'' in \emph{2003 {{Conference}} on {{Computer Vision}} and {{Pattern Recognition Workshop}}}, vol.~1, Jun. 2003, pp. 4--4. [Online]. Available: \url{https://ieeexplore.ieee.org/abstract/document/4624518}
\BIBentrySTDinterwordspacing

\bibitem[Kim et~al.(2021)Kim, Arshad, Yoo, Hong, Kim, and Kim]{9412372}
J.~E. Kim, M.~Z. Arshad, S.~J. Yoo, J.-H. Hong, J.~Kim, and Y.~M. Kim, ``3d pots configuration system by optimizing over geometric constraints,'' in \emph{International Conference on Pattern Recognition}, 2021, pp. 2398--2405.

\bibitem[Kim et~al.(2024)Kim, Lee, and Joo]{joo24}
J.~Kim, I.~Lee, and K.~Joo, ``Fracture assembly with segmentation and iterative registration,'' in \emph{IEEE International Conference on Acoustics, Speech and Signal Processing}.\hskip 1em plus 0.5em minus 0.4em\relax IEEE, 2024, pp. 6945--6949.

\bibitem[Lamb et~al.(2023)Lamb, Palmer, Molloy, Banerjee, and Banerjee]{FantaticBreaks}
N.~Lamb, C.~Palmer, B.~Molloy, S.~Banerjee, and N.~K. Banerjee, ``Fantastic {{Breaks}}: {{A Dataset}} of {{Paired 3D Scans}} of {{Real-World Broken Objects}} and {{Their Complete Counterparts}},'' in \emph{Proceedings of the {{IEEE}}/{{CVF Conference}} on {{Computer Vision}} and {{Pattern Recognition}}}, 2023, pp. 4681--4691.

\bibitem[Leitch(2013)]{leitchReconstructingHistoryPottery2013}
V.~Leitch, ``Reconstructing history through pottery: The contribution of {{Roman N African}} cookwares,'' \emph{Journal of Roman Archaeology}, vol.~26, pp. 281--306, Jan. 2013.

\bibitem[{Leordeanu} and {Hebert}(2005)]{loerdeanu05}
M.~{Leordeanu} and M.~{Hebert}, ``A spectral technique for correspondence problems using pairwise constraints,'' in \emph{IEEE International Conference on Computer Vision}, vol.~2, 2005, pp. 1482--1489 Vol. 2.

\bibitem[Li et~al.(2020)Li, Geng, and Zhou]{liPairwiseMatching3D2020}
Q.~Li, G.~Geng, and M.~Zhou, ``Pairwise {{Matching}} for {{3D Fragment Reassembly Based}} on {{Boundary Curves}} and {{Concave-Convex Patches}},'' \emph{IEEE Access}, vol.~8, pp. 6153--6161, 2020.

\bibitem[Liu et~al.(2023)Liu, Guo, Jiang, Liu, Zhang, and Yan]{liuPuzzleNetBoundaryAwareFeature2023}
H.-Y. Liu, J.-W. Guo, H.-Y. Jiang, Y.-C. Liu, X.-P. Zhang, and D.-M. Yan, ``{{PuzzleNet}}: {{Boundary-Aware Feature Matching}} for {{Non-Overlapping 3D Point Clouds Assembly}},'' \emph{Journal of Computer Science and Technology}, vol.~38, no.~3, pp. 492--509, Jun. 2023.

\bibitem[Lowe(2004)]{lowe04}
D.~G. Lowe, ``Distinctive image features from scale-invariant keypoints,'' \emph{Int. J. Comput. Vision}, vol.~60, no.~2, p. 91–110, Nov. 2004.

\bibitem[Lu et~al.(2023)Lu, Sun, and Huang]{jigsaw}
J.~Lu, Y.~Sun, and Q.~Huang, ``Jigsaw: {{Learning}} to {{Assemble Multiple Fractured Objects}},'' \emph{Advances in Neural Information Processing Systems}, vol.~36, pp. 14\,969--14\,986, Dec. 2023.

\bibitem[Mara and Sablatnig(2006)]{mara06}
H.~Mara and R.~Sablatnig, ``Orientation of fragments of rotationally symmetrical 3d-shapes for archaeological documentation,'' in \emph{Proceedings of the Third International Symposium on 3D Data Processing, Visualization, and Transmission}, ser. 3DPVT '06.\hskip 1em plus 0.5em minus 0.4em\relax Washington, DC, USA: IEEE Computer Society, 2006, pp. 1064--1071.

\bibitem[{McBride} and {Kimia}(2003)]{mcbride03}
J.~C. {McBride} and B.~B. {Kimia}, ``Archaeological fragment reconstruction using curve-matching,'' in \emph{IEEE/CVF Conference on Computer Vision and Pattern Recognition Workshop}, vol.~1, 2003, pp. 3--3.

\bibitem[Papaioannou et~al.(2017)Papaioannou, Schreck, Andreadis, Mavridis, Gregor, Sipiran, and Vardis]{papaioannou17}
G.~Papaioannou, T.~Schreck, A.~Andreadis, P.~Mavridis, R.~Gregor, I.~Sipiran, and K.~Vardis, ``From reassembly to object completion: A complete systems pipeline,'' \emph{J. Comput. Cult. Herit.}, vol.~10, no.~2, Mar. 2017.

\bibitem[Pottmann et~al.(1999)Pottmann, Peternell, and Ravani]{pottmann99}
H.~Pottmann, M.~Peternell, and B.~Ravani, ``An introduction to line geometry with applications,'' \emph{Computer-Aided Design}, vol.~31, no.~1, pp. 3--16, 1999.

\bibitem[Qi et~al.(2017)Qi, Yi, Su, and Guibas]{PointNet}
C.~R. Qi, L.~Yi, H.~Su, and L.~J. Guibas, ``{{PointNet}}++: {{Deep Hierarchical Feature Learning}} on {{Point Sets}} in a {{Metric Space}},'' Jun. 2017.

\bibitem[Qian et~al.(2022)Qian, Li, Peng, Mai, Hammoud, Elhoseiny, and Ghanem]{PointNext}
G.~Qian, Y.~Li, H.~Peng, J.~Mai, H.~Hammoud, M.~Elhoseiny, and B.~Ghanem, ``{{PointNeXt}}: {{Revisiting PointNet}}++ with {{Improved Training}} and {{Scaling Strategies}},'' \emph{Advances in Neural Information Processing Systems}, vol.~35, pp. 23\,192--23\,204, Dec. 2022.

\bibitem[Qin et~al.(2022)Qin, Yu, Wang, Guo, Peng, and Xu]{GeoTransformer}
Z.~Qin, H.~Yu, C.~Wang, Y.~Guo, Y.~Peng, and K.~Xu, ``Geometric {{Transformer}} for {{Fast}} and {{Robust Point Cloud Registration}},'' in \emph{Proceedings of the {{IEEE}}/{{CVF Conference}} on {{Computer Vision}} and {{Pattern Recognition}}}, 2022, pp. 11\,143--11\,152.

\bibitem[Rusu and Cousins(2011)]{Rusu_ICRA2011_PCL}
R.~B. Rusu and S.~Cousins, ``{3D is here: Point Cloud Library (PCL)},'' in \emph{{IEEE International Conference on Robotics and Automation}}, Shanghai, China, May 9-13 2011.

\bibitem[{Savitzky} and {Golay}(1964)]{SGmethod}
A.~{Savitzky} and M.~J.~E. {Golay}, ``{Smoothing and differentiation of data by simplified least squares procedures},'' \emph{Analytical Chemistry}, vol.~36, pp. 1627--1639, 1964.

\bibitem[Sch\"{o}nberger and Frahm(2016)]{schoenberger16}
J.~L. Sch\"{o}nberger and J.-M. Frahm, ``{Structure-from-Motion Revisited},'' in \emph{IEEE/CVF Conference on Computer Vision and Pattern Recognition}, 2016.

\bibitem[Sell{\'a}n et~al.(2022)Sell{\'a}n, Chen, Wu, Garg, and Jacobson]{breakingbad}
S.~Sell{\'a}n, Y.-C. Chen, Z.~Wu, A.~Garg, and A.~Jacobson, ``Breaking bad: A dataset for geometric fracture and reassembly,'' \emph{Advances in Neural Information Processing Systems}, vol.~35, pp. 38\,885--38\,898, 2022.

\bibitem[Son et~al.(2013)Son, Almeida, and Cooper]{son13}
K.~Son, E.~B. Almeida, and D.~B. Cooper, ``Axially symmetric 3d pots configuration system using axis of symmetry and break curve,'' in \emph{IEEE/CVF Conference on Computer Vision and Pattern Recognition}, June 2013.

\bibitem[{The CGAL Project}(2020)]{cgal20}
{The CGAL Project}, \emph{{CGAL} User and Reference Manual}, 5th~ed.\hskip 1em plus 0.5em minus 0.4em\relax {CGAL Editorial Board}, 2020.

\bibitem[Tsesmelis et~al.(2024)Tsesmelis, Palmieri, Khoroshiltseva, Islam, Elkin, Shahar, Scarpellini, Fiorini, Ohayon, Alali, Aslan, Morerio, Vascon, Gravina, Napolitano, Scarpati, Zuchtriegel, Sp{\"u}hler, Fuchs, James, {Ben-Shahar}, Pelillo, and Bue]{RePAIR}
T.~Tsesmelis, L.~Palmieri, M.~Khoroshiltseva, A.~Islam, G.~Elkin, O.~I. Shahar, G.~Scarpellini, S.~Fiorini, Y.~Ohayon, N.~Alali, S.~Aslan, P.~Morerio, S.~Vascon, E.~Gravina, M.~C. Napolitano, G.~Scarpati, G.~Zuchtriegel, A.~Sp{\"u}hler, M.~E. Fuchs, S.~James, O.~{Ben-Shahar}, M.~Pelillo, and A.~D. Bue, ``Re-assembling the past: {{The RePAIR}} dataset and benchmark for real world {{2D}} and {{3D}} puzzle solving,'' Nov. 2024.

\bibitem[Wang et~al.(2021, 01, 22)Wang, Zang, Liang, Dong, Fan, and Yang]{wangProbabilisticMethodFractured2021}
H.~Wang, Y.~Zang, F.~Liang, Z.~Dong, H.~Fan, and B.~Yang, ``A {{Probabilistic Method}} for {{Fractured Cultural Relics Automatic Reassembly}},'' \emph{Journal on Computing and Cultural Heritage}, vol.~14, no.~1, pp. 5:1--5:25, 2021, 01, 22.

\bibitem[Wang et~al.(2024{\natexlab{b}})Wang, Karaev, Rupprecht, and Novotny]{vggsfm}
J.~Wang, N.~Karaev, C.~Rupprecht, and D.~Novotny, ``Vggsfm: Visual geometry grounded deep structure from motion,'' in \emph{IEEE/CVF Conference on Computer Vision and Pattern Recognition}, 2024, pp. 21\,686--21\,697.

\bibitem[Wang et~al.(2024{\natexlab{a}})Wang, Chen, and Furukawa]{wang2024puzzlefusionpp}
Z.~Wang, J.~Chen, and Y.~Furukawa, ``Puzzlefusion++: Auto-agglomerative 3d fracture assembly by denoise and verify,'' \emph{arXiv preprint arXiv:2406.00259}, 2024.

\bibitem[Willis et~al.(2003)Willis, Orriols, and Cooper]{willis03}
A.~Willis, X.~Orriols, and D.~B. Cooper, ``Accurately estimating sherd 3d surface geometry with application to pot reconstruction,'' in \emph{Proceedings of the 2003 Conference on Computer Vision and Pattern Recognition Workshop}, vol.~1, June 2003, pp. 5--5.

\bibitem[Willis and Cooper(2004)]{willis04}
A.~R. Willis and D.~B. Cooper, ``Bayesian assembly of 3d axially symmetric shapes from fragments,'' in \emph{Proceedings of the 2004 IEEE Computer Society Conference on Computer Vision and Pattern Recognition, 2004. CVPR 2004.}, vol.~1, June 2004, pp. I--I.

\bibitem[Ye et~al.(2023)Ye, {Zhu-Tian}, Chu, Li, Luo, Li, Geng, and Wu]{yePuzzleFixerVisualReassembly2023}
S.~Ye, C.~{Zhu-Tian}, X.~Chu, K.~Li, J.~Luo, Y.~Li, G.~Geng, and Y.~Wu, ``{{PuzzleFixer}}: {{A Visual Reassembly System}} for {{Immersive Fragments Restoration}},'' \emph{IEEE Transactions on Visualization and Computer Graphics}, vol.~29, no.~1, pp. 429--439, Jan. 2023.

\bibitem[Yoo* et~al.(2025)Yoo*, Liu*, Arshad, Kim, Kim, Aloimonos, Ferm¨uller, Joo, Kim, and Hyeong]{supmat_journal}
S.~J. Yoo*, S.~Liu*, M.~Z. Arshad, J.~Kim, Y.~M. Kim, Y.~Aloimonos, C.~Ferm¨uller, K.~Joo, J.~Kim, and H.~J. Hyeong, ``Supplementary materials: Structure-from-sherds++: Robust incremental 3d reassembly of axially symmetric pots from unordered and mixed fragment collections,'' \url{https://github.com/SeongJong-Yoo/Pottery-Hierarchy-Clear}, 2025.

\bibitem[Yoo et~al.(2025)Yoo, Liu, Arshad, Kim, Kim, Aloimonos, Fermüller, Joo, Kim, and Hong]{YooandLiu2024SfS}
S.~J. Yoo, S.~Liu, M.~Z. Arshad, J.~Kim, Y.~M. Kim, Y.~Aloimonos, C.~Fermüller, K.~Joo, J.~Kim, and J.~H. Hong, ``Structure-from-sherds++: Robust incremental 3d reassembly of axially symmetric pots from unordered and mixed fragment collections,'' 2025.

\bibitem[{Zach} et~al.(2010){Zach}, {Klopschitz}, and {Pollefeys}]{zach10}
C.~{Zach}, M.~{Klopschitz}, and M.~{Pollefeys}, ``Disambiguating visual relations using loop constraints,'' in \emph{2010 IEEE Computer Society Conference on Computer Vision and Pattern Recognition}, 2010, pp. 1426--1433.

\bibitem[{Zhang} et~al.(2015){Zhang}, {Yu}, {Manhein}, {Waggenspack}, and {Li}]{zhang15}
K.~{Zhang}, W.~{Yu}, M.~{Manhein}, W.~{Waggenspack}, and X.~{Li}, ``3d fragment reassembly using integrated template guidance and fracture-region matching,'' in \emph{IEEE International Conference on Computer Vision}, 2015, pp. 2138--2146.

\end{thebibliography}

\begin{IEEEbiography}[{\includegraphics[width=1in,height=1.25in,clip,keepaspectratio]{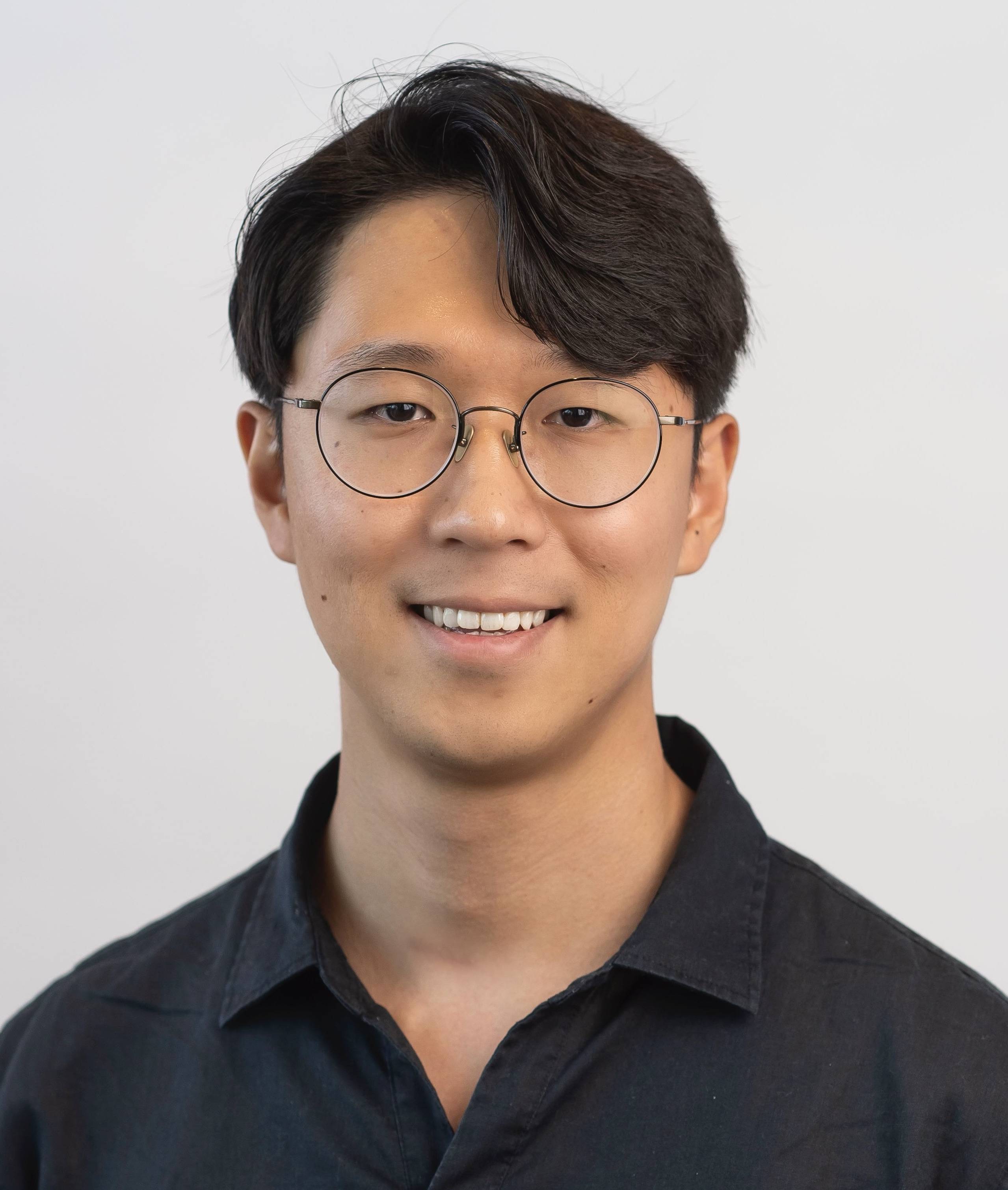}}]{Seongjong Yoo} received the B.S and M.S degrees in mechanical engineering from Soongsil University, South Korea, in 2017 and 2019, respectively. 
After graduation, he served alternative military service at Korea Institute of Science and Technology for three years. 
He is currently a Ph.D. student in computer science department at University of Maryland, MD, USA. 
His research interests include computer vision, multi-modal perception and digital humans. 
\end{IEEEbiography}

\begin{IEEEbiography}
[{\includegraphics[width=1in,height=1.25in,clip,keepaspectratio]{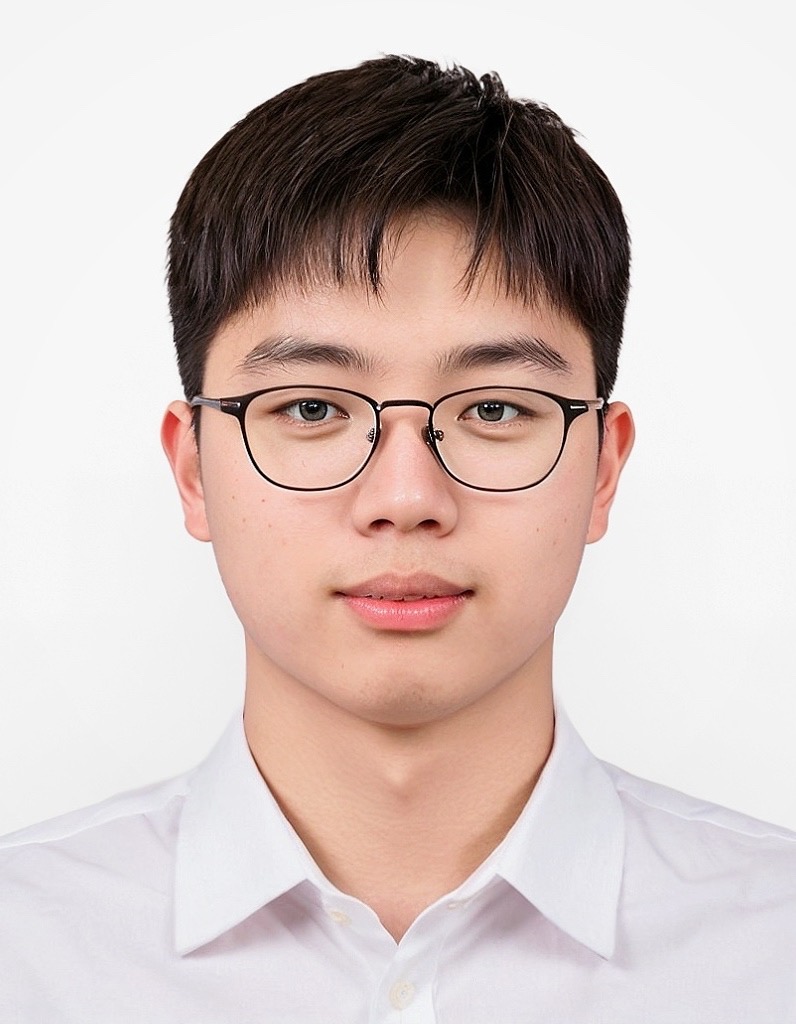}}]{Sisung Liu}
 received the B.Sc. degree in Applied Physics from Hanyang University in 2023. 
 Currently, he is pursuing a Ph.D. degree in the Department of Artificial Intelligence at Hanyang University.
 His research interests include neural geometry estimation and information forensics. \end{IEEEbiography}

\begin{IEEEbiography}
[{\includegraphics[width=1in,height=1.25in,clip,keepaspectratio]{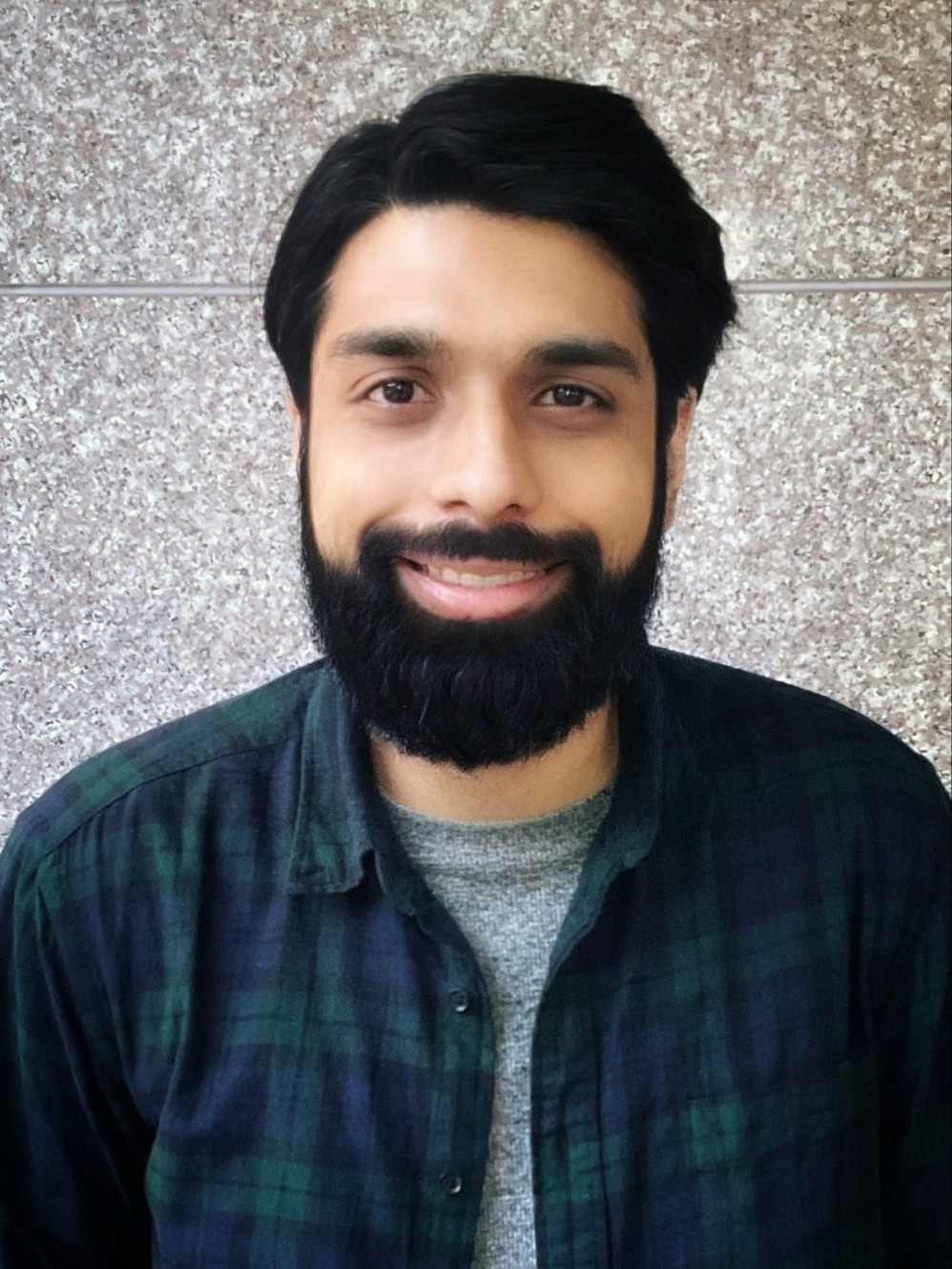}}]{Muhammad Zeeshan Arshad} completed his Master's and Ph.D. degrees at Myongji University, Yongin, South Korea. He has worked as a postdoctoral fellow at the Center for Artificial Intelligence (CAI) at the Korea Institute of Science and Technology (KIST), Seoul, and is currently a postdoctoral fellow at the University of Ottawa. His research has transitioned from monitoring, fault detection, and diagnostics using machine data to physiological data, with a focus on applying artificial intelligence (AI) in healthcare. His key areas of interest include algorithm design, time series modeling, and the development of ML/DL models, particularly in their application to gait analysis.
\end{IEEEbiography}

\begin{IEEEbiography}[{\includegraphics[width=1in,height=1.25in,clip,keepaspectratio]{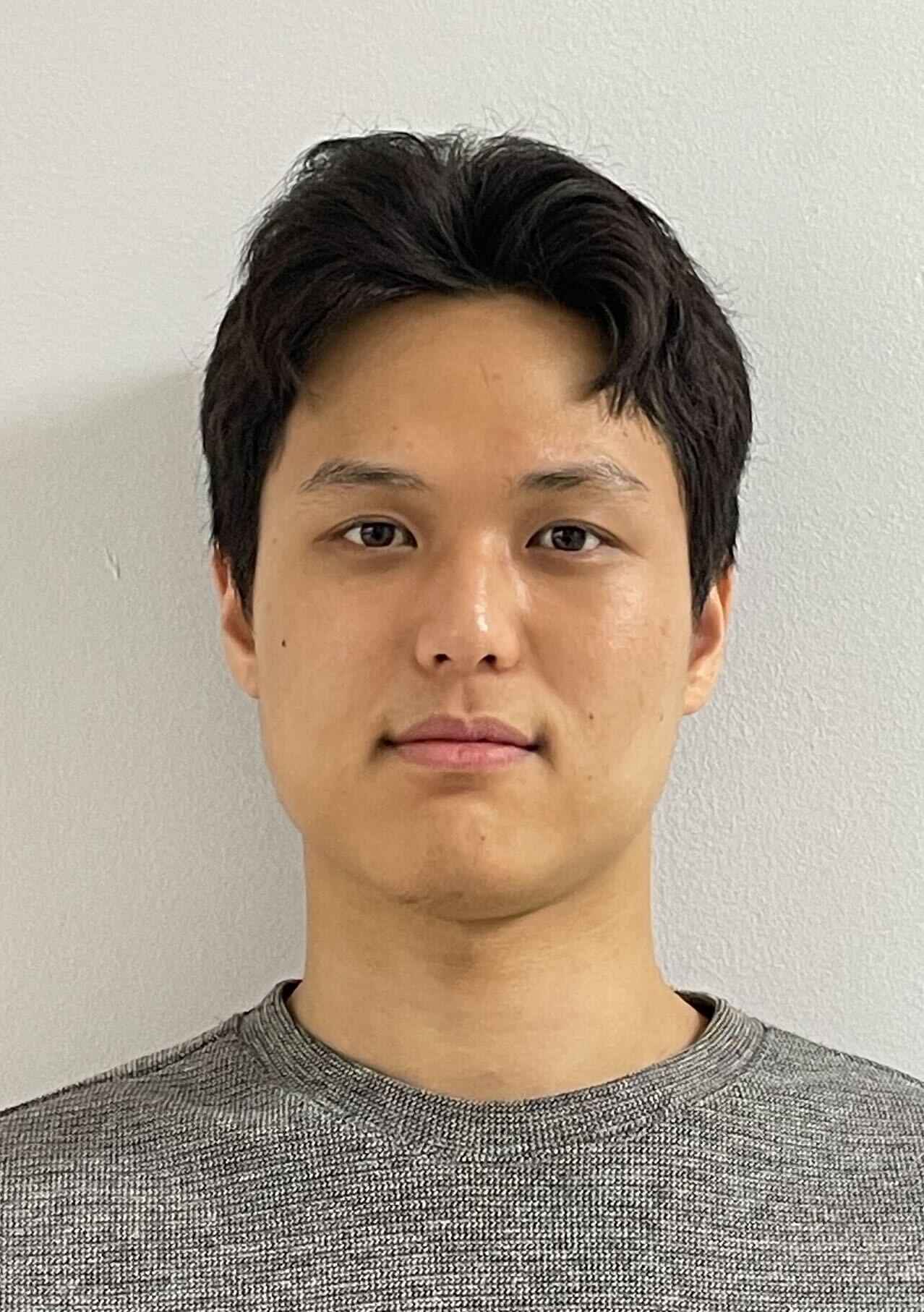}}]{Jinhyeok Kim} is a M.S. student of the Artificial Intelligence Graduate School at UNIST, South Korea. He received B.E. degree in the Department of Computer Science from UNIST, South Korea in 2023. His research interests include generative model and neural rendering for 3d understanding.
\end{IEEEbiography}

\begin{IEEEbiography}[{\includegraphics[width=1in,height=1.25in,clip,keepaspectratio]{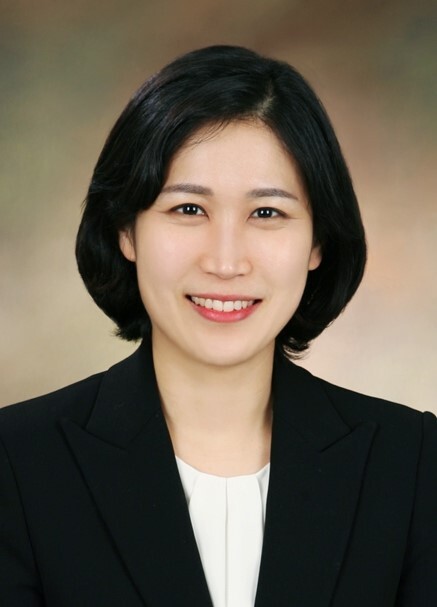}}]{Young Min Kim}
 is an Associate Professor in the Department of Electrical and Computer Engineering at Seoul National University, Seoul, Korea. She received a B.S. from Seoul National University in 2006 and an M.S. and Ph.D. in electrical engineering from Stanford University in 2008 and 2013, respectively. Before joining SNU, she was a Senior Research Scientist at the Korea Institute of Science and Technology (KIST). Her research interest lies in 3D vision, where she combines computer vision, graphics, and robotics algorithms to solve practical problems.
\end{IEEEbiography}

\begin{IEEEbiography}[{\includegraphics[width=1in,height=1.25in,clip,keepaspectratio]{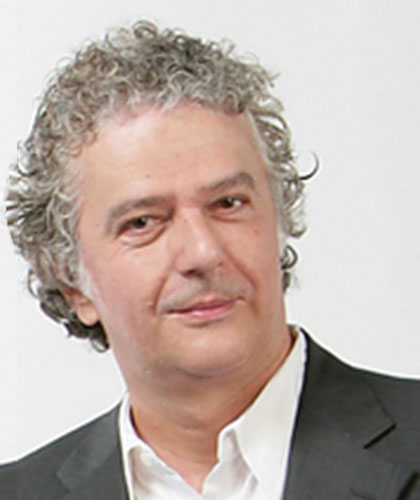}}]{Yiannis Aloimonos} currently is a professor of computational vision and intelligence with the the Department of Computer Science, University of Maryland, College Park, and the director of the Computer Vision Laboratory at the Institute for Advanced Computer Studies (UMIACS). He is interested in Active Perception and the modeling of vision as an active, dynamic process for real time robotic systems. For the past five years, he has been working on bridging signals and symbols, specifically on the relationship of vision to reasoning, action, and language.
\end{IEEEbiography}

\begin{IEEEbiography}[{\includegraphics[width=1in,height=1.25in,clip,keepaspectratio]{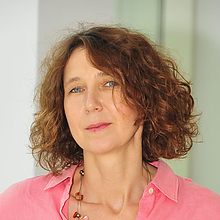}}]{Cornelia Ferm\"{u}ller} received an  MS degree from the Graz University of Technology and a PhD degree from the Vienna University of Technology, Austria in applied mathematics. She is a research scientist at the University of Maryland Institute for Advanced Computer Studies. Her research interest has been understanding the principles of active vision systems and developing biological-inspired methods, especially in motion. Her recent work has focused on human action interpretation and the development of event-based motion algorithms.
\end{IEEEbiography}

\begin{IEEEbiography}[{\includegraphics[width=1in,height=1.25in,clip,keepaspectratio]{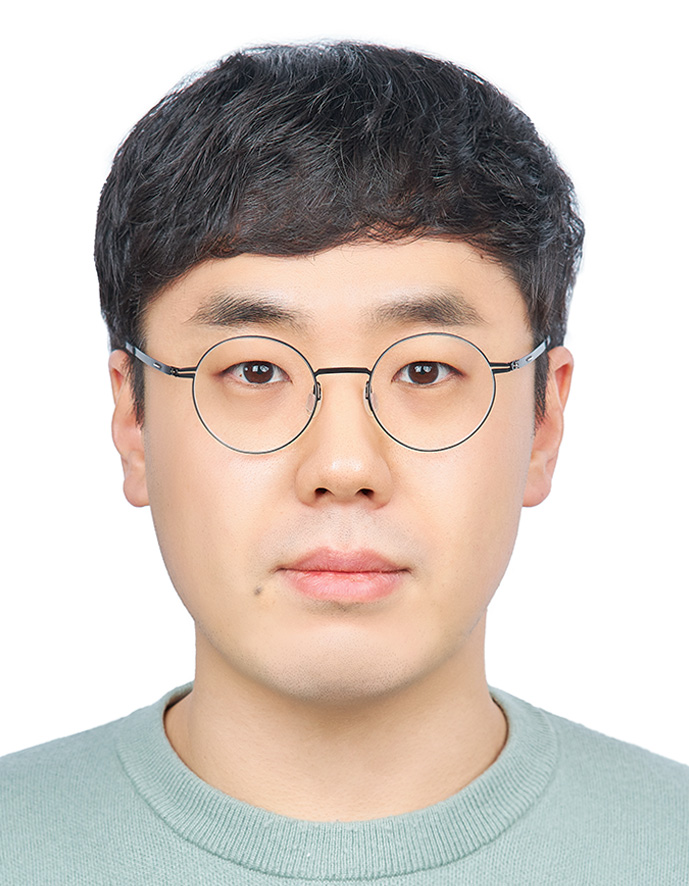}}]{Kyungdon Joo} is an associate professor of the Artificial Intelligence Graduate School and the Department of Computer Science at UNIST, South Korea.
He received the B.E. degree in School of Electrical and Computer Engineering from University of Seoul, South Korea in 2012, and the M.S. and Ph.D. degrees in Robotics Program and School of Electrical Engineering from KAIST, South Korea in 2014 and 2019, respectively.
Before joining UNIST, he was a postdoctoral researcher at CMU RI, US.
He was a member of ``Team KAIST,'' which won the first place in DARPA Robotics Challenge Finals 2015. 
He was a research intern at Oculus Research (Facebook Reality Labs), Pittsburgh in 2017.
His research interests include computer vision and robotics.
\end{IEEEbiography}

\begin{IEEEbiography}[{\includegraphics[width=1in,height=1.25in,clip,keepaspectratio]{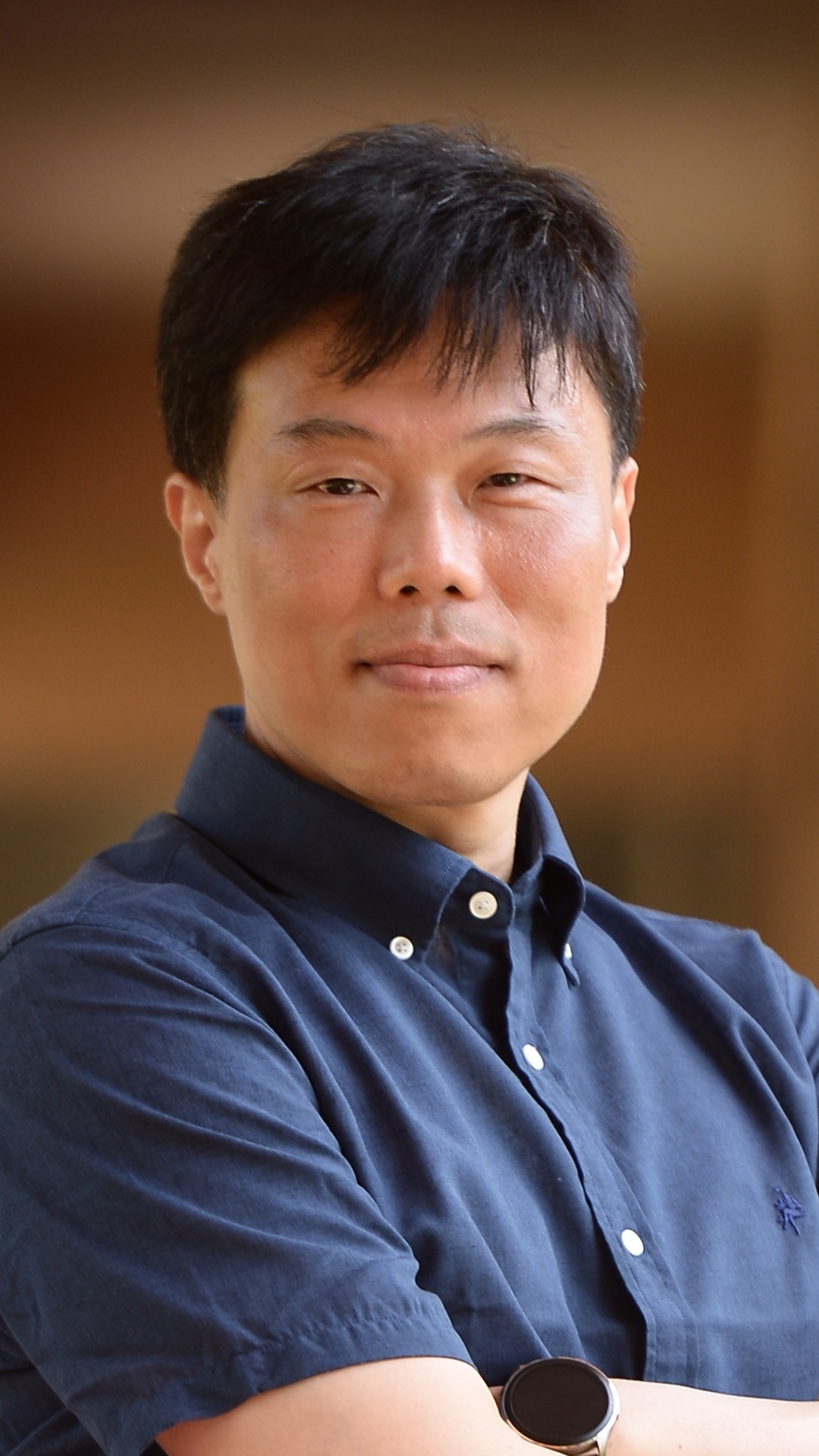}}]{Jinwook Kim} received the B.S., M.S., and Ph.D. degrees in Mechanical and Aerospace Engineering from Seoul National University, Seoul, Korea, in 1995, 1997, and 2002, respectively. He is currently a Principal Research Scientist at Intelligence and Interaction Research Center, Korea Institute of Science and Technology (KIST). His research interests include human motion analysis, healthcare technology, and assistive systems for elderly care, leveraging virtual reality, computer graphics, simulation, and machine learning.
\end{IEEEbiography}

\begin{IEEEbiography}[{\includegraphics[width=1in,height=1.25in,clip,keepaspectratio]{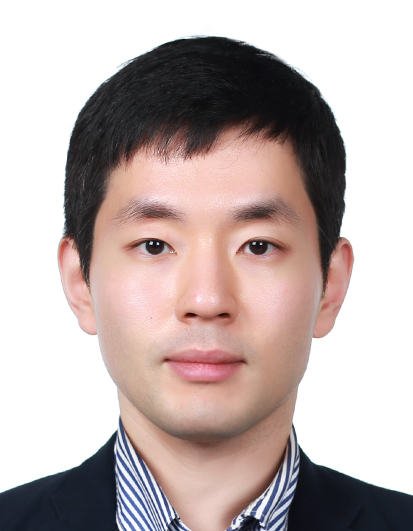}}]{Je Hyeong Hong}
received B.A. and M.Eng. degrees in Engineering (Electrical and Information Sciences) from the University of Cambridge, UK in 2011, and subsequently received a Ph.D. degree in Engineering (Computer Vision) from the University of Cambridge, UK, in 2018 after collaborating with Microsoft and Toshiba Research Europe.
He served alternative military service in South Korea as a postdoctoral researcher at the Korea Institute of Science and Technology (KIST) between 2018 and 2021.
Currently, Je Hyeong is an Assistant Professor in the Department of Electronic Engineering at Hanyang University in South Korea, a position he has held since 2021.
He has published several papers at prestigious conferences such as CVPR, ECCV, ICCV and ICASSP.
His main research interests include computer vision, machine learning and optimization.
\end{IEEEbiography}


\clearpage
\renewcommand{\thesection}{\Roman{section}}
\renewcommand{\thefigure}{\arabic{figure}}
\renewcommand{\thetable}{\Roman{table}}
\renewcommand{\theequation}{\arabic{equation}}
\renewcommand{\thealgorithm}{\arabic{algorithm}}

\setcounter{section}{0}
\setcounter{figure}{0}
\setcounter{table}{0}
\setcounter{algorithm}{0}
\twocolumn[
    \begin{center}
        \suppmaketitle  
    \end{center}
]

\vspace{-13mm}

\section{Appendix A - Cluster merging}\label{sec:appendix_A}
\label{sec:cluster_merging}
\begin{table}[ht]
\centering
\caption{Parameters used in cluster merging algorithm.}
\begin{tabular}{lcl}
\hline\hline
Parameter      & Value & Description \\ \hline
$k_1$          & 10    & Neighbors for boundary evaluation \\
$k_2$          & 30    & Neighbors for normal estimation \\ \hline
$\rho_d$       & 1.0   & Distance threshold for boundary pairs \\
$\rho_n$       & 0.95  & Normal sim. threshold \\
$\rho_c$       & 0.1   & Curvature diff. threshold \\
$\rho_{in,n}$  & 0.75  & Normal sim. threshold (internal) \\
$\rho_{in,c}$  & 0.25  & Curvature diff. threshold (internal) \\ \hline\hline
\end{tabular}
\end{table}

After applying the region growing algorithm, a single continuous surface may become over-segmented into multiple clusters. 
To address this issue, we employ a cluster merging algorithm.
In this algorithm, we first define a cluster pair $(C_i, C_j)$, which consists of two distinct clusters. 
The purpose of defining such a pair is to assess whether the clusters belong to the same continuous surface by examining adjacent boundary points between them.

For each cluster pair, boundary point pairs are detected within the specified distance threshold $\rho_d$ as described in Alg.~\ref{alg:cluster_merge} (Step 1).
In this step, the boundary status of each point is evaluated by considering $k_1$ neighboring points and at least three boundary point pairs must be detected.
If the number of boundary point pairs is less than three, the cluster pair is considered to contain noisy or unstable boundary information and is excluded from merging.
Next, for each boundary point, an average normal is computed using $k_2$ neighbors to estimate the local surface normal vector. 
If the dot product between the average normal vectors is at least $\rho_n$, and the difference in mean curvature is below $\rho_c$, the boundary points are deemed similar, indicating that the boundaries of the two clusters likely belong to the same continuous surface (Alg.~\ref{alg:cluster_merge}, Step 2).

Since boundary conditions alone may not fully overcome noise or inaccuracies near the boundaries, additional reliable internal information is incorporated.
As shown in Alg.~\ref{alg:cluster_merge} (Step 3), each cluster selects a representative internal point $p_i^{\mathrm{in}}$, which is the median point sorted by perpendicular distance to the principal axis, and the normals and curvatures of these internal points are compared using thresholds $\rho_{in,n}$ for normals and $\rho_{in,c}$ for curvatures.
The clusters are merged only if both the boundary and internal conditions are satisfied.
Once the conditions are met, cluster $C_j$ is merged into $C_i$ (Alg.~\ref{alg:cluster_merge}, Step 4), and this process is repeated until no further mergers can be performed (Alg.~\ref{alg:cluster_merge}, Line 28).

\algrenewcommand\algorithmicindent{1.0em}

\begin{algorithm}[t]
\caption{Cluster merging algorithm}
\label{alg:cluster_merge}
\begin{algorithmic}[1] 

\Require{
  Clusters $\{C_1, \ldots, C_n\}$; \\
  Point cloud with normals; \\
  Thresholds: $\rho_d, \rho_n, \rho_c, \rho_{\mathrm{in},n}, \rho_{\mathrm{in},c}$; \\
  KNN sizes: $k_1, k_2$.
}
\Ensure{Merged clusters $\{C'_1,\ldots,C'_m\}$.}

\Repeat
  \For{$i \gets 1$ to $n-1$}
    \For{$j \gets i+1$ to $n$}
      \If{$C_j$ is merged} 
        \State \textbf{continue} 
      \EndIf
      
      \Statex \(\triangleright\) \textit{Step 1: Boundary pairs below distance $\rho_d$}
      \State $B \gets \mathrm{FindBoundaryPairs}(C_i, C_j, \rho_d)$
      \If{$|B| < 3$}
        \State \textbf{continue} \Comment{Not enough boundary points}
      \EndIf
      
      \Statex \(\triangleright\) \textit{Step 2: Normal similarity and curvature}
      \State $\mathit{normalOK} \gets \mathrm{CheckNormalSimilarity}(B, \rho_n)$
      \State $\mathit{curvOK} \gets \mathrm{CheckCurvature}(B, \rho_c)$
      \If{\textbf{not} ($\mathit{normalOK} \wedge \mathit{curvOK}$)}
        \State \textbf{continue}
      \EndIf
      
      \Statex \(\triangleright\) \textit{Step 3: Inner point verification}
      \State $p_i^{\mathrm{in}} \gets \mathrm{SelectInnerPoint}(C_i)$
      \State $p_j^{\mathrm{in}} \gets \mathrm{SelectInnerPoint}(C_j)$
      \If{\textbf{not} $\mathrm{VerifyMerge}(p_i^{\mathrm{in}}, p_j^{\mathrm{in}}, \rho_{\mathrm{in},n}, \rho_{\mathrm{in},c})$}
        \State \textbf{continue}
      \EndIf
      
      \Statex \(\triangleright\) \textit{Step 4: Merge $C_j$ into $C_i$}
      \State $C_i \gets C_i \cup C_j$
      \State \textbf{Clear} $C_j$; \textbf{Mark} $C_j$ as merged
    \EndFor
  \EndFor
\Until{\textit{no more merges can be made}}

\State \(\triangleright\) Remove empty clusters
\State Remove all $C_k$ where $C_k = \emptyset$

\State \Return $\{C'_1,\ldots,C'_m\}$
\end{algorithmic}
\end{algorithm}

\section{Appendix B - ICP refinement}\label{sec:appendix_B}
\label{sec:ICP_Refinement}

After obtaining initial candidate pairs through descriptor matching, false matches may still exist.  
To eliminate these, we apply the Iterative Closest Point (ICP) algorithm to iteratively optimize point-to-point or point-to-line alignments, thereby refining the local alignment between sherds. 

\pparagraph{Pairwise cost function.}
\begin{align}
    J_R(A, B):=\sum_{\smash{\mathclap{(i,j)\in\Omega^{AB}}}} 
    \rho_d(d_{ij}(\m T^A, \m T^B)) + \lambda \rho_e(e^2_{ij}(\m T^A, \m T^B))
    \eqlabel{eq:J_R}
\end{align}
Equation~\eqref{eq:J_R} encourages two sherds, $A$ and $B$, to align their shared fracture surfaces (edge lines) by matching corresponding points. 
Let $\Omega^{AB}=\{(i, j)| \v q^A_i \in A, \v q^B_j \in B \}$ as the correspondence index set between sherd $A$ and sherd $B$.
Here, $\rho_d(\cdot)$ and $\rho_e(\cdot)$ are Cauchy kernels, and $\lambda$ is a positive scalar that balances the weight of the normal misalignment term, which we set to 0.4 in our experiments. 
The function $d_{ij}$ represents the geometric error (e.g. point-to-point or point-to-line distance) between the corresponding points
and $e_{ij}$ measures the squared difference between their surface normals.

At first ICP iteration, we use geometric error function as a point-to-point (P2P) alignment ($d_{ij}(\m T^{A}, \m T^{B})=d_{ij}^{\text{P2P}}(\m T^{A}, \m T^{B})$:
\begin{align}
d_{ij}^{\mathrm{P2P}}(\m T^A, \m T^B)&:= d(\v p^A_i, \hat{\v n}^A_i, \v p^B_j, \hat{\v n}^B_j, \m T^A, \m T^B)  \\ \nonumber
&=
\bigl\|
m_{ij}(\m T^A, \m T^B)
\bigr\|_2^2
\end{align}

\begin{align}
m_{ij}(\m T^A, \m T^B)&:= m_{ij}(\v p^A_i, \hat{\v n}^A_i, \v p^B_j, \hat{\v n}^B_j, \m T^A, \m T^B)  \\ \nonumber
&= \m R^A \v p^A_i + \v t^A - \m R^B \v p^B_j - \v t^B
\end{align}

where $\m R^A, \m R^B \in \mathbb{R}^{3 \times 3}$ are rotation matrices that transform sherds $A$ and $B$, $\v t^A, \v t^B \in \mathbb{R}^3$ are translation vectors, and $\v p_i^A, \v p_j^B$ are points on each sherd's edge line.

Once the sherds have been roughly aligned after the first iteration of ICP, we switch the P2P alignment to point-to-line (P2L) alignment. 
Empirically, P2P model often adjusts precise geometric difference under good initial condition. 
Specifically, 
\begin{align}
&d_{ij}^{\mathrm{P2L}}(T_A, T_B) := d(\v p^A_i, \hat{\v n}^A_i, \v p^B_j, \hat{\v n}^B_j, \m T^A, \m T^B)  \\ \nonumber
&=
\bigl\| 
\m R^A \hat{\v l}_i^{A} \cdot m_{ij}(\m T^A, \m T^B) \bigr\|_2^2
+ \bigl\| \m R^B \hat{\v l}_j^{B} \cdot m_{ij}(\m T^A, \m T^B) \bigr\|_2^2 \\ \nonumber 
&+ \bigl\| \m R^A \hat{\v n}_i^{A} \cdot m_{ij}(\m T^A, \m T^B) \bigr\|_2^2 
+ \bigl\|\m R^B \hat{\v n}_j^{B} \cdot m_{ij}(\m T^A, \m T^B) \bigr\|_2^2
\end{align}
where $\hat{\v l}_i^{A}$ denotes the fracture surface normal of the edge line on sherd $A$. 
To clarify, the fracture surface normal (\(\hat{\v l}_i\)) of an edge line point is distinct from its the surface normal (\(\hat{\v n}_i\)).
Specifically, the fracture surface normal (\(\hat{\v l}_i\)) is a normalized vector orthogonal to the local fracture surface of the edge line point \(\v p_i\).
It can be computed by taking the cross product between the point's surface normal (\(\hat{\v n}_i\)) and the tangent vector:
\begin{align}
\hat{\v l}_i^{A}
~:=~
\frac{
\hat{\v n}_i^{A} 
\,\times\,
\bigl(\v p_{i+1}^{A} - \v p_i^{A}\bigr)
}{
\bigl\|
\hat{\v n}_i^{A} 
\,\times\,
\bigl(\v p_{i+1}^{A} - \v p_i^{A}\bigr)
\bigr\|_2
}
\eqlabel{fracture_surface_normal}
\end{align}
This distinction resolves the local minima caused by simple geometric characteristic of ceramic broken sherds. 
Because it ensures that the fracture surface normal is orthogonal to the local surface geometry of the fracture edge, aiding in alignment accuracy during optimization, as shown in Fig. 9 in main paper.

\begin{align}
e_{ij}(T_A,\,T_B)
~:=~
\bigl\|\,
R_A\,\hat{n}_i^{A}
~-~
R_B\,\hat{n}_j^{B}
\bigr\|_2^2
\eqlabel{eq:norm}
\end{align}
Equation~\eqref{eq:norm} ensures the surface normals of the two sherds remain consistent when they are matched.

\pparagraph{Individual cost function.}
\begin{align}
    J_{I}(A) := \mu\!\!\sum_{i\in\Omega_{A}}\!\!\rho_{f}\big(f_{i}^{2}(\m T^A)\big)
    + \nu\!\!\sum_{i\in\Psi_{A}}\!\!\rho_{g}\big(g_{i}^{2}(\m T^A,r^{\mathrm{rim}},h^{\mathrm{rim}})\big)
    \eqlabel{eq:J_I}
\end{align}

where $\Omega_A$ is the set of points used for axis alignment, and $\Psi_A$ is the set of rim points (e.g., the top edge of the pottery).
$\mu$ and $\nu$ is a positive scalar that balance the weight of the axis alignment term and the rim-consistency term, respectively, both of which we set to 0.4 in our experiments.
\( f_i \) aligns each sherd to the global symmetry axis, extending the geometric error approach proposed by Cao and Mumford~\cite{cao02} and further developed in PotSAC~\cite{hong19}. This method robustly handles the normal sign ambiguity for inner and outer surfaces by considering both inward and outward normals. Specifically, the deviation of the curvature center from the symmetry axis can be measured as:
\begin{align}
&\epsilon_i^s(\v u, \hat{\v v}) := \bigl(c_i^s(\v u, \hat{\v v}) - \v u\bigr) \times \hat{\v v} \\ \nonumber
& c_i^s(\v u, \hat{\v v}) := \v p_i - \frac{\| (\v p_i - \v u) \times \hat{\v  v}\|_2}{\|\hat{\v v}\times \hat{\v n}_i^s \|_2}\hat{\v n}_i^s
\end{align}
where $s \in \{+, -\}$ indicates the normal direction (inward or outward), $\hat{\v v}$ and $\v u$ are the direction and offset of the symmetric axis, respectively. To account for this ambiguity, a log-sum-exp smoothing function is applied:
\begin{align}
f_i := -\tfrac{1}{t} \ln \left( \sum_s \exp\Big(-t\,\|\epsilon_i^s\|_2^2\Big)  \right)
\end{align}
where \( \|\epsilon_i^+\|^2 \) and \( \|\epsilon_i^-\|^2 \) represent the squared deviations for inward and outward normals, respectively. This approach ensures robust alignment of sherds to the global symmetry axis, even in cases where the normal directions are flipped.

The rim-consistency term $g_i$ enforces that the points belonging to the rim adhere to rim constraints.
Specifically,
\begin{align}
g_i(\m T^A,r^{\mathrm{rim}},h^{\mathrm{rim}}) =
&\bigl\|\,r^{\mathrm{rim}}-r\bigl(R^A\,p_i^{A} + t^A\bigr)\bigr\|_2^2 \notag \\
&+\bigl\|\,h^{\mathrm{rim}}-h\bigl(R^A\,p_i^{A} + t^A\bigr)\bigr\|_2^2
\label{eq:rim_function}
\end{align}
where $r^{\mathrm{rim}}$ and $h^{\mathrm{rim}}$ are radius and height of rim based on axis of symmetric, computed as the median value of all rim points. 
The function $r(\v p)$ and $h(\v p)$ compute radius and height of $\v p$, respectively. 
This constraints ensures that all rim points lie on the same circular lip after transformation.

\begin{algorithm}[t]
	\caption{Algorithm for checking overlaps}
	\label{alg:3doverlap}
	\begin{algorithmic}[1]
		\Statex {\textbf{Inputs}: Edge lines for sherds $A$ ($\{\v p^A\}$) and $B$ ($\{\v p^B\}$)}
		\State Find a set of correspondence index set $\Omega^{AB}$ between  $\{\v p^A\}$ and $\{\v p^B\}$
		\State Select a region (i.e. a range of correspondence pairs) to test potential overlap between the two edge lines based on one of two conditions ($^*$) listed in Sec.~\ref{sec:appendix_C}.
		\For{each correspondence pair ($(\v p^A_i, \v p_j^B) \in \Omega^{AB}$) in the investigated region}
		    \If{$(\v p^A_i, \v p^B_j)$ satisfies at least one of the two conditions ($^\dagger$)}
		        \State Area of overlap $S_{overlap}$ $\leftarrow$ $S_{overlap}$ + $\l2{\v p_j^A - \v p_j^B}$
		    \EndIf
		\EndFor
		\Statex {\textbf{Outputs}: Area of overlap}
	\end{algorithmic}
\end{algorithm}

\begin{algorithm}[t]
    \caption{Checking the consistency of the profile curve}
    \label{alg:pc_check}
    \begin{algorithmic}[1]
        \State {\textbf{Inputs}: edge line points ($\{\v p\}$) across  matched sherds}
        \State result $\leftarrow$ true
        \State Align the edge points $\{\v p\}$ to the axis of symmetry ($z^+$ is now in the axis direction)
        
        \State Compute the radii of edge line points ($\{r\}$) from the axis-aligned $x$ and $y$ coordinate values.

        \State Sort edge line points in ascending order of $z$ values.
        \State Segment line points $\{\v p \}$ into bins every $w$ (7 mm) in $z$.
        \For{each bin of line points}
            \State Fit a line using orthogonal regression.
            \State Compute standard deviation ($\sigma$) of orthogonal distance errors between the line of best fit and the edge line points.
            \If{$\sigma > \delta_d$}
                \State result $\leftarrow$ fail
                \State break
            \EndIf
        \EndFor
        \State {\textbf{Output}: result}
    \end{algorithmic}
\end{algorithm}

\section{Appendix C - Geometric verification}\label{sec:appendix_C}
To ensure the accurate reassembly of pottery fragments, it is essential to verify whether the edge lines of the fragments align correctly. This process involves two key steps: overlap checking (Algorithm~\ref{alg:3doverlap}) and profile curve consistency verification (Algorithm~\ref{alg:pc_check}).

\pparagraph{Checking potential overlaps.}
The overlap check identifies corresponding edge points between two fragments and evaluates their potential overlap in 3D space. The two conditions with asterisks (*) in Algorithm~\ref{alg:3doverlap} are designed to select the potential overlapping regions: 
\begin{enumerate}
    \item $ \l2{\v p_j^A - \v p_j^B} < d$
    \item $\v n_j^A \cdot \v n_j^B > 0$ and $\left| \frac{(\v p_j^A-\v p_j^B)}{\| \v p_j^A- \v p_j^B\|_2} \cdot \v n_j^A\right| > \cos\theta$
\end{enumerate}
The first condition filters out candidate points by measuring their Euclidean distance, ensuring that only closely aligned points are considered.
The second condition examines the possible overlapping alongside the radial direction as shown in Fig~\figref{radial_overlap}.
Once the region of interest is detected, we further verify additional conditions ($\dagger$).
First, whether one fragment is partially embedded inside another  $\big(\hat {\v l}_j^A \cdot \hat {\v l}_j^B > 0 \big)$. This occurs when one piece is almost submerged into another piece.
Second, when two fragments overlap in an unexpected manner $\big( \hat {\v l}_j^A \cdot (\v p_j^B - \v p_j^A) < 0$ and $\hat {\v l}_j^B \cdot (\v p_j^B - \v p_j^A) > 0 \big)$. This occurs when two pieces match as expected but with some overlaps.
If the total overlapped area $S_{overlap}$ exceeds $50 \text{mm}^2$, the pair is discarded from the match list.

\pparagraph{Checking the profile curve.}
However, overlap checking alone cannot eliminate all false positive matches, necessitating an additional profile curve consistency verification step. The profile curve represents the projection of the pot’s edge onto the $rz$-plane, reflecting its axial symmetry.
A correctly assembled pot should exhibit a continuous, smooth profile curve.
To evaluate this, the edge points are first aligned to the symmetry axis, and their radial distances are computed. The points are then segmented into bins at $7\text{mm}$ intervals along the $z$-axis, and within each bin, a line is fitted using orthogonal regression. 
The standard deviation of the orthogonal distance errors between the fitted line and the actual edge points is then measured. 
If this deviation exceeds a predefined threshold $7\text{mm}$, the match is deemed inconsistent and rejected.

By integrating both verification methods, the algorithm effectively reduces errors that may arise from purely distance-based matching. 
Fig.~\figref{radial_overlap} illustrates an example of a detected radial overlap, while Fig.~\figref{pc_pass_fail} demonstrates a case where the overlap test fails to identify a false positive configuration, but the profile curve consistency check successfully detects the mismatch.
The combination of these two algorithms significantly enhances the reliability and accuracy of pottery fragment reassembly.

\begin{figure}[t!]
    \centering
    \includegraphics[width=0.9\linewidth]{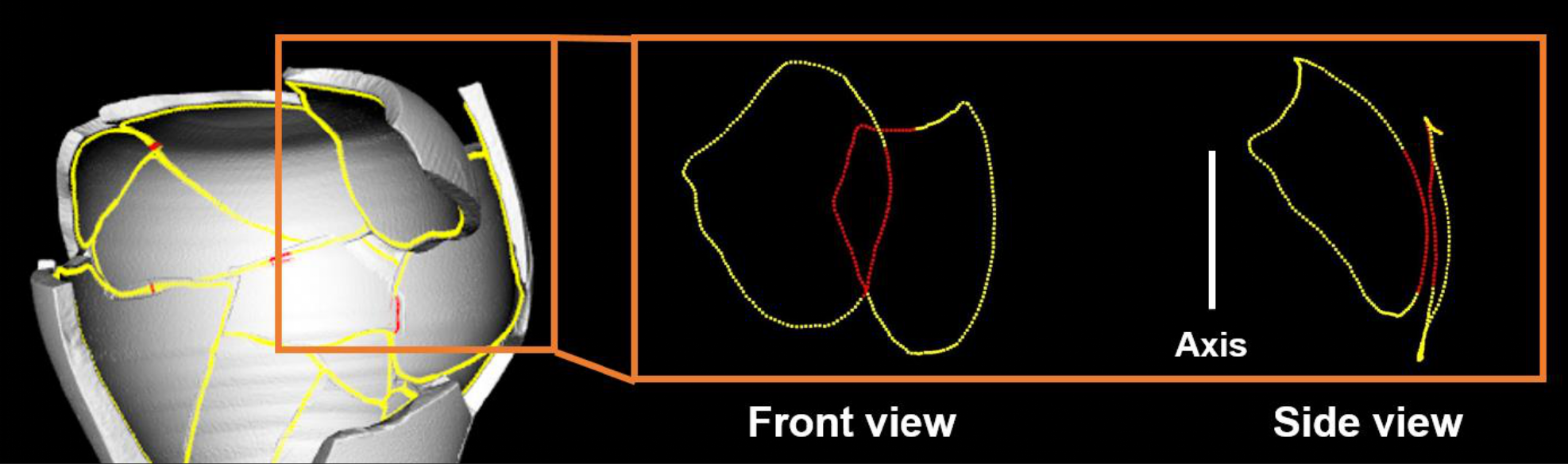}
    \caption{An example of the radial overlap case (which is false and thus discarded) from Sec.~\ref{sec:appendix_C}. 
    Red line denotes the overlapped region.}
    \figlabel{radial_overlap}
\end{figure}

\begin{figure}[t!]
	\centering
    \captionsetup[subfloat]{labelfont=footnotesize,textfont=footnotesize}
	\subfloat[][Correct configuration]{\includegraphics[width=0.45\linewidth]{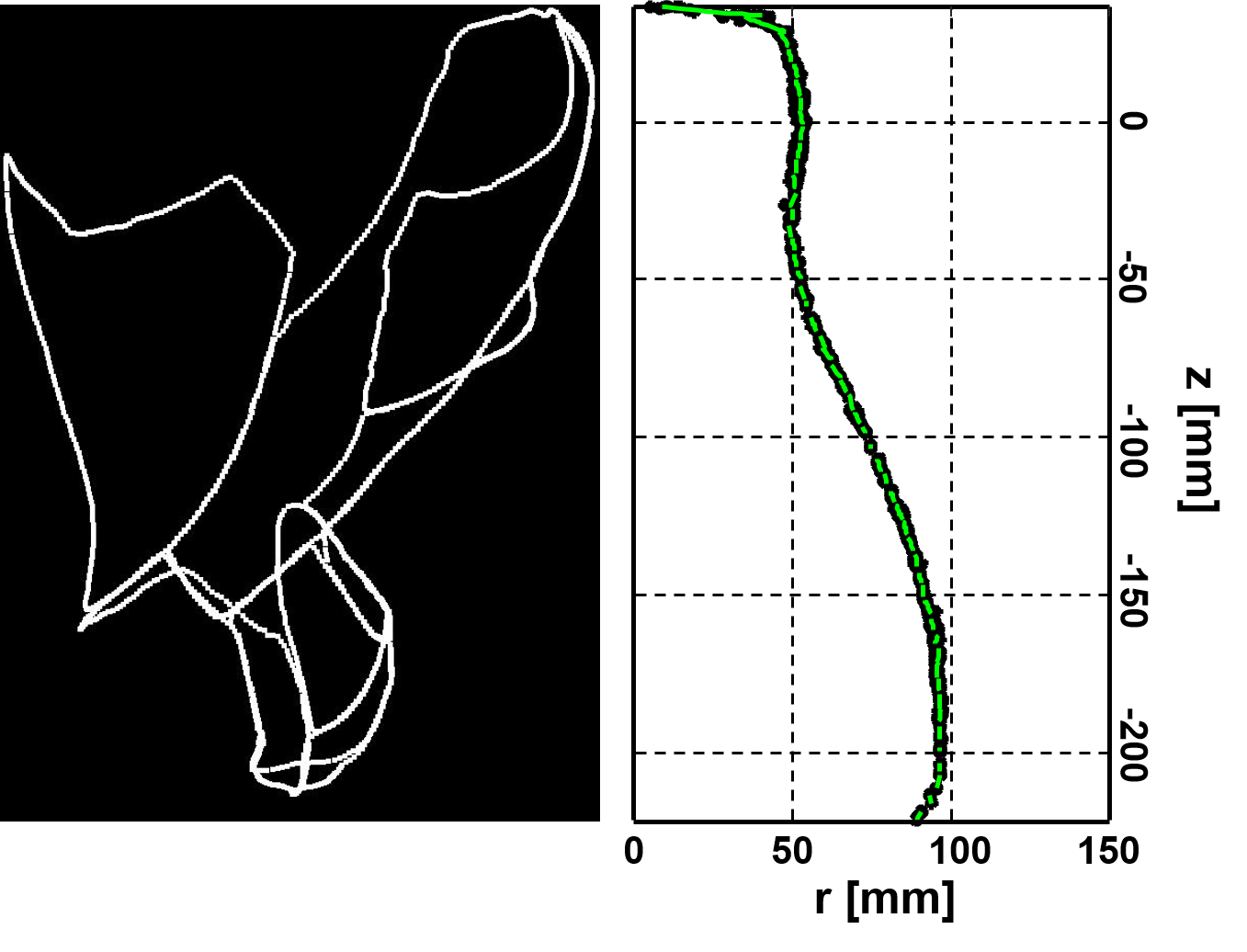}\figlabel{overlap}}
	~
	\subfloat[][Incorrect case detected by the profile curve test]{\includegraphics[width=0.45\linewidth]{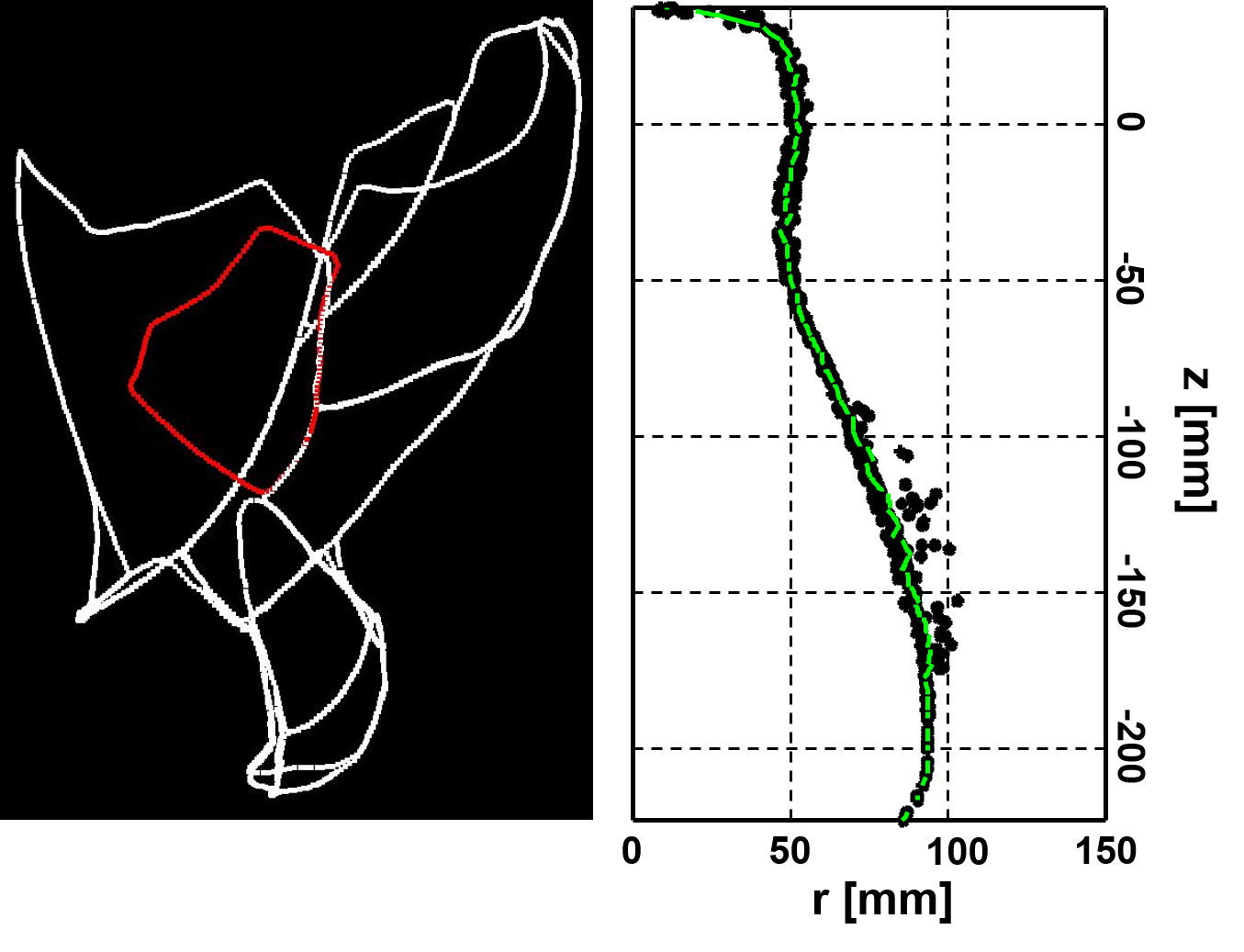}\figlabel{pc_check}}	
    ~
	\caption{
		The left side of (a) and (b) show the edge lines of matched sherds, and the right side of (a) and (b) show the corresponding profile curves
    (with locally fitted lines in green). In (b), the overlap test fails to find a false positive configuration but the profile curve test detects it.
	}
	\figlabel{pc_pass_fail}
\end{figure}

\begin{table}[t]
\centering
\caption{Parameters used in geometric verification.}
\begin{tabular}{lcl}
\hline\hline
Parameter      & Value & Description \\ \hline
$d$          & 5mm  & Distance threshold  \\
$\theta$          & $30^\circ$    & Overlapping direction threshold \\ 
$S_{overlap}$    & $50 \text{mm}^2$  & Total overlapped area \\ \hline
$\omega$       & 7mm  & Interval along the $z$-axis (bin) \\
$\delta_d$       &  7mm  &  Orthogonal distance error threshold \\ \hline\hline
\end{tabular}
\end{table}

\section{Appendix D - Additional experimental visualization of reassembly }\label{sec:appendix_D}
We present a visual comparison of reassembly results corresponding to the quantitative evaluations in Tables V and VI in the main paper~\cite{YooandLiu2024SfS}. 
Fig. \figref{Result_SFS} and \figref{Result_BB} show pottery reassembled using different methods, including Jigsaw~\cite{jigsaw}, FRASIER~\cite{joo24}, PuzzleFusion++~\cite{wang2024puzzlefusionpp}, and our proposed \AlgName~(with $k=5$, $b=3$). 
Each row depicts a pot reconstructed from 2 to 20 fragments, aligning with the fragment range used during training for deep learning-based models. 
The first column provides ground truth models for reference, while the last column presents the results of our method, demonstrating improved reassembly quality over existing approaches.

\begin{figure*}[!h]
	\centering
	\includegraphics[width=0.7\textwidth]{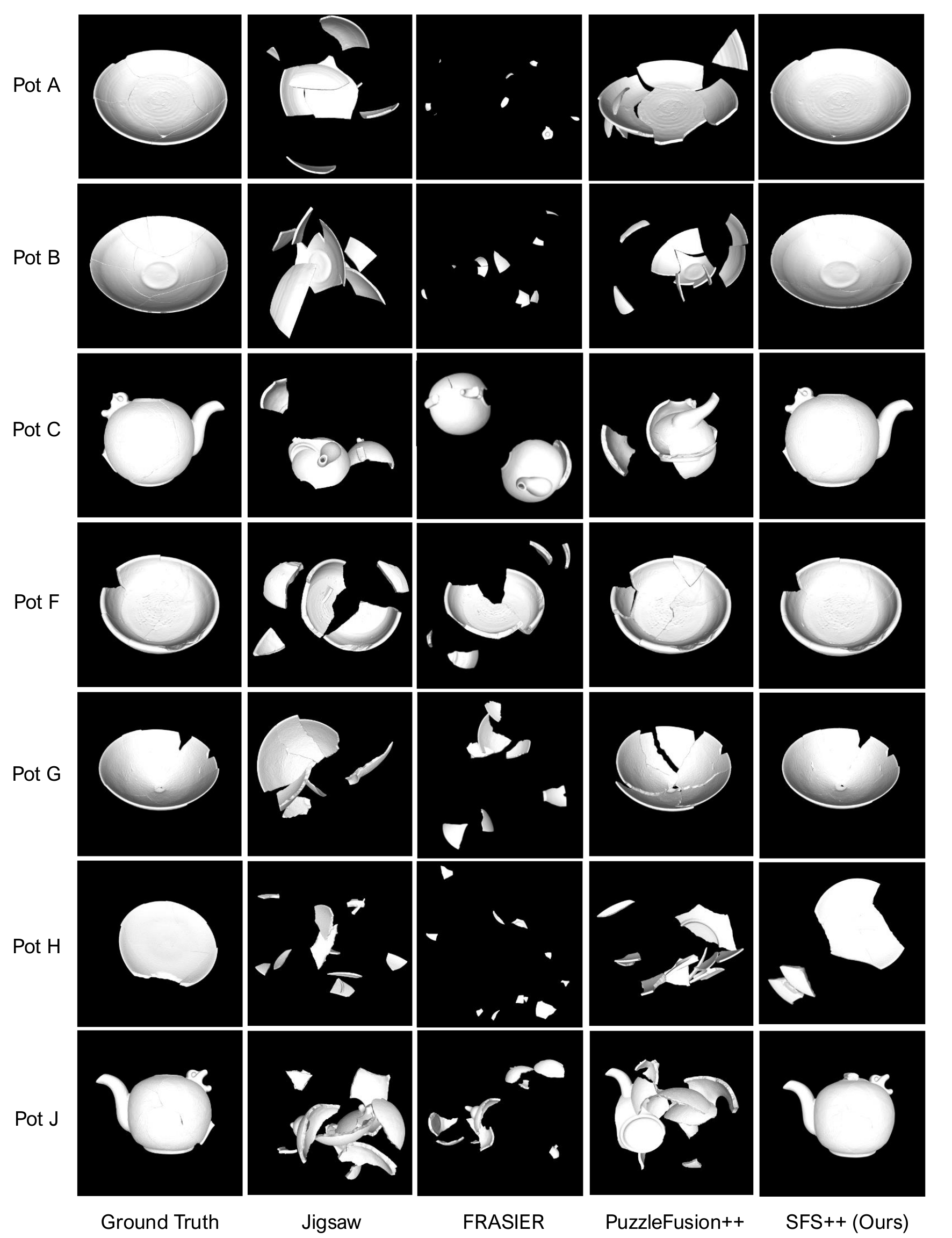}
	\caption{
    Comparison of visualized results from different models on our dataset. 
    Each row represents a reassembly result for pots with 2 to 20 pieces, as this range reflects the data used during training for deep learning models. 
    The last column shows our method (\AlgName with $k=5$ and $b=3$), demonstrating improved reassembly quality compared to other models like Jigsaw, FRASIER, and PuzzleFusion++.
    Ground truth models are provided for reference in the first column. 
}

	\figlabel{Result_SFS}
\end{figure*}

\begin{figure*}[!h]
	\centering
	\includegraphics[width = 0.7\textwidth]{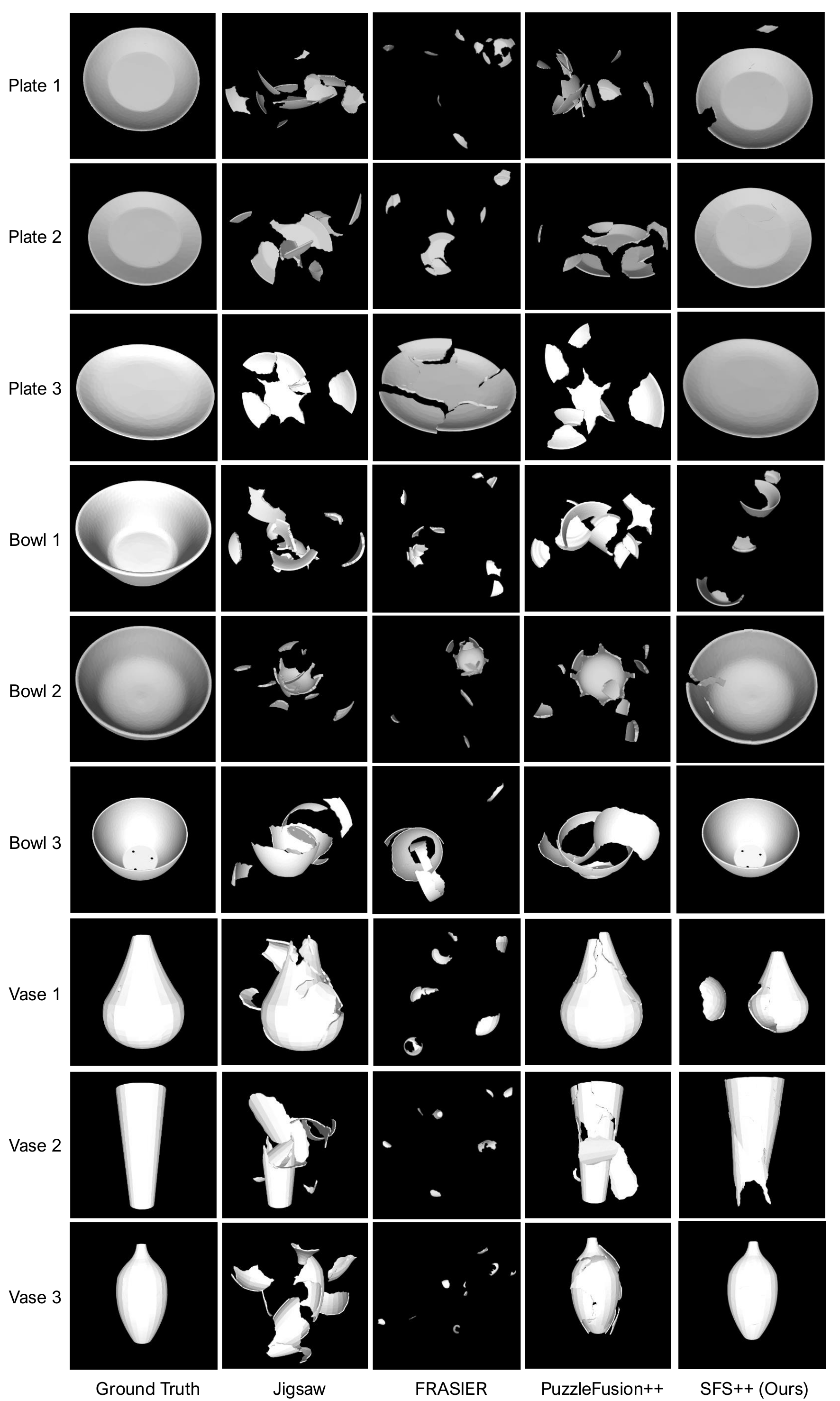}
	\caption{
   Comparison of visualized results from different models on Breaking bad dataset. 
    Each row represents a reassembly result for pots with 2 to 20 pieces, as this range reflects the data used during training for deep learning models. 
    The last column shows our method (\AlgName with $k=5$ and $b=3$), demonstrating improved reassembly quality compared to other models like Jigsaw, FRASIER, and PuzzleFusion++.
    Ground truth models are provided for reference in the first column. 
}

	\figlabel{Result_BB}
\end{figure*}

\ifCLASSOPTIONcaptionsoff
  \newpage
\fi

\end{document}